\documentclass{article}
\usepackage[utf8]{inputenc}
\usepackage{amsthm}
\usepackage{xcolor}
\usepackage{graphicx,verbatim,array,multicol, subfigure, color, lscape, mathrsfs, dsfont,url}
\graphicspath{{figures/}}
\usepackage{psfrag, amsmath, amsfonts, epsfig, fancybox, setspace, soul, amsthm, longtable, threeparttable, threeparttablex, makecell, xcolor, pdflscape, tabu, booktabs, natbib, amssymb}
\usepackage[normalem]{ulem}
\usepackage{cancel,float}
\usepackage[utf8]{inputenc}

\input{common-defs}
\newtheorem{assumption}{Assumption}
\newtheorem{condition}{Condition}

\def\logit{\operatorname{logit}}
\def\eif{\text{EIF}}
\def\bm{\boldsymbol{m}}
\def\bba{\bar\ba}
\def\en{\epsilon_n}
\def\ofl{\underline{\omega}}
\def\bbA{\bar\bA}
\def\bI{\mathbb{I}}
\def\bbDelta{\bar{\bDelta}}
\def\bblambda{\bar\blambda}
\def\P{\mathbf{P}}
\def\hcKen{\hat\cK_{\en}}
\def\cKen{\cK_{\en}}
\newcommand{\PP}{\Prob}

\usepackage{mathpazo}        
\usepackage[margin=0.75in]{geometry}
\usepackage{setspace}
\usepackage{titlesec}
\usepackage[colorlinks=true, linkcolor=blue, citecolor=blue, urlcolor=blue]{hyperref}

\title{Vector incremental treatment effects}
\author{Denis Agniel$^{1}$, Max Rubinstein$^{2}$, Larry Han$^{3}$, Anthony Yu$^{4}$,\\ Molly T. Finnerty$^{5}$, Marcela V. Horvitz-Lennon$^{6}$, Sharon-Lise T. Normand$^{7}$\\[0.75em]
{\small $^{1}$RAND, Santa Monica \quad $^{2}$RAND, Pittsburgh \quad $^{3}$Brown University \quad $^{4}$RAND, Washington, DC}\\
{\small $^{5}$New York State Office of Mental Health \quad $^{6}$RAND, Boston \quad $^{7}$Harvard Medical School}}
\date{}

\begin{document}

\maketitle

\begin{abstract}
Causal questions involving multiple binary treatments that act together are poorly served by standard causal approaches. With $m$ treatments there are $2^m$ potential outcomes, many treatment combinations are rarely or never observed, and defining effects by fixing the entire population to a single treatment combination can require implausible positivity assumptions. To address these issues, we extend incremental propensity score interventions to vector-valued binary treatments by defining causal effects in terms of counterfactual distributions that scale the odds of each treatment component by a user-specified factor $\bdelta$. These distributions are highly customizable and may assign unique odds changes to each treatment. Because these marginal odds constraints do not uniquely determine the joint counterfactual distribution, we define the target distribution as the minimum-divergence projection of the observed law onto a set defined by the constraints, using a user-specified separable divergence and preserving the observed conditional support. The resulting parameter $\psi(\bdelta) = \E[Y^{\Q_{\bdelta}}]$ is identified under consistency and conditional mean exchangeability---crucially, this approach does not require every one of the $2^m$ treatment combinations to have positive conditional probability. For a specified class of smooth divergences, we derive the efficient influence function and construct cross-fitted one-step estimators that are $\sqrt{n}$-consistent and asymptotically normal under regularity conditions that imply that marginal treatment probabilities are either exactly 0, exactly 1, or bounded away from those values. We relax this assumption by specifying a slightly different estimand and deriving a corresponding estimator, and prove its asymptotic normality under a convergence-rate condition on the estimated active set. We also extend a sensitivity analysis to this setting to allow researchers to assess the robustness of estimates to unmeasured confounding. Applying the framework to $159{,}466$ New York State Medicaid beneficiaries with serious mental illness, we estimate how healthcare costs would change under uniform improvements in the odds of six recommended quality-of-care processes. We find that increasing recommended care is associated with lower total, inpatient, and mental-health cost and higher outpatient cost, consistent with a substitution toward ambulatory care.
\end{abstract}

\section{Introduction}\label{sec:intro}
Researchers are often interested in causal effects when many binary treatments or exposures act together---for example, when policies shift several factors at once, when interactions among treatments matter, or when the intervention is best described as a vector of binary components rather than a single (possibly multi-valued) variable. The same structure arises when a construct of interest (e.g. quality of health care) is not directly measurable and we only observe a series of binary indicators. We may then ask how downstream outcomes such as mortality or comorbidity would change if that construct improved. Such questions often arise in epidemiology, policy evaluation, and the social sciences, yet standard causal contrasts are poorly suited to this setting.

With $m$ binary treatments there are up to $2^m$ potential outcomes and a proliferation of possible average treatment effects. With so many contrasts the preferred target of inference may be unclear. Marginal structural models are often used to reduce the dimension of moderate-dimensional treatments, but the resulting estimates may be difficult to interpret and may not match the scientific question. Moreover, defining causal quantities in terms of counterfactuals where the entire population adopts a single treatment combination is often implausible, implying positivity violations at the population level. Furthermore, even absent population-level positivity violations, for a finite sample, many of the $2^m$ treatment combinations may be rarely or never observed. This leads to limited within-sample overlap (a practical positivity violation), and makes standard nonparametric identification and estimation difficult.

To address these issues, we extend incremental propensity score interventions \citep{kennedy_nonparametric_2019, bonvini_incremental_2023} to allow for a vector of binary treatments. We define causal effects by contrasting the mean outcome under a counterfactual distribution in which each treatment's odds are scaled by a user-specified factor, relative to the observed distribution. Unlike in the single-treatment setting, specifying how treatment odds are scaled does not uniquely identify a counterfactual treatment distribution. From among the distributions satisfying the user-specified constraints, we select the closest one to the observed treatment distribution, where closeness is determined by a user-specified divergence (e.g.\ Kullback--Leibler). 

To be precise, consider the simplest version of this problem: two binary treatments $A_1, A_2$ with a strong positive dependence, say $\PP_0(A_1{=}A_2{=}0)=0.600$, $\PP_0(A_1{=}0,A_2{=}1)=0.050$, $\PP_0(A_1{=}1,A_2{=}0)=0.050$, $\PP_0(A_1{=}A_2{=}1)=0.300$ (both marginal probabilities equal $0.35$). Suppose we want to ask what would happen if the odds of $A_1$ doubled and the odds of $A_2$ tripled---an odds-scaling vector $\bdelta=(2,3)$, giving target marginals $\zeta_1=0.5185$, $\zeta_2=0.6176$. Scaling each treatment's marginal odds pins down two numbers but not the full $2\times2$ joint distribution consistent with those marginals: infinitely many joint laws share the same two marginals. Table~\ref{tab:distributions_clean} contrasts two natural choices. The distribution \emph{closest to $\PP_0$} in forward Kullback--Leibler divergence, among those satisfying the odds constraints, preserves the observed odds ratio exactly $\left(\frac{\PP_0(A_2 = 1 | A_1 = 1)/\PP_0(A_2 = 0 | A_1 = 1)}{\PP_0(A_2 = 1 | A_1 = 0)/\PP_0(A_2 = 0 | A_1 = 0)}\text{ = 72, unchanged}\right)$, and hence the strong association between $A_1$ and $A_2$: the conditional probability $\PP_0(A_2{=}1\mid A_1{=}1)$ moves only from $0.857$ to $0.960$. The \emph{independence} distribution satisfying the same marginal constraints collapses this association almost entirely (odds ratio $1$), to $0.618$---an artifact of how we select among the infinite possible distributions, arguably not of the intervention $\bdelta$ itself. This is the central technical problem the paper solves: pinning down a single, well-justified joint counterfactual distribution from among the infinitely many laws consistent with a vector of marginal odds constraints. Table~\ref{tab:distributions_clean} also reports the projections under three further admissible divergences---reverse KL, $\chi^2$, squared Hellinger---that we use in Section~\ref{sec:choosing-divergence} to build intuition for the implications of this choice.

\begin{table}[h]
\centering
\small
\caption{Observed distribution and its divergence-projection under $\bdelta = (2,3)$, for all four admissible divergences and the independence baseline.}
\begin{tabular}{|c|c|c|c|c|c|c|}
\hline
$(A_1, A_2)$ & $\PP_0$ & Forward KL & Reverse KL & $\chi^2$ & Hellinger & Independence \\
\hline
$(0,0)$ & 0.6000 & 0.3615 & 0.3560 & 0.3738 & 0.3581 & 0.1841 \\
$(0,1)$ & 0.0500 & 0.1200 & 0.1255 & 0.1076 & 0.1234 & 0.2974 \\
$(1,0)$ & 0.0500 & 0.0208 & 0.0264 & 0.0085 & 0.0242 & 0.1983 \\
$(1,1)$ & 0.3000 & 0.4977 & 0.4922 & 0.5100 & 0.4943 & 0.3203 \\
\hline
$\PP_0(A_2 = 1 \mid A_1 = 1)$ & 0.8571 & 0.9598 & 0.9492 & 0.9836 & 0.9533 & 0.6176 \\
Cross-product odds ratio & 72.00 & 72.00 & 52.96 & 208.07 & 59.25 & 1.00 \\
\hline
\end{tabular}
\label{tab:distributions_clean}
\end{table}

We define the counterfactual treatment law $\Q_{\bdelta}$ as the distribution satisfying the odds constraints that is closest to $\PP_0$ in a user-specified divergence, and we take as our target parameter the mean outcome this law would induce,
\begin{align}
    \psi(\bdelta) = \E[Y^{\Q_{\bdelta}}],\label{eqn:intro-estimand}
\end{align}
the counterfactual mean outcome under uniform (or component-specific) proportional shifts in treatment odds, selected to disturb the observed joint law as little as possible beyond the shift itself. Notice that when there is only a single treatment ($m=1$), the odds constraint alone pins down a unique distribution, so no divergence minimization is needed, and $\psi(\delta)$ reduces exactly to the incremental propensity score effect of \citet{kennedy_nonparametric_2019}. 

Our paper therefore makes four contributions.
\begin{enumerate}
    \item We define $\psi(\bdelta)$ via divergence-projection onto the set of laws satisfying vector-valued marginal odds constraints, characterize when a feasible projection exists, and show it is a multiplicative tilt of the observed propensity score indexed by at most $m+1$ multipliers (Proposition~\ref{prop:kkt}). 
    \item We identify $\psi(\bdelta)$ without positivity and derive its efficient influence function (Propositions~\ref{prop:identification} and~\ref{prop:eif}).
    \item We establish $\sqrt n$-consistency and asymptotic normality of a cross-fitted one-step estimator (Theorem~\ref{thm:onestep}), which handles marginals exactly at $0$ or $1$ by restricting to the active set and requires the others to stay a fixed distance from the boundary.
    \item We extend a sensitivity analysis to bound $\psi(\bdelta)$ under a bounded counterfactual risk ratio, with influence functions for the bounds (Proposition~\ref{prop:bounds}).
\end{enumerate}

By contrast, the existing literature that considers complex or multi-valued treatments is often framed via marginal structural models \citep{robins2000marginal}, propensity score weighting for multiple treatments \citep{li2019multiple}, factorial-design analogues \citep{pashley2023fractional}, or regression-based contrasts (see \citet{wang_causal_2024} for a survey). Those approaches typically require positivity and/or focus on parameters of a specified model. Incremental propensity score interventions \citep{kennedy_nonparametric_2019, bonvini_incremental_2023} avoid positivity by shifting odds rather than setting treatment to a fixed value. We extend incremental effects to a vector of treatments and uniquely pin down the joint counterfactual distribution by minimizing a divergence subject to the marginal odds constraints. Longitudinal and resource-constrained extensions of IPSIs are studied in \citet{sarvet_longitudinal_2023}, and estimand choice for multicomponent exposures has also been considered previously \citep{rudolph2026}. Multivariate incremental effects for continuous mixtures have also been proposed under a specific exponential tilt form \citep{huang2026}. Modified treatment policies (MTPs) and stochastic interventions also relax positivity by shifting the treatment rather than setting it \citep{munoz2012population, haneuse2013estimation, diaz2023lmtp, levis2024stochastic}, and two recent papers extend them to vector-valued treatments. \citet{jiang2025mvtp} define modified vector-valued treatment policies for continuous treatments and estimate effects via energy balancing weights. \citet{koo2026partial} study MTPs with multivariate and mixture exposures and give partial-identification bounds. Unlike MTPs, our odds-scaling interventions are  specified through the marginal treatment odds.

The rest of the paper is laid out as follows. Section~\ref{sec:setting} defines the setting, shows how to construct and compute $\Q_{\bdelta}$, and introduces the choice among admissible divergences at the estimand level. Section~\ref{sec:identification} covers non-parametric identification and derivation of the efficient influence function, and Section~\ref{sec:estimation} establishes the root-n consistency and asymptotic normality of the cross-fitted one-step estimator. Section~\ref{sec:sensitivity} considers sensitivity to unmeasured confounding, while Section~\ref{sec:sims} reports simulation results, and Section~\ref{sec:example} applies the method to estimate the effect of increasing/decreasing healthcare quality on costs among New York Medicaid beneficiaries with serious mental illness. We give some concluding remarks in Section~\ref{sec:conclusion}. 
\section{Setting and approach}\label{sec:setting}

Suppose that for each unit we observe a vector of binary treatments $\bA = (A_1, ..., A_m) \in \cA := \{0, 1\}^m$, an outcome $Y \in \R$, and pre-treatment covariates $\bX \in \cX \subseteq \R^p$, and we observe $n$ i.i.d. copies of $\bO = (\bX, \bA, Y) \sim \PP_0$. For any distribution $\PP$ of $\bA$ given $\bX = \bx$, write $\PP(\bx)$ for the conditional distribution $\PP(\cdot \mid \bX = \bx)$. Under $\PP_0$, the probability mass function is $\pi_\ba(\bx) = \PP_0(\bA = \ba \mid \bX = \bx)$, and the marginal treatment probabilities are $m_k(\bx) = \PP_0(A_k = 1 \mid \bX = \bx)$ for $k = 1,\dots,m$. Let $\bpi(\bx) = (\pi_\ba(\bx))_{\ba \in \cA}$ and $\bm(\bx) = (m_1(\bx),\dots,m_m(\bx))$. We define $\cP(\bx)$ as the set of all possible conditional distributions of $\bA$ given $\bX = \bx$, $S(\bx) = \supp(\PP_0(\cdot\mid\bx))$ the observed conditional treatment support, and $\mu_\ba(\bx) = \E[Y \mid \bX = \bx, \bA = \ba]$. Appendix~\ref{app:notation} collects the notation used throughout.

For any distribution $\Q(\cdot\mid\bx) \in \cP(\bx)$ with pmf $q_\ba(\bx)$ and marginals $\bzeta(\bx)$, the counterfactual mean under $\Q$ is $\E[Y^\Q] = \int_{\cX}\sum_{\ba\in\cA}\E[Y(\ba)\mid\bX=\bx]q_\ba(\bx)\,d\PP_0(\bx)$. For $\bdelta=(\delta_1,\ldots,\delta_m)\in(0,\infty)^m$, define the target marginal probability for component $k$ as the transformed marginal

\begin{align}
\zeta_{k,\bdelta}(\bx) = T_{\delta_k}\{m_k(\bx)\} := \frac{\delta_k m_k(\bx)}{1+(\delta_k-1)m_k(\bx)}, \qquad k=1,\ldots,m, \label{marginal-odds}
\end{align}
so that whenever $0<m_k(\bx)<1$, $\zeta_{k,\bdelta}(\bx)/\{1-\zeta_{k,\bdelta}(\bx)\} = \delta_k\, m_k(\bx)/\{1-m_k(\bx)\}$: $\delta_k$ multiplies the observed conditional odds of component $k$.\footnote{Unlike the odds ratio itself, $T_{\delta_k}$ is well-defined at the boundary: if $m_k(\bx)=0$ then $\zeta_{k,\bdelta}(\bx)=0$, and if $m_k(\bx)=1$ then $\zeta_{k,\bdelta}(\bx)=1$, for every finite $\delta_k>0$. We collect these into $\bzeta_{\bdelta}(\bx)=\{\zeta_{1,\bdelta}(\bx),\ldots,\zeta_{m,\bdelta}(\bx)\}$.}

Since $\bzeta_{\bdelta}(\bx)$ does not by itself determine a joint law for $\bA$, we restrict attention to joint distributions that satisfy the target marginal constraints while remaining within the observed conditional support $S(\bx)$:
\begin{align}
\cP_{\bdelta}^S(\bx) = \left\{\Q(\bx)\in\cP(\bx): q_\ba(\bx)=0 \text{ for } \ba\notin S(\bx),\ \textstyle\sum_{\ba\in\cA}a_kq_\ba(\bx)=\zeta_{k,\bdelta}(\bx),\ k=1,\dots,m\right\}.\label{feasible-set}
\end{align}
The first restriction prevents the intervention from assigning positive probability to a treatment configuration that does not occur at $\bX=\bx$, while the second requires it to attain the target marginal for each component. Restricting to $S(\bx)$ means some requested marginal shifts may not be feasible even though $T_{\delta_k}$ itself is always well-defined (we discuss this further in Section~\ref{sec:computing-q}). As illustrated in Table~\ref{tab:distributions_clean}, when $m\ge2$, $\cP_{\bdelta}^S(\bx)$ is typically not a singleton: it is an infinite collection of joint laws sharing the same $m$ marginal constraints. We must therefore select among them.

\subsection{Selecting $\Q_{\bdelta}(\bx)$}\label{sec:selecting-q}

When $S(\bx)=\cA$, the member of $\cP_{\bdelta}^S(\bx)$ that encodes independence between marginal treatment probabilities sets $q_\ba(\bx) = \prod_{k=1}^m \zeta_{k,\bdelta}(\bx)^{a_k}\{1-\zeta_{k,\bdelta}(\bx)\}^{1-a_k}$. While simple to construct, as Table~\ref{tab:distributions_clean} shows, this distribution typically lies far from $\PP_0(\cdot\mid\bx)$ because it discards the joint dependence structure of $\bA$ entirely. We instead select, at each $\bx$, the member of $\cP_{\bdelta}^S(\bx)$ closest to $\PP_0(\cdot\mid\bx)$ under a user-specified divergence $\D$, i.e. a function $\D:\cP(\bx)\times\cP(\bx)\to[0,\infty)$ that is non-negative, zero only at equality, and convex in its first argument:
\begin{align}
    \Q_{\bdelta}(\bx) = \Q_{\bdelta}(\cdot | \bx) = \argmin_{\Prob(\bx) \in \cP_{\bdelta}^S(\bx)} \D\left\{\Prob(\bx); \PP_0(\cdot|\bx)\right\}.\label{Q}
\end{align}
This construction preserves the aspects of $\PP_0$ not directly constrained by $\bdelta$, and Table~\ref{tab:distributions_clean} demonstrates one reason why this matters: the solutions that are minimized with respect to a divergence keep $\PP_0(A_2{=}1\mid A_1{=}1)$ closer to its observed value (and preserve other aspects of the joint distribution). In particular, the KL-divergence solution ensures that the odds ratios are exactly equal to the original distribution.

\subsection{Computing $\Q_{\bdelta}$}\label{sec:computing-q}

Finding $\Q_{\bdelta}(\bx)$ directly requires optimizing over a probability vector with up to $2^m$ free coordinates, infeasible for even moderate $m$. We restrict to divergences with the following separable form, which both simplifies computation and allows the derivation of the efficient influence function in terms of a closed, pointwise-tilt form.

\begin{condition}[Regular separable divergence]\label{cond:f-div}
    $\D$ takes the separable form
\[
\D\{\Q(\bx), \PP_0(\bx)\} = \sum_{\ba \in S(\bx)} \pi_\ba(\bx) \phi\left\{ q_\ba(\bx)/\pi_\ba(\bx) \right\},
\]
where $\phi:[0,\infty)\to\R\cup\{+\infty\}$ is proper, lower semicontinuous, strictly convex, finite and twice continuously differentiable on $(0,\infty)$, with $\phi(1)=\dot\phi(1)=0$ and $0<\ddot\phi(u)<\infty$ for $u>0$. (The normalization $\dot\phi(1)=0$ is without loss of generality, since adding an affine multiple of $u-1$ to $\phi$ leaves $\D$ unchanged, as both $\Q(\bx)$ and $\PP_0(\bx)$ are probability distributions.)
\end{condition}
Under Condition~\ref{cond:f-div}, \eqref{Q} is a convex program with affine constraints, so any feasible point satisfying the KKT conditions is optimal.
\begin{proposition}[KKT form]\label{prop:kkt}
Under Condition~\ref{cond:f-div}, let $h(t) = \argmax_{u\ge0}\{tu-\phi(u)\}$ be the unique multiplicative tilt function. Suppose that for some $\bdelta$ and $\bx$ there is a finite $\blambda\in\R^{m+1}$ (with the intercept implicit in $\ba^\top\blambda := \lambda_0+\sum_k a_k\lambda_k$) such that
\begin{align}
    q_{\ba}(\bx; \bdelta) = \pi_\ba(\bx) h\left\{\transpose{\ba}\blambda\right\}, \qquad \ba \in S(\bx),\label{q-delta-form}
\end{align}
and $q_{\ba}(\bx;\bdelta)=0$ for $\ba\notin S(\bx)$, satisfy \eqref{marginal-odds}--\eqref{feasible-set}. Then $q(\bx;\bdelta)$ is the unique solution to \eqref{Q} at $\bx$, and $\blambda$ are its Lagrange multipliers.
\end{proposition}
For forward KL divergence, $\phi(u)=u\log u - u + 1$ and $h(t)=e^t$, so \eqref{q-delta-form} takes the concrete multiplicative-exponential-tilt form
\begin{align}
    q_{\ba}(\bx;\bdelta) = \pi_\ba(\bx)\exp\{\ba^\top\blambda\},\label{eqn:kl-tilt-concrete}
\end{align}
i.e. an exponential tilt of the observed propensity score. For Pearson $\chi^2$ ($\phi(u)=(u-1)^2/2$), the convex-conjugate tilt is $h(t)=\max(0,1+t)$: past $t=-1$ the constrained solution clips to zero exactly, yielding $q_\ba(\bx; \bdelta) = 0$.

However, a solution need not exist for every $(\bdelta,\bx)$. Throughout the remainder of this paper, we therefore assume such a tilt exists:
\begin{condition}[Existence of tilt]\label{cond:tilt-existence}
For a given $\bdelta$ and each $\bx\in\cX$ there exists a finite $\blambda\in\R^{m+1}$ satisfying the premise of Proposition~\ref{prop:kkt}, which we write as $\blambda(\bpi(\bx);\bdelta)$.
\end{condition}
To see why this condition can fail, suppose $m=2$, $\PP_0(A_1{=}1,A_2{=}0\mid\bx)=0$, $m_1(\bx)=0.2$, $m_2(\bx)=2/3$. Any $\Q_{\bdelta}(\bx)$ sharing this support must satisfy $\Q_{\bdelta}(A_1{=}1\mid\bx)\le\Q_{\bdelta}(A_2{=}1\mid\bx)$. With $\bdelta=(6,0.5)$, \eqref{marginal-odds} demands $\Q_{\bdelta}(A_1{=}1\mid\bx)=0.6 > \Q_{\bdelta}(A_2{=}1\mid\bx)=0.5$: infeasible for every divergence. 

More generally, letting $\cM\{S(\bx)\}$ denote the marginal polytope of the observed support (in other words, $\operatorname{conv}\{\ba:\ba\in S(\bx)\}$), $\bzeta_{\bdelta}(\bx)\in\cM\{S(\bx)\}$ is necessary for Condition~\ref{cond:tilt-existence} and sufficient for divergences whose tilt can equal zero at the boundary (e.g.\ Pearson $\chi^2$). For other divergences, a slightly stronger requirement that the target marginals $\bzeta_{\bdelta}$ lie on the relative interior of the polytope is sufficient to ensure the condition. Feasibility  therefore depends only on the target marginal vector and the observed support, but whether that feasibility extends to a finite-multiplier solution still depends on which divergence is chosen. 

\subsection{The divergence as part of the estimand}\label{sec:choosing-divergence}

While $\bdelta$ determines the marginals, the divergence fills out the rest of the estimand. One way to select a divergence is to look at what characteristics it is likely to produce in the resulting counterfactual distribution. Forward KL is the unique member of the family with a composition property: applying the odds-scaling projection twice, $\Q_{\bdelta'}(\Q_{\bdelta}(\PP_0)) = \Q_{\bdelta'\bdelta}(\PP_0)$ (with $\bdelta'\bdelta$ the componentwise product), so that doubling the odds and then doubling again is the same as quadrupling them once. This ensures ``double, then double again'' and ``quadruple'' denote the same counterfactual law. No other divergence in Condition~\ref{cond:f-div} has this property. On the other hand, Pearson $\chi^2$ may set $q_\ba(\bx)$ to 0 even when $\pi_\ba(\bx) \neq 0$ and will in general more aggressively empty out small $\PP_0$ cells. As an example of this behavior, Table \ref{tab:divergence-extreme} revisits the example in Table \ref{tab:distributions_clean} but considers a more extreme $\bdelta = (2, 3.2)$. In this example, the $\chi^2$-based $\Q_{\bdelta}$ assigns more mass to $(0, 0)$ and $(1,1)$ than Forward KL and less to $(0, 1)$ and nearly none to $(1, 0)$, yielding an odds ratio of 1331 (compared to the original 72). Conversely, Reverse KL and Hellinger divergences yield counterfactual distributions that assign more mass to $(0, 1)$ and $(1, 0)$ and less to $(0, 0)$ and $(1,1)$. The four candidate laws share the same target marginals by construction, yet encode  different implicit beliefs about how $A_1$ and $A_2$ remain associated after the intervention.
\begin{table}[h]
\centering
\small
\caption{The same $\PP_0$ at the more aggressive target $\bdelta=(2,3.2)$: divergence choice now moves the odds ratio by up to $18\times$.}
\label{tab:divergence-extreme}
\begin{tabular}{|c|c|c|c|c|c|c|}
\hline
$(A_1, A_2)$ & $\PP_0$ & Forward KL & Reverse KL & $\chi^2$ & Hellinger & Independence \\
\hline
$(0,0)$ & 0.6000 & 0.3489 & 0.3423 & 0.3660 & 0.3448 & 0.1768 \\
$(0,1)$ & 0.0500 & 0.1325 & 0.1392 & 0.1155 & 0.1367 & 0.3047 \\
$(1,0)$ & 0.0500 & 0.0183 & 0.0250 & 0.0012 & 0.0225 & 0.1904 \\
$(1,1)$ & 0.3000 & 0.5002 & 0.4935 & 0.5173 & 0.4961 & 0.3281 \\
\hline
Cross-product odds ratio & 72.00 & 72.00 & 48.60 & 1330.5 & 55.70 & 1.00 \\
\hline
\end{tabular}
\end{table}

Whether the estimand $\psi(\bdelta)$ itself moves with the divergence, rather than just the fitted law, is governed by an exact identity (Appendix~\ref{app:additivity}): the two coincide when $m=1$ or, more generally, when the outcome model $\mu_\ba$ is affine on the support, and otherwise the gap is a specific linear functional of $\mu_\ba$ that can be large or small depending on $\mu_\ba$. In addition to changing the estimand, the choice of divergence can impact numerical stability of the estimator. We defer discussion of these numerical issues to Section \ref{sec:choosing-divergence-practice}.

\section{Identification and efficient influence function}\label{sec:identification}
In this section, we identify $\psi(\bdelta)$ and derive its EIF, which yields a single one-step estimator. Under regularity conditions stated below, this estimator is $\sqrt n$-consistent and asymptotically normal. One of the regularity conditions implies that non-boundary treatment marginals $m_k(\bX)$ must be bounded away from the boundary $\{0,1\}$. Appendix~\ref{sec:adaptive} gives an alternative estimand, at the cost of targeting a different quantity, for settings where that cannot be assumed. 

\subsection{Identification}

We require only two of the three standard causal inference assumptions to identify $\psi(\bdelta) = \E[Y^{\Q_{\bdelta}}]$. Specifically, we require
\begin{assumption}[Consistency]
    $Y(\ba) = Y$ if $\bA = \ba, \quad \ba \in \cA$.\label{consistency}
\end{assumption}
That is, the observed outcome equals the potential outcome under the treatment vector actually received (no systematic measurement error or interference). We also require 
\begin{assumption}[Ignorability]
    $Y(\ba) \indep \bA | \bX,\quad \ba \in \cA$.\label{ignorability}
\end{assumption}
Conditional on $\bX$, treatment is as-if random with respect to potential outcomes. With many treatments this assumes no unmeasured confounding for all $2^m$ potential outcomes. While this will hold by design in a randomized trial, this assumption may be strong in some observational settings. We therefore propose a sensitivity analysis to relax this assumption in Section \ref{sec:sensitivity}.

Consistency and ignorability are standard assumptions in the causal inference literature. We notably do not require the standard positivity assumption, which means that we allow $\pi_\ba(\bx) = 0$ for some $\ba \in \cA, \bx \in \cX.$ Where $\pi_\ba(\bx)=0$ the regression $\mu_\ba(\bx)$ is not defined. Throughout we adopt the convention $\mu_\ba(\bx)\pi_\ba(\bx):=0$ and, for tilts that clip at the boundary, $\dot h(\ba^\top\blambda)\pi_\ba(\bx):=0$ off $S(\bx)$, so that sums written over $\ba\in\cA$ agree numerically with sums over $\ba\in S(\bx)$. The intervention never places mass off $S(\bx)$, since \eqref{feasible-set} forces $q_\ba(\bx)=0$ there. This is particularly important in this setting where the number of positivity violations (or near violations) will necessarily be large. The space of potential treatments may be as large as $|\cA| = 2^m$. When $m$ is not small, many treatment vectors will never be observed in reasonable sample sizes, leading to practical positivity violations, even in settings with $\pi_\ba(\bx) > 0$ for all $\ba$. 

When $m_k(\bx)\in\{0,1\}$ for some $k$, the $k$-th marginal constraint in \eqref{marginal-odds} is already satisfied trivially (there is nothing left to scale), so the corresponding multiplier $\lambda_k$ may be fixed at $0$ and omitted from $\blambda$ without changing $\Q_{\bdelta}(\bx)$. Write
\[
\cK(\bx) := \{k: m_k(\bx)\in(0,1)\}
\]
for this \emph{population active set} of non-degenerate coordinates at $\bx$. In a slight abuse of notation, let $\blambda_\cK$, $h_{\cK} = h(\ba_{\cK}^\top\blambda_\cK) = h(\ba^\top\blambda_{\cK})$ denote the multiplier vector and tilt restricted to these coordinates, where in the final equality $\ba$ is implicitly understood to have the same restriction. 

Because $\zeta_{k,\bdelta}(\bx)=m_k(\bx)$ whenever $m_k(\bx)\in\{0,1\}$, for every finite $\delta_k>0$, the constraints indexed by $k\notin\cK(\bx)$ in \eqref{feasible-set} are implied by the support restriction $q_\ba(\bx)=0$ off $S(\bx)$ together with the normalization, and may be deleted from the program without changing its solution. Accordingly, for a coordinate set $\cS\subseteq\{1,\ldots,m\}$ write
\[
\bdelta_{\cS} := (\delta_k)_{k\in\cS}
\]
for the sub-vector of $\bdelta$ it indexes, and let $\psi(\bdelta_{\cS})$ denote the estimand obtained from \eqref{Q} subject to \eqref{marginal-odds}--\eqref{feasible-set} with the marginal constraints imposed for $k\in\cS$ only: the deleted marginals are left free to move under the divergence projection, so $\blambda_\cS$ solves a $(1+|\cS|)$-dimensional system and $\lambda_k=0$ for $k\notin\cS$. Since deleting a redundant constraint changes neither the solution nor its value, $\Q_{\bdelta_{\cK(\bx)}}(\bx)=\Q_{\bdelta}(\bx)$ at every $\bx$, and hence $\psi(\bdelta_{\cK})=\psi(\bdelta)$ exactly.

These assumptions allow us to identify $\psi(\bdelta)$ from the observed data. 
\begin{proposition}\label{prop:identification}
For a fixed $\bdelta$, under Assumptions \ref{consistency} and \ref{ignorability} and Condition~\ref{cond:tilt-existence} (applied to $\bdelta_{\cK(\bx)}$ at each $\bx$), we may identify $\psi(\bdelta)$ as
\begin{align}
    \psi(\bdelta) = \E\left[\sum_{\ba \in \cA} \mu_{\ba}(\bX)\pi_\ba(\bX) h(\ba^\top\blambda_{\cK})\right] = \E\left[Y\sum_{\ba \in \cA} \bI\{\bA = \ba\} h(\ba^\top\blambda_{\cK})\right].
\end{align}
\end{proposition}
The proof is in Appendix \ref{app:proofs}. This proposition shows two different ways of identifying $\psi(\bdelta)$. The first identifying expression operates through the outcome function $\mu_{\ba}(\bx)$, while the second operates through a bounded tilt weight on the observed outcome, with no inverse probability appearing, which is why positivity is not needed. These alternative ways of identifying $\psi(\bdelta)$ suggest that we may develop a doubly robust estimator which takes advantage of both of these paths to identification. We do so through the efficient influence function (EIF), which we derive in Section \ref{sec:eif}. 

Notice that, as in the scalar case, identification of $\psi(\bdelta)$ needs no positivity restriction on $\pi_\ba(\bx)$. Unlike the scalar case, which requires no feasibility condition, once $m\ge2$ Condition~\ref{cond:tilt-existence} is necessary: it can fail because target margins for treatment components that are logically or empirically linked can be jointly unattainable within the observed support. Importantly, this is a well-posedness condition on the requested target law, not a positivity-type identification condition. Thus, beyond consistency and ignorability, only this feasibility condition is needed to identify $\psi(\bdelta)$. Obtaining $\sqrt n$-inference additionally requires controlling how close the observed marginals $\bm(\bX)$ come to the boundary $\{0,1\}$, which we discuss in Remark~\ref{rmk:thm-scope} and Appendix~\ref{sec:margin-implications}.

\subsection{Influence function}\label{sec:eif}

The efficient influence function (EIF) is a fundamental quantity in modern causal inference, facilitating the construction of efficient estimators and understanding their theoretical properties \citep{kennedy2024semiparametric}. However, to ensure that it exists and is well-behaved, we impose four conditions. Loosely, the first two conditions ensure that the Jacobian for the system of equations encoding the marginal odds constraints \eqref{marginal-odds} is invertible, while the second two control its size.

\begin{condition}[Interiority]\label{assumption:interior}
For a given $\bdelta$ and each $\bx \in \cX$, the solution $\Q_{\bdelta}(\cdot|\bx)$ to \eqref{Q} satisfies $q_{\ba}(\bx;\bdelta) > 0$ for all $\ba \in \supp(\PP_0(\cdot|\bx))$. 
\end{condition}

Condition \ref{assumption:interior} is strictly stronger than Condition~\ref{cond:tilt-existence}: it requires not only that the tilt representation exists (finite $\blambda$) but that $q_\ba(\bx) > 0$ for all $\ba$ in the support, or equivalently, that $\bzeta_{\bdelta}(\bx)$ lies in the \emph{relative interior} of the marginal polytope, not merely inside it. Fortunately, interiority holds automatically for any divergence with $\lim_{u\rightarrow 0+}\dot\phi(u) = -\infty$ (e.g., KL, reverse KL, squared Hellinger), and is therefore a relatively weak condition. We also require that there are no redundant active constraints,

\begin{condition}[No redundant active constraints]\label{cond:nac}
For each $\bx \in \cX$, the intercept-augmented vectors restricted to the active coordinates, $\{\ba_{\cK(\bx)} : \ba \in \cA,\ \pi_\ba(\bx) > 0\}$, span $\mathbb{R}^{d(\bx)}$, where $d(\bx) = 1 + |\cK(\bx)|$. 
\end{condition}

\noindent This condition rules out deterministic linear relationships among the active treatment indicators conditional on $\bX = \bx$. This again is a relatively weak condition. Nevertheless, both conditions are formally required to derive Proposition \ref{prop:pointwiseinvert} below:

\begin{proposition}\label{prop:pointwiseinvert}
    Define the constraint system formalizing \eqref{marginal-odds} as $\bF(\bpi,\bdelta_{\cK};\blambda_{\cK}) = 0$, where,
\begin{align*}
    F_0(\bpi,\bdelta_{\cK};\blambda_{\cK}) &:= \sum_{\ba\in\cA} \pi_\ba(\bx)h(\ba^\top\blambda_{\cK}) - 1, \\
    F_k(\bpi,\bdelta_{\cK};\blambda_{\cK}) &:= \sum_{\ba\in\cA}a_k\pi_\ba(\bx)h(\ba^\top\blambda_{\cK}) - \frac{\delta_km_k(\bx)}{\delta_km_k(\bx) + 1 - m_k(\bx)}, \qquad k \in \cK(\bx),
\end{align*}
    and let $\bOmega(\blambda_{\cK}) = \bOmega\left\{\blambda(\bpi(\bx)),\bpi(\bx);\bdelta_{\cK}\right\} :=\frac{\partial F}{\partial \blambda_\cK} = \sum_{\ba\in\cA}\pi_\ba(\bx)\dot h(\ba^\top\blambda_{\cK})\bba_\cK\bba_\cK^\top$ be the constraint Jacobian defined with respect to the intercept and the active indices $\cK(\bx)$.
    Let $\omega_{\min}(\bXi)$ denote the minimum eigenvalue of some matrix $\bXi$. Under Conditions~\ref{cond:f-div}, \ref{assumption:interior}, and~\ref{cond:nac}, $\omega_{\min}(\bOmega) > 0$.
\end{proposition}

Proposition \ref{prop:pointwiseinvert} shows that Conditions~\ref{assumption:interior} and~\ref{cond:nac} are sufficient to ensure that the constraint Jacobian $\bOmega(\blambda_{\cK})$ is invertible for each $\bx$. This is important because the EIF is a function of the inverted matrix $\bOmega(\blambda_{\cK})$. These conditions therefore guarantee that the expression for the EIF given in Proposition \ref{prop:eif} below is well-defined. Proposition~\ref{prop:eif} also requires smoothness of the tilt and a uniform lower bound on the smallest eigenvalue of the Jacobian $\bOmega$. We state both for a generic active set $\cS$ and propensity $\bpi^*$, since Section~\ref{sec:estimation} applies them at estimated propensities and estimated active sets as well. Write $\bOmega(\bblambda',\bpi^*)=\sum_{\ba\in\cA}\pi^*_\ba(\bx)\dot h(\bba_\cS^\top\bblambda')\bba_\cS\bba_\cS^\top$ for the Jacobian of Proposition~\ref{prop:pointwiseinvert} with $\cK$, $\blambda_\cK$, and $\bpi$ replaced by $\cS$, $\bblambda'$, and $\bpi^*$.

\begin{condition}[Tilt regularity]\label{cond:tilt-regularity}
For active set $\cS(\bx)$ and propensity $\bpi^*$, let $\Lambda_\cS(\bx)$ be a convex set of multipliers containing $\bblambda_\cS(\bpi';\bdelta_\cS)$ for every $\bpi'$ on the segment joining $\bpi(\bx)$ and $\hat\bpi(\bx)$.
The scalar map $t\mapsto h(t)$ is twice continuously differentiable on $\{\bba_\cS^\top\bblambda':\bblambda'\in\Lambda_\cS(\bx),\ \bx\in\cX,\ \ba\in\cA\}$, with
\[
\sup_{\ba,\bx}\ \sup_{\bblambda'\in\Lambda_\cS(\bx)}\max\{|h|,|\dot h|,|\ddot h|\}(\bba_\cS^\top\bblambda')\le C_h<\infty.
\]
\end{condition}

\begin{condition}[Uniform nonsingularity]\label{cond:unif-bound}
For $(\cS,\bpi^*)$, there exists $\ofl>0$ such that
\[
\inf_{\bx\in\cX}\ \inf_{\bblambda'\in\Lambda_\cS(\bx)}\omega_{\min}[\bOmega\{\bblambda',\bpi^*(\bx)\}]\ge\ofl,
\]
with $\Lambda_\cS(\bx)$ as in Condition~\ref{cond:tilt-regularity}. In particular, this holds at $\bblambda_\cS(\bpi^*(\bx);\bdelta_\cS)$ itself.
\end{condition}

These conditions control the size of quantities that arise in the EIF. Condition \ref{cond:tilt-regularity} places restrictions on how big the tilt and its derivatives may become, and Condition \ref{cond:unif-bound} ensures that the minimum eigenvalue of the Jacobian is bounded below uniformly in $\bx$, which strengthens the pointwise positivity of Proposition~\ref{prop:pointwiseinvert}. Proposition~\ref{prop:eif} uses both conditions only at $(\cK,\bpi)$, where $\Lambda_\cK(\bx)$ need only contain $\bblambda_\cK(\bpi(\bx))$. The estimator of Section~\ref{sec:estimation} also evaluates them at $\hat\bpi$ on the same active set, so there $\Lambda_\cK(\bx)$ must contain the multipliers along the path from $\bpi(\bx)$ to $\hat\bpi(\bx)$ (Remark~\ref{rmk:segment-conditions}).

Proposition~\ref{prop:eif} also requires that the outcome and outcome regression be bounded, which makes $\varphi$ square-integrable and licenses differentiating under the integral sign.
\begin{assumption}[Bounded outcome and outcome regression]\label{asmpt:outcome-bounded}
There exists $M<\infty$ such that $|Y|\le M$ almost surely and $\sup_{\bx\in\cX}|\mu_\ba(\bx)|\le M$ for every $\ba\in\cA$.
\end{assumption}

\begin{proposition}\label{prop:eif}
For a fixed $\bdelta$, under Assumptions~\ref{consistency}--\ref{asmpt:outcome-bounded}, Conditions~\ref{cond:tilt-regularity}--\ref{cond:unif-bound} at $(\cK,\bpi)$, Conditions~\ref{cond:f-div}--\ref{cond:nac}, and treating $\cK(\bx)$ as known for all $\bx$, the (uncentered) efficient influence function of $\psi(\bdelta)$ is
\begin{align}\label{eqn:eifsimple}
\varphi(\bO;\bdelta) &= h(\bA^\top\blambda_{\cK})Y- \sum_{\ba\in\cA}Z_{\ba}(\bX)^\top\inv{\bOmega(\blambda_{\cK})} g(\bA_{\cK};\blambda_{\cK}, \bpi(\bX)),
\end{align}
where $Z_{\ba}(\bX) = \mu_\ba(\bX)\pi_\ba(\bX)\dot{h}(\ba^\top\blambda_{\cK})\ba_{\cK}$, $g(\bA_{\cK};\blambda_{\cK}, \bpi(\bX)) = h(\bA^\top\blambda_{\cK})\bA_{\cK} - \E_{\Q_{\bdelta}}[\bA_{\cK} \mid \bX] - \bDelta^2_{\bm}(\bX)(\bA_{\cK} - \E[\bA_{\cK} \mid \bX])$, and $\bDelta^2_{\bm}(\bX)$ is the diagonal matrix with zero in the intercept slot and $\delta_k/\{\delta_k m_k(\bx) + 1 - m_k(\bx)\}^2$ for active $k\in\cK(\bx)$. Since the model is nonparametric, the tangent space is all of $L^2_0(\PP_0)$ and the influence function is unique, hence efficient.

\end{proposition}

\begin{remark}[Known-active-set interpretation of the EIF]\label{rmk:oracle}
The EIF above is the EIF for $\psi(\bdelta_{\cK})$ with $\cK(\bx)$ treated as known for all $\bx$. Pathwise differentiability holds only once the coordinate set is fixed pointwise in $\bx$: if $\cK$ is replaced by an active set selected as a functional of the propensity score $\bpi$ or its estimate $\hat\bpi$, the resulting Lagrange multipliers are generally a nonsmooth functional of the law through the selected set. Section~\ref{sec:estimation} shows that, under the conditions of Theorem~\ref{thm:onestep}, the estimated active set recovers $\cK$ with probability tending to one, so inference for $\psi(\bdelta)$ is unaffected. The same derivation applies to any fixed measurable active set $\cS(\bx)$: under Conditions~\ref{cond:tilt-existence}--\ref{cond:unif-bound} for the reduced system on $\cS$ (with the other hypotheses of Proposition~\ref{prop:eif} unchanged), $\varphi(\bO;\eta,\bdelta_\cS)$ is the uncentered efficient influence function of $\psi(\bdelta_\cS)$. Section~\ref{sec:estimation} and the appendix use this at $\cS\in\{\cK_\epsilon,\hat\cK_\epsilon,\cKen,\hcKen\}$, each treated as fixed conditional on the training sample.
\end{remark}

\section{Estimation and inference}\label{sec:estimation}
In this section, we construct an estimator of $\psi(\bdelta)$ and establish its asymptotic normality. We write $\P f = \E_{\PP_0}[f(\bO)]$ for the population expectation of a measurable $f$, and $\P_n f = n^{-1}\sum_{i=1}^n f(\bO_i)$ for its empirical counterpart. The $L_2(\PP_0)$ norm is $\|f\|_2=\{\P f^2\}^{1/2}$, and the sup norm is $\|f\|_\infty=\sup_{\bx\in\cX}|f(\bx)|$. For smooth $h$, $\dot h,\ddot h$ denote first and second derivatives. We use $a_n\lesssim b_n$ for $a_n\le Cb_n$ (universal $C$), $\cO_p,o_p$ for stochastic boundedness and convergence to zero in probability, and $\rightsquigarrow$ for convergence in distribution. The nuisance functions $\hat\eta=(\hat\bpi,\hat\bmu)$ are fit on an independent sample or by cross-fitting. We write $\bO_1^n$ for the sample on which they are fit and, for a fitted function $\hat f$, $\P[\hat f]=\E[\hat f(\bO)\mid\bO_1^n]$, treating $\hat f$ as fixed.

For an active set $\cS(\bx)$, write $\varphi(\bO;\eta,\bdelta_\cS)$ for the influence function of Proposition~\ref{prop:eif} evaluated at $\cS$, with $\bdelta_\cS$ and $\blambda_\cS$ as in Section~\ref{sec:identification}. Let $b_k(\bx)=\min\{m_k(\bx),1-m_k(\bx)\}$ be the distance of the $k$-th marginal from the boundary and $b^\star(\bx)=\min_{k:b_k(\bx)>0}b_k(\bx)$ the distance of the nearest active marginal, with $b^\star(\bx)=\infty$ when $\cK(\bx)=\emptyset$.

The population active set $\cK(\bx)=\{k:b_k(\bx)>0\}$ is unknown, so we estimate it by thresholding the plug-in marginals $\hat m_k(\bx):=\sum_{\ba:a_k=1}\hat\pi_\ba(\bx)$ at a threshold $\en\in(0,1/2)$ that shrinks with $n$,
\[
\hcKen(\bx) := \{k: \en < \hat m_k(\bx) < 1-\en\},
\]
and write $\cKen(\bx):=\{k:\en<m_k(\bx)<1-\en\}$ for its population counterpart. Conditions~\ref{cond:tilt-regularity} and~\ref{cond:unif-bound} are imposed at the true active set $\cK$, so the threshold is not a regularization device. Its only role is to make $\hcKen=\cK$ with probability tending to one (Lemma~\ref{lem:exact-recovery}). The estimator is
\[
\hat\psi(\bdelta)=\P_n\varphi(\bO;\hat\eta,\bdelta_{\hcKen}).
\]

\subsection{Asymptotic normality}\label{sec:step0}
Write $\varphi_\cK=\varphi(\bO;\eta,\bdelta_\cK)$ and $\varphi_{\cKen}=\varphi(\bO;\eta,\bdelta_{\cKen})$, with hats for the versions in which $\hat\eta$ replaces $\eta$ and $\hcKen$ for the estimated set. Since $\varphi_\cK=\varphi(\bO;\eta,\bdelta)$ (constraints for $k\notin\cK(\bX)$ are redundant), the error of $\hat\psi(\bdelta)$ decomposes as

\begin{align}
\hat\psi(\bdelta)-\psi(\bdelta)
&= (\P_n-\P)\varphi_\cK \notag\\
&\quad+ \underbrace{\P[\varphi_{\cKen}-\varphi_\cK] + (\P_n-\P)[\varphi_{\cKen}-\varphi_\cK]}_{\text{(i) threshold crossing}} \notag\\
&\quad+ \underbrace{\P[\hat\varphi_{\hcKen}-\hat\varphi_{\cKen}] + (\P_n-\P)[\hat\varphi_{\hcKen}-\hat\varphi_{\cKen}]}_{\text{(ii) active-set estimation}} \notag\\
&\quad+ \underbrace{\P[\hat\varphi_{\cKen}-\varphi_{\cKen}] + (\P_n-\P)[\hat\varphi_{\cKen}-\varphi_{\cKen}]}_{\text{(iii) nuisance}}.\label{eqn:step0}
\end{align}
The leading term, $(\P_n-\P)\varphi_\cK$, is a centered empirical average, so it is $\cO_p(n^{-1/2})$ and asymptotically normal by the central limit theorem. Term (i) is zero for large $n$, term (ii) is zero with probability tending to one, and term (iii) is $\cO_p(R_n)+o_p(n^{-1/2})$, where $R_n$ is the second-order product rate defined in Theorem~\ref{thm:onestep}. The bounds use Conditions~\ref{cond:tilt-regularity} and~\ref{cond:unif-bound} of Section~\ref{sec:eif} together with Assumption~\ref{asmpt:outcome-bounded}.

In particular, Condition~\ref{cond:unif-bound} at $\cK$ keeps every active marginal at least $\ofl/C_h$ from $\{0,1\}$ (Appendix~\ref{sec:margin-implications}), so $\cKen=\cK$ identically once $\en<\ofl/C_h$ and term (i) is zero. Term (ii) vanishes because the same margin lets us recover $\cK$ from the estimated marginals with probability tending to one (Lemma~\ref{lem:exact-recovery}). Writing $\rho_n:=\sup_{k,\bx}|\hat m_k(\bx)-m_k(\bx)|$ for the sup-norm error in the estimated marginals, this requires only $\rho_n=o_p(\en)$. Term (iii) is the nuisance remainder, a standard quantity in many modern causal estimators. Conditions~\ref{cond:tilt-regularity}--\ref{cond:unif-bound} make $h$ and $\bOmega^{-1}$ Lipschitz-stable in the propensity score (Lemma~\ref{lem:Omega-stability}), so nuisance error enters through a second-order product-rate bound rather than a first-order one (Lemma~\ref{lem:second-order}).

\begin{theorem}[Asymptotic normality of the one-step estimator]\label{thm:onestep}
For the estimated-active-set estimator $\hat\psi(\bdelta)=\P_n\varphi(\bO;\hat\eta,\bdelta_{\hcKen})$, fix $\bdelta>0$ componentwise and assume Conditions~\ref{cond:f-div}--\ref{cond:unif-bound}, with Conditions~\ref{cond:tilt-regularity} and~\ref{cond:unif-bound} imposed at $(\cK,\bpi^*)$ for $\bpi^*\in\{\bpi,\hat\bpi\}$, Assumptions~\ref{consistency}--\ref{asmpt:outcome-bounded}, $\Var[\varphi(\bO;\eta,\bdelta)]>0$, sample splitting or cross-fitting, $L_2(\PP_0)$-consistency of $\hat\mu_\ba$ for every $\ba$, and that $\hat\bpi(\bx)$ takes values in the simplex over $\cA$ for every $\bx$. Let $\en\to0$ satisfy $\max_\ba\|\hat\pi_\ba-\pi_\ba\|_\infty=o_p(\en)$. Define the second-order product rate
\[
    R_n:=\max_\ba\|\hat\pi_\ba-\pi_\ba\|_2\max_\ba\|\hat\mu_\ba-\mu_\ba\|_2+\max_\ba\|\hat\pi_\ba-\pi_\ba\|_2^2,
\]
and assume $R_n=o_p(n^{-1/2})$. Define $\hat\sigma^2=\P_n[\{\varphi(\bO;\hat\eta,\bdelta_{\hcKen})-\hat\psi(\bdelta)\}^2]$. Then
\[
\sqrt n\{\hat\psi(\bdelta)-\psi(\bdelta)\}\rightsquigarrow\cN(0,\Var[\varphi(\bO;\eta,\bdelta)]),
\qquad
\hat\sigma^2\overset{\PP}{\to}\Var[\varphi(\bO;\eta,\bdelta)],
\]
so for each fixed $\alpha\in(0,1)$ the Wald interval $\hat\psi(\bdelta)\pm z_{1-\alpha/2}\hat\sigma/\sqrt n$ has asymptotic coverage $1-\alpha$ for $\psi(\bdelta)$.
\end{theorem}

While we have a product of nuisance errors in the remainder term, it would not be precisely correct to call this estimator doubly robust. Because $\psi(\bdelta)$ depends on $\bpi$ through the tilt $h(\ba^\top\blambda)$, not only through inverse-probability weighting, $\hat\bpi$ must be consistent for $\bpi$ on its own regardless of how well $\hat\bmu$ performs. Unlike standard doubly-robust estimators, this one is not consistent when $\bpi$ is inconsistently estimated but $\bmu$ is correct. The theorem separately assumes $L_2(\PP_0)$-consistency of $\hat\mu_\ba$ because the product-rate condition gives $L_2$-consistency of $\hat\pi_\ba$ but not of $\hat\mu_\ba$, and both are needed.

\begin{remark}[Limits of the regularity conditions]\label{rmk:regularity-limits}
Condition~\ref{cond:unif-bound} is one sufficient route to $\sqrt n$-inference on $\psi(\bdelta)$, not obviously a necessary one. Informally, the estimator may remain regular even where $\bOmega$ degenerates on some coordinates, provided the part of the influence function that would otherwise blow up is suitably \emph{aligned} with the directions of degeneracy, so that the correction term does not ``see'' the ill-conditioned directions. We do not develop this idea into a formal condition: whether it is genuinely weaker than Condition~\ref{cond:unif-bound}, and whether it could be made verifiable, remain open questions.
\end{remark}

When $\cK(\bx)$ is itself known and measurable, as with the clinical eligibility criteria of Section~\ref{sec:example}, no thresholding is needed and the argument for Theorem~\ref{thm:onestep} applies without the recovery step.

\begin{corollary}[CLT at a known active set]\label{cor:onestep-fixed}
Let $\cK(\cdot)$ be a known measurable function of $\bx$ with $\cK(\bx)=\{k:b_k(\bx)>0\}$ for every $\bx$, and let $\hat\psi(\bdelta_{\cK(\cdot)})=\P_n\varphi(\bO;\hat\eta,\bdelta_{\cK(\cdot)})$. Assume the hypotheses of Theorem~\ref{thm:onestep} other than the one involving $\en$. Then $\sqrt n\{\hat\psi(\bdelta_{\cK(\cdot)})-\psi(\bdelta)\}\rightsquigarrow\cN(0,\Var[\varphi(\bO;\eta,\bdelta)])$. The variance estimator $\hat\sigma^2=\P_n[\{\varphi(\bO;\hat\eta,\bdelta_{\cK(\cdot)})-\hat\psi(\bdelta_{\cK(\cdot)})\}^2]$ is consistent for $\Var[\varphi(\bO;\eta,\bdelta)]$ by Lemma~\ref{lem:variance-consistency}, so the corresponding Wald intervals attain asymptotic coverage $1-\alpha$.
\end{corollary}
Terms (i) and (ii) of \eqref{eqn:step0} do not arise here, and only the second-order nuisance term (iii) remains. If $\cK(\bx)$ takes finitely many values $\cK_1,\ldots,\cK_J$ with population weights $w_j=\PP_0[\cK(\bX)=\cK_j]$, then $\hat\psi(\bdelta_{\cK(\cdot)})$ is the pooled average $\sum_j\hat w_j\hat\psi_j(\bdelta_{\cK_j})$ of within-stratum one-step estimates $\hat\psi_j(\bdelta_{\cK_j})=n_j^{-1}\sum_{i:\cK(\bX_i)=\cK_j}\varphi(\bO_i;\hat\eta,\bdelta_{\cK_j})$, where $n_j$ is the number of observations in stratum $j$, $\hat w_j=n_j/n$, each observation uses the same fitted $\hat\eta$ (its fold's fit, under cross-fitting) in every stratum, and $\hat w_j\hat\psi_j:=0$ when the stratum is empty. The two expressions are equal for every $n$ (Appendix~\ref{app:proofs}), so the pooled average inherits the CLT. This known-active-set case is the one used in Section~\ref{sec:example} where each observation is only eligible for a certain subset of treatments, and $\cK_j$ denotes the eligibility set.

\begin{remark}[Scope of Theorem~\ref{thm:onestep}]\label{rmk:thm-scope}
Condition~\ref{cond:unif-bound} at $\cK$ already keeps every active marginal a fixed distance from $\{0,1\}$ (Appendix~\ref{sec:margin-implications}). The shrinking threshold recovers $\cK$ from estimated marginals and does not accommodate true marginals that approach the boundary. If Condition~\ref{cond:unif-bound} at $\cK$ is not realistic in a given situation, Appendix~\ref{sec:adaptive} gives an alternative estimator that targets a fixed-threshold estimand $\psi(\bdelta_{\cK_\epsilon})$ and may allow such near-boundary marginals. This estimand differs from $\psi(\bdelta)$ by a bias of known order under the margin condition (Condition~\ref{cond:margin}).
\end{remark}

\subsection{Choosing a divergence in practice}\label{sec:choosing-divergence-practice}

Section~\ref{sec:choosing-divergence} discussed theoretical reasons for choosing among divergences but deferred discussions of numerical reasons for choosing divergences. Condition~\ref{cond:unif-bound}, through which the divergence most directly enters the theory, can fail in two structurally different ways, governed by different features of $\PP_0$ and of the tilt derivative $\dot h$ in the tilt ratio $h=q_\ba/\pi_\ba$. We outline these two ways below, and give more detail in Appendix~\ref{app:regularity}.

On the one hand, as $m_k(\bx)$ approaches the $\{0, 1\}$ boundary (the case Theorem~\ref{thm:onestep} excludes, as discussed in Remark~\ref{rmk:thm-scope}), the corresponding values of $\pi_{\ba}(\bx)$ also go to 0 or 1. This then sends $\omega_{\min}(\bOmega)\to0$ at the same order for every divergence (Appendix~\ref{app:regularity}). Thus, this violation of Condition \ref{cond:unif-bound} does not give us an avenue for selecting among the divergences. On the other hand, when the marginal stays away from the boundary, but $\zeta_{\bdelta}$ nevertheless approaches $\{0, 1\}$, $q_\ba(\bx)$ will approach the boundary while $\pi_\ba(\bx)$ does not. The ill-conditioning of $\bOmega$ is governed by $\dot h$ in this case, a property of the divergence alone, not $\PP_0$. Divergences with more well-behaved $\dot h$ may maintain moderate conditioning of $\bOmega$ when other divergences allow $\mathrm{cond}(\bOmega)$ to blow up.

For example, return again to Table~\ref{tab:distributions_clean}, where we posit the treatment probability vector $\PP_0=(0.60,0.05,0.05,0.30)$. Fixing $\delta_1=2$ and increasing $\delta_2$ pushes $\zeta_2=T_{\delta_2}(0.35)\to1$. Since $S(\bx)=\cA$ has full support, $\cM\{S(\bx)\}=[0,1]^2$ and this sweep never becomes infeasible (Condition~\ref{cond:tilt-existence} holds throughout). It approaches the boundary of the closed polytope, forcing $q_{(0,0)}(\bx)\to0$ while $\pi_{(0,0)}(\bx)=0.6$ stays fixed. Table~\ref{tab:omega-sweep} reports $\mathrm{cond}(\bOmega)$ for all four divergences.
\begin{table}[h]
\centering
\caption{$\mathrm{cond}(\bOmega)$ as $\delta_2$ increases with $\delta_1=2$ fixed, on Table~\ref{tab:distributions_clean}'s $\PP_0$.}
\label{tab:omega-sweep}
\begin{tabular}{ccccc}
\hline
$\delta_2$ & Forward KL & Reverse KL & $\chi^2$ & Hellinger \\
\hline
3   & 28.4  & 32.6      & 27.0 & 27.5 \\
10  & 31.2  & 345.9     & 54.3 & 87.2 \\
30  & 79.1  & 3543.3    & 54.3 & 453.6 \\
100 & 252.4 & 41383.7   & 54.3 & 2791.0 \\
\hline
\end{tabular}
\end{table}
By $\delta_2=100$ ($\zeta_2=0.982$), $\chi^2$'s conditioning is flat at $54.3$---it stopped changing once the clipped cell first hit zero, near $\delta_2=10$ --- while reverse KL's has degraded by a factor of $762$ relative to $\chi^2$. Forward KL and Hellinger sit in between. This suggests $\chi^2$ may provide some robustness to boundary issues, but this comes at the expense of complying with the regularity conditions (Condition \ref{assumption:interior} requires no clipped cells), so our theoretical results would not hold and its performance is more uncertain. Forward KL stays more well behaved than the other approaches and may be considered a reasonable middle ground.

\section{Sensitivity analysis}\label{sec:sensitivity}

We now consider relaxing another key causal assumption: ignorability. Ignorability (Assumption~\ref{ignorability}) is what lets us equate $\psi(\bdelta)$ with the identified expression of Proposition~\ref{prop:identification}. When it fails, that expression instead equals some other quantity $\tilde\psi(\bdelta)$, potentially arbitrarily biased for $\psi(\bdelta)$.

We therefore propose a sensitivity analysis to evaluate the robustness of the findings to unmeasured confounding. We limit our analysis to settings where $Y$ is binary or bounded and rescaled to fall within zero and one, which is again a common setting for many applied problems. To do this, we adopt and extend the framework from \cite{luedtke2015statistics}, which considers the maximum deviation of the counterfactual outcome regression function from the observed regression function. To be precise, in place of no unmeasured confounding we instead define the quantity,
\[
\mu_\ba(\bx,\ba') = \E[Y(\ba) \mid \bA= \ba', \bx],
\]
and assume bounded counterfactual risk ratio:

\begin{assumption}[Bounded counterfactual risk ratio]\label{asmpt:bounded}
The outcome $Y$ takes values in $[0,1]$, and for some $\Gamma\ge1$, $\Gamma^{-1}\mu_\ba(\bx,\ba)\le\mu_\ba(\bx,\ba')\le\min\{\Gamma\,\mu_\ba(\bx,\ba),\,1\}$ for all $\ba,\ba'\in S(\bx)$, $\bx\in\cX$, where $\mu_\ba(\bx,\ba')=\E[Y(\ba)\mid\bA=\ba',\bx]$. \end{assumption}
This bound on $Y$ ensures that $\mu_\ba(\bx,\ba') \leq 1$. It costs no generality for a bounded outcome: if $Y\in[y_{\min},y_{\max}]$ for known $y_{\min}<y_{\max}$, the assumption holds for the rescaled outcome $(Y-y_{\min})/(y_{\max}-y_{\min})$, and the resulting bounds on $\psi(\bdelta_\cK)$ return to the original units as $y_{\min}+(y_{\max}-y_{\min})\times(\text{bound})$. Because $\Gamma$ bounds a \emph{ratio} of conditional means of the rescaled outcome, it is unchanged by multiplying $Y$ by a constant but not by shifting it: $\Gamma$ is measured from $y_{\min}$, so robustness values are comparable only for a fixed $y_{\min}$. When the outcome has a natural floor (e.g.\ costs, $y_{\min}=0$), that floor is the natural choice. Sections~\ref{sec:identification}--\ref{sec:estimation} already assume $Y$ and $\mu_\ba$ uniformly bounded. What is added here is the choice of origin and scale.

This assumption is quite weak, and is effectively guaranteed to hold for some value of $\Gamma$ as long as $\mu_\ba(\bx,\ba) = \mu_\ba(\bx)$ is uniformly bounded away from zero. $\Gamma=1$ recovers exact conditional mean exchangeability. Larger $\Gamma$ permits greater departures from it.
Under Assumption~\ref{asmpt:bounded}, $\psi(\bdelta_\cK)$ admits explicit lower and upper bounds $\psi_{\ell,\Gamma}(\bdelta_\cK)\le\psi(\bdelta_\cK)\le\psi_{u,\Gamma}(\bdelta_\cK)$ (Proposition~\ref{prop:bounds}, Appendix~\ref{app:sensitivity-eif}), derived treating $\cK$ as known. The lower bound is pathwise differentiable and root-$n$ estimable under the same conditions as Theorem~\ref{thm:onestep} (or Corollary~\ref{cor:onestep-fixed} when $\cK$ is known). The upper bound is \emph{not} pathwise differentiable, because of the indicator it contains, so no one-step estimator of it exists as stated---a mild margin condition on the density of $\Gamma\mu_\ba(\bX)$ (Appendix~\ref{app:asymptotic-theory}, Theorem~\ref{thm:master-expansion}) recovers one. In practice, $\Gamma$ can be increased from $1$ until the bound for the causal contrast of interest includes zero, giving a robustness value on the risk-ratio scale.

Interpreting these bounds requires a key conceptual change. The observed conditional treatment law
$\pi_{\ba}(\bx)$ is not itself a primitive intervention mechanism. It is a mixture over the latent assignment mechanism under the conditional distribution of the unmeasured confounders given $\bX$, namely
$\PP(\bA=\ba\mid \bx,\bU=\bu)$. Because this mixture averages over $\bU$, the observed propensity does not itself determine a unique intervention mechanism. Accordingly, the estimand $\psi(\bdelta)$ should instead be interpreted as a stochastic intervention: a policy that assigns treatment according to a prespecified distribution $\Q(\cdot\mid \bx)$, under an intervention in which treatment is independent of $\bU$ given $\bX$. Thus, $\Q_{\bdelta}$ is not a literal modification of the observed assignment mechanism, but an externally imposed treatment policy. To see the consequence of this, notice that under this interpretation, $\psi(\bdelta=1)$ need not equal $\E[Y]$. Heuristically, this follows because the policy at $\bdelta=1$ corresponds to an $\bX$-only randomized assignment law rather than the observed confounded mechanism. Future work could consider a sensitivity analysis that preserves the original intervention-based interpretation of the estimand.

\section{Simulations}\label{sec:sims}
We conducted two simulation studies to evaluate the proposed estimator and associated inference. Study 1 validates the implementation across divergences and tests robustness to nuisance estimation quality. Study 2 examines how the estimator behaves as we vary the sample size, the number of treatments, and the degree of overlap. We summarize performance by three metrics. The first is the bias $\bar{\hat\psi}(\bdelta) - \psi_0$ relative to the true parameter $\psi_0$. The second is the empirical standard error, meaning the standard deviation of $\hat\psi(\bdelta)$ across replications. The third is the coverage of the nominal $95\%$ Wald interval $\hat\psi \pm 1.96\,\hat\sigma/\sqrt{n}$. Here $\hat\sigma = \{\hat\Var[\varphi(\bO;\bdelta)]\}^{1/2}$ estimates the standard deviation of the EIF. Covariates and outcome errors are Gaussian, so the bounded-outcome assumption and the uniform margin in Condition~\ref{cond:unif-bound} hold only approximately. The simulations examine the estimator when these are mildly violated.

\subsection{Study 1: Divergences and nuisance estimation}
Study 1 tests whether the estimator is unbiased in finite samples, achieves the semiparametric efficiency bound with correctly specified nuisances, and remains robust as the nuisance convergence rate degrades. We generated data with $n = 1000$ observations, $m = 4$ binary treatments $A_1, \ldots, A_4$, and covariates $(X_1, X_2) \sim \cN(0,1)$. Treatment propensities were sequential:
\begin{align*}
\logit P(A_1 \mid \bX) &= 0.5X_1 + 0.3X_2, \\
\logit \PP(A_k \mid \bX, A_{<k}) &= 0.5X_1 + 0.3X_2 + 0.4\sum_{j<k} A_j, \quad k = 2,3,4.
\end{align*}
The outcome included main effects and all pairwise treatment interactions:
\[
Y = 1.0 + 0.3X_1 + 0.4X_2 + \sum_{k=1}^4 0.5 A_k + \sum_{k<j} 0.2 A_k A_j + \xi, \quad \xi \sim \cN(0, 1).
\]
The target marginals were selected to correspond to a 50\% increase in the odds of each treatment. We computed true parameter values via Monte Carlo integration ($10^6$ observations), yielding divergence-specific targets: $\psi_0 = 2.872$ (KL), $2.879$ (reverse KL), $2.854$ ($\chi^2$), and $2.876$ (Hellinger).

We tested five versions of the estimator across four divergences (KL, reverse KL, $\chi^2$, Hellinger). Two of them estimate the nuisance functions from data, one with a parametric model and one with XGBoost. The remaining three isolate the effect of the nuisance \emph{convergence rate}: they either plug in the true nuisances (``oracle'') or add mean-zero Gaussian noise to the oracle nuisances at a controlled rate. Because these three are consistent estimators that differ only in how fast their $L_2$ error vanishes, they let us probe the product-rate condition of Theorem~\ref{thm:onestep} directly, rather than through any particular working model.
\begin{enumerate}
    \item \textbf{Oracle:} True propensities and conditional outcome means from the known DGP.
    \item \textbf{Parametric (correct):} Correctly specified logistic models for sequential propensities and linear model for outcome ($L_2$ error $\cO_p(n^{-1/2})$).
    \item \textbf{XGBoost:} Gradient-boosted trees for both propensity and outcome models (no knowledge of true functional forms).
    \item \textbf{Consistent (fast rate):} Oracle nuisances perturbed by mean-zero Gaussian noise so that $L_2$ error is $\cO_p(n^{-0.3})$. The product condition then gives $R_n = o_p(n^{-1/2})$, so the rate is sufficient for $\sqrt{n}$-inference.
    \item \textbf{Consistent (slow rate):} The same construction at $\cO_p(n^{-0.1})$, giving $R_n = \cO_p(n^{-0.2})$, which is \emph{not} $o_p(n^{-1/2})$. This deliberately violates the rate condition.
\end{enumerate}

\begin{figure}[htbp]
\centering
\includegraphics[width=0.85\textwidth]{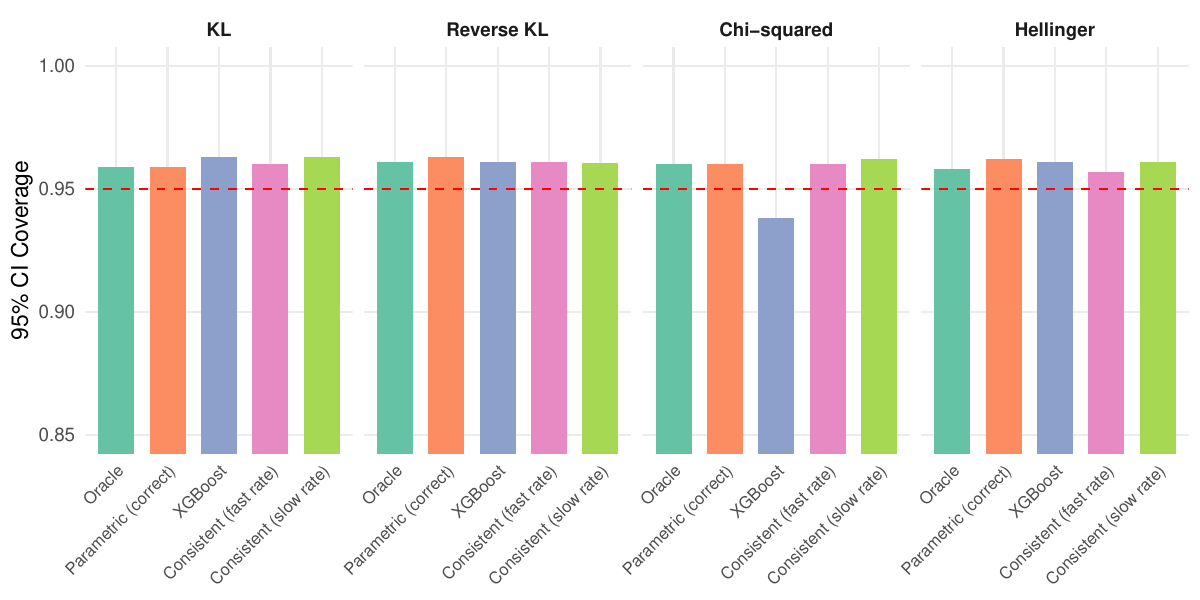}
\caption{Study 1: 95\% confidence interval coverage by nuisance estimation method and divergence. Dashed line at 0.95. All methods achieve nominal or near-nominal coverage across all four divergences. The lowest is $0.938$ (XGBoost with the $\chi^2$ divergence), and all others lie between $0.957$ and $0.963$. Each cell has $1000$ replications (Monte Carlo standard error $0.007$).}
\label{fig:study1_coverage}
\end{figure}

Across all methods and divergences the estimator was essentially unbiased, with every estimate within $0.93\%$ of the true parameter and coverage at or near the nominal $0.95$ (Figure~\ref{fig:study1_coverage}). The oracle and correctly specified parametric methods showed bias below $0.1\%$ under KL, reverse-KL, and Hellinger divergences and about $0.4\%$ under $\chi^2$. XGBoost exhibited a modest finite-sample bias of roughly $0.7\%$ ($0.57$--$0.93\%$ across divergences).

The most notable case is the slow-rate consistent method (item 5 above). Its $L_2$ nuisance error decays at only $\cO_p(n^{-0.1})$, so the product remainder is $\cO_p(n^{-0.2})$ and \emph{violates} the $o_p(n^{-1/2})$ rate condition of Theorem~\ref{thm:onestep}. Even so, it showed negligible bias ($+0.09\%$ under KL, at most $0.44\%$ under $\chi^2$) and near-nominal coverage, suggesting that the rate condition is sufficient but not necessary for this specific simulation design. In coverage terms (Figure~\ref{fig:study1_coverage}), all methods attained nominal or near-nominal CI coverage, with the lowest coverage, $0.938$, for XGBoost under $\chi^2$.

\subsection{Study 2: Scaling, dimensionality, and overlap}\label{sec:study2}

Study 2 examines how the estimator behaves as three practically important factors vary: the sample size $n$, the number of treatments $m$, and the degree of overlap. We generated data with $m$ binary treatments assigned sequentially,
\[
\logit \PP(A_k \mid \bX, A_{<k}) = \tfrac{1}{\kappa}\Big(0.5X_1 + 0.3X_2 + 0.4\textstyle\sum_{j<k} A_j\Big), \quad k = 1, \ldots, m,
\]
where $(X_1, X_2) \sim \cN(0,1)$. The overlap parameter $\kappa$ is an inverse scale on the propensity coefficients: larger $\kappa$ shrinks the coefficients toward zero (propensities near $0.5$, good overlap), while the smallest value, $\kappa = 1.2$, inflates them most (propensities nearest $\{0,1\}$) and yields the strongest imbalance. The outcome combined main effects and adjacent-pair interactions, normalized by $m$ so that the signal magnitude is comparable across dimensions. We crossed $n \in \{200, 500, 1000, 2000, 5000\}$, $m \in \{2,3,4,5,6\}$, and $\kappa \in \{1.2, 1.5, 2, 3, 5\}$ with target $\bdelta = (1.5, \ldots, 1.5)$ under KL divergence, running 1000 replications per design cell (each cell being one $(n, m, \kappa)$ combination). At each cell we compared two estimators, both using gradient-boosted (XGBoost) nuisances: the \emph{active-set} estimator of Theorem~\ref{thm:onestep} with a vanishing threshold, and the \emph{standard} one-step estimator over the full treatment set (Corollary~\ref{cor:onestep-fixed} with $\cK(\bx)=\{1,\ldots,m\}$). True parameters were computed by Gauss--Hermite tensor quadrature over $(X_1, X_2) \sim \cN(0,1)^2$, ranging from $\psi_0 = 1.413$ ($m=2$, $\kappa=5$) to $\psi_0 = 1.493$ ($m=6$, $\kappa=1.2$). Full specifications are in Appendix~\ref{app:simdetails}.

\begin{figure}[htbp]
\centering
\includegraphics[width=0.8\textwidth]{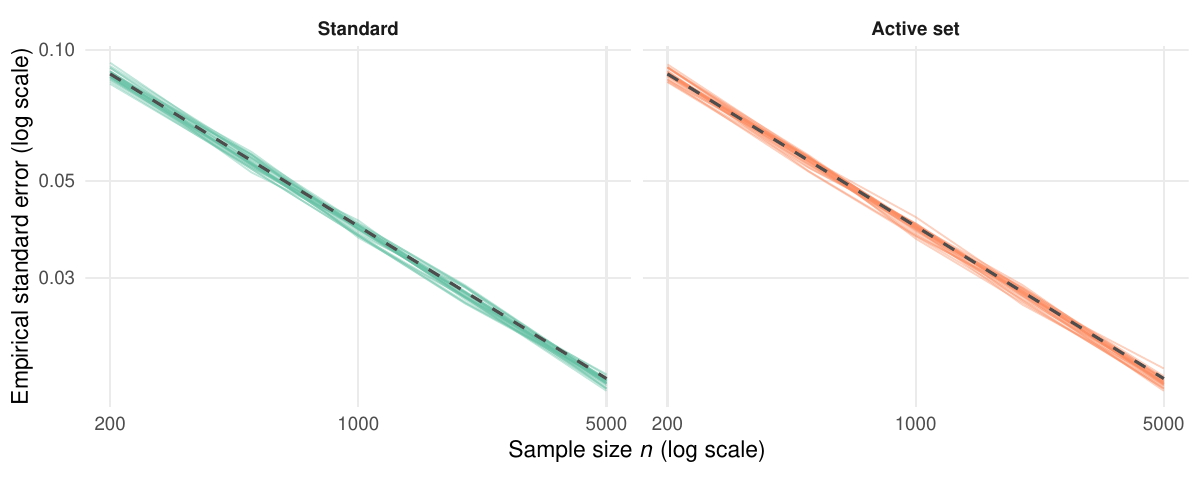}
\caption{Study 2: Empirical standard error versus sample size on a log--log scale, one line per fully-crossed $(m, \kappa, \text{method})$ combination ($50$ lines, with no averaging across settings), faceted by estimator. Within each panel every $(m, \kappa)$ line tracks the dashed reference of slope $-0.5$, confirming the $n^{-1/2}$ convergence rate uniformly across dimension and overlap (per-line slope summary in the text).}
\label{fig:study2_scaling}
\end{figure}

To assess the $n^{-1/2}$ convergence rate, Figure~\ref{fig:study2_scaling} plots empirical standard error against sample size on a log--log scale, one line for each of the $50$ fully-crossed $(m, \kappa, \text{method})$ combinations. Every line tracks the slope-$-0.5$ reference (median slope $-0.504$, range $[-0.523, -0.485]$), confirming the rate uniformly across the design. Bias was small throughout and governed more by sample size than by the number of treatments. The median absolute relative bias fell with sample size (from $1.6\%$ at $n = 200$ to $0.2\%$ at $n = 5,000$, with worst cells at $2.8\%$ and $0.4\%$, respectively) and rose modestly with dimension (from $0.3\%$ at $m = 2$ to roughly $0.9\%$ at $m = 6$). Within every cell the two estimators differed by at most $0.8$ percentage points. At $n=200$ the threshold $\epsilon_n\approx0.12$ drops the coordinates of some observations when $\kappa=1.2$, dropping fewer as $\epsilon_n\to0$.

\begin{figure}[htbp]
\centering
\includegraphics[width=0.8\textwidth]{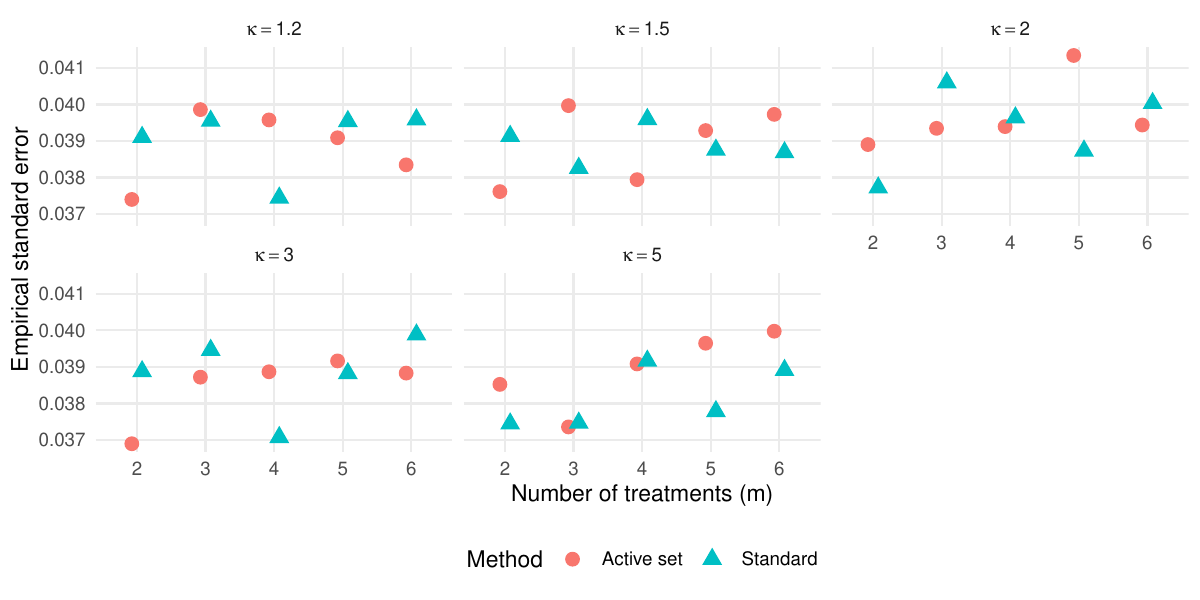}
\caption{Study 2: Empirical standard error by number of treatments ($m$) and overlap parameter ($\kappa$) at $n = 1000$. Standard errors are stable across dimension and nearly identical for the standard and active-set estimators.}
\label{fig:study2_m_beta}
\end{figure}

Figure~\ref{fig:study2_m_beta} shows standard error by number of treatments $m$ at fixed $n = 1000$. Because the outcome signal is normalized by $m$, standard errors were essentially flat in dimension ($0.037$ to $0.041$ across $m = 2$ through $6$) and nearly indistinguishable across the two methods, indicating that the tilting and active-set machinery added no appreciable variance as the treatment vector grows.

Wald coverage was near nominal across the full range of dimensions studied. Reporting each of the $250$ design-cell and estimator combinations separately, per-cell coverage had median $0.949$ at $m = 2$ and stayed near nominal through $m = 6$ (median $0.946$), with the full range across all cells $0.922$--$0.963$ (the lowest cell at $m = 5$, $n = 200$, $\kappa = 1.2$). The estimated standard errors were well calibrated to the empirical spread in every cell (per-cell ratio of the estimated $\hat\sigma/\sqrt n$ to the empirical standard error between $0.95$ and $1.10$, median $1.02$), confirming that coverage is governed by the (small) residual finite-sample bias rather than by any failure of variance estimation.

\section{Quality of care and outcomes among New York Medicaid patients with serious mental illness}\label{sec:example}

We illustrate the vector incremental approach using an administrative dataset spanning years 2016-2024 on New York State Medicaid beneficiaries with serious mental illness. Since quality of mental health care is not directly measurable, it is captured through a battery of binary quality indicators reflecting recommended care processes. We seek to determine how downstream outcomes would change if quality improved across the board.

We analyze $n = 159{,}466$ beneficiaries with serious mental illness who were eligible for at least one of $m = 6$ quality-of-care indicators drawn from the Medicaid Adult Core Set and were continuously enrolled in Medicaid for the three years surrounding the date of their first qualifying diagnosis of serious mental illness (the index date): the $12$ months before the index date, the $12$ months after it (the treatment window), and months $13$--$24$ after it (the outcome window). A qualifying diagnosis requires either at least one primary acute inpatient discharge or at least two primary or secondary outpatient or emergency-department claims on distinct service dates within a $12$-month period, with a qualifying ICD-10 code for schizophrenia or schizophrenia-spectrum disorders (F20, F22, F24, F25, F28, F29), bipolar disorder (F30, F31), or severe major depressive disorder (F32.3, F33.2, F33.3, or other F32--F33 codes accompanied by a severity indicator). The analysis is structured around three periods anchored by the index date. Baseline covariates are measured in the year before the index date, the quality indicators are measured over the year following the index date (the \emph{treatment window}), and the outcome accrues over the subsequent year (months $13$--$24$ after the index date). 

The quality indicators are process measures of recommended care: adherence to antipsychotic medications for individuals with schizophrenia, follow-up after hospitalization for mental illness (within 7 and within 30 days), antidepressant medication management (acute and continuation phases), and cervical cancer screening. Each indicator $A_k$ equals one if the recommended care was delivered, and zero otherwise. For the follow-up after hospitalization measures, where a beneficiary may have had multiple hospitalizations during the year, the indicator equals one if follow-up care was received after half or more of their discharges during the treatment year. Eligibility is determined by clinical criteria and varies across beneficiaries, so each beneficiary is eligible for a subset of the six indicators. We accommodate this by stratifying estimation on the observed eligibility pattern and aggregating across strata (Corollary~\ref{cor:onestep-fixed}). The outcome $Y$ is total healthcare cost (in dollars), covering both mental health and general medical care and excluding pharmacy, accrued over the $12$-month follow-up window. We decompose it into inpatient cost (hospital, nursing, and residential settings) and outpatient cost (professional, clinic, institutional, and emergency department services, together with ancillary services such as equipment, radiology, transport, and laboratory), an additive partition, and, as a non-additive overlay, mental-health cost: claims with a primary diagnosis of a mental disorder other than a substance use disorder (ICD-10 chapters F0 and F2--F9) or delivered in a specialty mental-health setting---psychiatric or mental-health facilities and clinics, New York's community mental-health programs (Personalized Recovery Oriented Services and Assertive Community Treatment), and crisis services, including mental-health emergency department visits---which span both inpatient and outpatient care and so do not sum to the total.\footnote{The corresponding setting labels in the claims data are Psychiatric, Mental Health, PROS, ACT, Crisis/CPTE, and the mental-health emergency department category. Substance use treatment settings (OASAS, Substance, Detox, Methadone) are excluded.} We adjust for $66$ covariates, including calendar time, demographics, psychiatric and medical comorbidities, baseline medication use, and prior healthcare utilization, including baseline engagement in mental-health outpatient care (Appendix~\ref{app:covariates} defines each group). 

Table~\ref{tab:descriptives} summarizes the six indicators. Eligibility is highly heterogeneous, ranging from $13.9\%$ of the cohort (follow-up after hospitalization) to $64.1\%$ (cervical cancer screening). Among eligible beneficiaries, the recommended care was delivered for between $44\%$ and $75\%$. This variation in eligibility is what motivates the pattern-stratified estimation of Corollary~\ref{cor:onestep-fixed}: the $15$ distinct eligibility patterns---each with at least $500$ beneficiaries---together cover the entire cohort, and estimation stratifies on these patterns and aggregates across strata by the pooled average of Corollary~\ref{cor:onestep-fixed}. The cost outcomes are right-skewed and, for inpatient cost, heavily zero-inflated ($70\%$ of beneficiaries incur no inpatient cost), with mean total cost of roughly \$33{,}000 over the follow-up window.

\begin{table}[ht]
\centering
\caption{The six quality-of-care indicators among $n = 159{,}466$ beneficiaries: number eligible, eligibility rate (\% of cohort), and prevalence of recommended care among eligible beneficiaries (\%). Eligibility is determined by clinical criteria.}
\label{tab:descriptives}
\begin{tabular}{lrrr}
\hline
Quality indicator & $N$ eligible & Eligible (\%) & Delivered $\mid$ eligible (\%) \\
\hline
Cervical cancer screening & 102,274 & 64.1 & 62.9 \\
Follow-up after hospitalization for mental illness (30 days) & 22,192 & 13.9 & 75.3 \\
Follow-up after hospitalization for mental illness (7 days) & 22,192 & 13.9 & 58.2 \\
Antidepressant medication management (continuation) & 41,092 & 25.8 & 44.2 \\
Antidepressant medication management (acute) & 41,092 & 25.8 & 57.4 \\
Adherence to antipsychotic medications (schizophrenia) & 51,243 & 32.1 & 60.4 \\
\hline
\end{tabular}
\end{table}

We consider odds-scaling factors $\bdelta = \delta \mathbf{1}$ that apply uniformly to all six indicators, and we target the counterfactual mean cost $\psi(\bdelta) = \E[Y^{\Q_{\bdelta}}]$ under the divergence-minimizing distribution $\Q_{\bdelta}$ of \eqref{Q}, using KL divergence. Here $\delta > 1$ encodes a counterfactual world in which the odds of receiving each recommended care process are uniformly higher than observed, and $\delta < 1$ a world in which they are lower. Setting $\delta = 1$ recovers the observed-population mean. We estimate $\psi(\bdelta)$ with the cross-fitted known-active-set one-step estimator ($3$ folds), estimating the propensity scores and outcome regressions by gradient boosting, and report results across a grid of $\delta$ ranging from $0.5$ (halving the odds of recommended care) to $2.0$ (doubling them). Because eligibility varies across beneficiaries, we stratify on the observed eligibility pattern and, within each stratum, tilt only the treatments for which the patient is eligible (Corollary~\ref{cor:onestep-fixed}). The eligibility pattern is determined by events in the treatment window, such as a psychiatric hospitalization for the follow-up measures. We condition on the observed pattern and assume ignorability given the baseline covariates and that pattern. This treats the pattern as unaffected by the care we vary. We checked the constraint Jacobian $\bOmega$ of each \emph{reduced}, within-stratum system, and on this cohort those systems suggest that uniform nonsingularity holds empirically. Because the eligibility-defined active sets are known, Corollary~\ref{cor:onestep-fixed} applies when they coincide with the population active sets, and no thresholding is required.

\begin{figure}[htbp]
\centering
\includegraphics[width=0.95\textwidth]{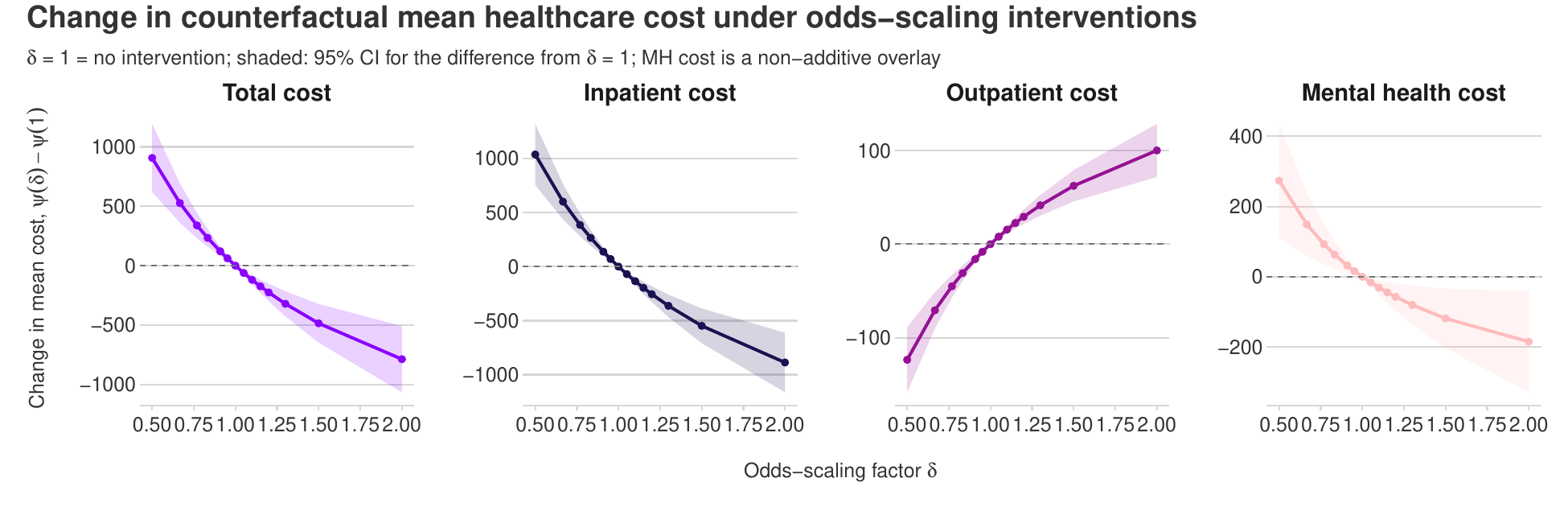}
\caption{Estimated change in counterfactual mean healthcare cost, $\E[Y^{\Q_{\bdelta}}]-\E[Y^{\Q_{\mathbf 1}}]$, as a function of the uniform odds-scaling factor $\delta$, for total, inpatient, and outpatient cost and, as a non-additive overlay, mental-health cost. Shaded bands are pointwise $95\%$ confidence intervals for the difference from $\delta=1$. The dashed line at $\delta = 1$ marks no intervention. Increasing the odds of recommended care lowers total, inpatient, and mental-health cost while raising outpatient cost.}
\label{fig:application}
\end{figure}

Figure~\ref{fig:application} and Table~\ref{tab:application} (Appendix~\ref{app:appresults}) summarize the estimated counterfactual mean costs. Increasing the odds of recommended care is associated with monotonically lower total cost: as $\delta$ rises from $0.5$ to $2.0$, estimated mean total cost falls from \$34{,}080 to \$32{,}387, a reduction of roughly \$1{,}700 (SE \$279; about $5\%$ of the observed-population mean of \$33{,}174 at $\delta = 1$). Of this, about \$790 (SE \$142; $2.4\%$) is the reduction from $\delta=1$ to $\delta=2$ and about \$910 (SE \$146) is the increase from $\delta=1$ down to $\delta=0.5$; the overall reduction from $\delta=0.5$ to $\delta=2$ has standard error \$279. This decline is driven by inpatient cost, which falls steeply from \$23{,}608 to \$21{,}683 over the same range. Outpatient cost moves in the opposite direction, rising from \$10{,}480 to \$10{,}703 as recommended care increases. Mental-health cost, the non-additive overlay, declines in parallel with inpatient cost, from \$12{,}574 to \$12{,}116. The overall pattern is consistent with a substitution mechanism in which improved ambulatory care processes---closer follow-up after hospitalization, better medication management, and routine screening---shift care away from costly inpatient admissions toward outpatient services, with a net reduction in total spending. Standard errors of the estimated mean cost at a single $\delta$ are about \$340--\$390 for total cost (Table~\ref{tab:application}, Panel~A), but estimates at different $\delta$ are strongly positively correlated, so the differences from $\delta=1$ are estimated far more precisely (Panel~B). Costs are unbounded and heavy-tailed, so the bounded-outcome assumption behind the theory does not hold exactly, and the reported standard errors should be considered asymptotic approximations. For reference, a bootstrap of the estimated influence function (holding the fitted nuisance functions fixed) gave standard errors within $3\%$ of the analytic ones, and percentile intervals that agree with the Wald intervals, for total cost at $\delta\in\{0.5,1,1.5,2\}$ and for the differences from $\delta=1$.

\section{Conclusion}\label{sec:conclusion}

We introduced incremental interventions for vector-valued binary treatments. The intervention multiplies the marginal odds of each treatment component by a prespecified factor, while selecting the resulting joint treatment distribution through a minimum-divergence projection of the observed law. Our construction resolves the non-uniqueness of a joint distribution with specified marginals and allows the intervention to preserve features of the observed dependence structure. We also avoid requiring full joint-treatment positivity.

We established identification of $\psi(\bdelta)=\E[Y^{\Q_{\bdelta}}]$ under consistency and ignorability and showed that the intervention distribution can be represented as a multiplicative tilt of the observed treatment mechanism. This facilitates derivation of the efficient influence function. We proposed a one-step estimator and established asymptotic normality under nuisance-rate, smoothness, and nonsingularity conditions. We handle exact boundary marginals by removing redundant coordinates and estimate the resulting active set with a shrinking threshold, which permits inference for the original estimand under suitable conditions. When the active set is known, as with the eligibility criteria in our application, no threshold is needed. An appendix extension defines a fixed-threshold estimand, a more stable but different target, for marginals that may approach the boundary. We also developed bounds under a bounded counterfactual risk-ratio model to assess sensitivity to violations of ignorability.

Several extensions warrant further investigation. First, it would be useful to characterize which interactions are preserved by common divergences and to develop scientifically interpretable ways to compare conclusions across divergence choices. Data-adaptive selection is also possible, but would define a data-dependent intervention and require accounting for the additional selection step in inference. Second, although the dual formulation reduces the optimization from as many as $2^m$ probabilities to at most $m+1$ multipliers, evaluating the constraint equations may still require summing over a large treatment space. Extensions that exploit graphical or other sparse structure in the joint treatment mechanism could make the approach practical for substantially larger $m$. A further extension could introduce an individual-level mechanism for changing one treatment vector into another. For example, an optimal transport-based formulation with a scientifically chosen cost function could favor interventions that alter fewer components or avoid implausible transitions. Finally, the same framework could be developed for categorical or ordinal treatment components and for longitudinal vector treatments. The longitudinal setting would require history-dependent marginal constraints, sequential support conditions, and repeated minimum-divergence projections \citep{sarvet_longitudinal_2023}. These extensions would broaden the range of interventions that can be studied while keeping the principle of shifting the treatment distribution in an interpretable direction and departing as little as possible from the observed joint law.

\section*{Acknowledgements and funding}
This work was supported by NIH grant R01MH136658. We thank Qingxian Chen and Junghye Jeong for their help assembling the data for analysis.

\section*{Code availability}
The \texttt{incremental} R package implementing the methods described in this paper is available at \url{https://github.com/denisagniel/incremental} (release v0.1.3; archived on Zenodo, \url{https://doi.org/10.5281/zenodo.23815244}).

\bibliography{vector-incremental}

@incollection{bonvini_incremental_2023,
	title = {Incremental causal effects: an introduction and review},
	shorttitle = {Incremental causal effects},
	booktitle = {Handbook of matching and weighting adjustments for causal inference},
	publisher = {Chapman and Hall/CRC},
	author = {Bonvini, Matteo and McClean, Alec and Branson, Zach and Kennedy, Edward H.},
	year = {2023},
	pages = {349--372},
}

@article{kennedy2020sharp,
  title={Sharp instruments for classifying compliers and generalizing causal effects},
  author={Kennedy, Edward H and Balakrishnan, Sivaraman and G’sell, Max},
  journal={The Annals of Statistics},
  volume={48},
  number={4},
  pages={2008--2030},
  year={2020}
}

@article{luedtke2015statistics,
  title={The statistics of sensitivity analyses},
  journal={U.C. Berkeley Division of Biostatistics Working Paper Series},
  author={Luedtke, Alexander R and Diaz, Ivan and van der Laan, Mark J},
  year={2015},
  publisher={bepress}
}

@article{kennedy_nonparametric_2019,
	title = {Nonparametric {Causal} {Effects} {Based} on {Incremental} {Propensity} {Score} {Interventions}},
	volume = {114},
	issn = {0162-1459, 1537-274X},
	url = {https://www.tandfonline.com/doi/full/10.1080/01621459.2017.1422737},
	doi = {10.1080/01621459.2017.1422737},
	language = {en},
	number = {526},
	journal = {Journal of the American Statistical Association},
	author = {Kennedy, Edward H.},
	month = apr,
	year = {2019},
	pages = {645--656},
}

@article{sarvet_longitudinal_2023,
	title = {Longitudinal {Incremental} {Propensity} {Score} {Interventions} for {Limited} {Resource} {Settings}},
	volume = {79},
	copyright = {http://creativecommons.org/licenses/by/4.0/},
	issn = {0006-341X, 1541-0420},
	url = {https://academic.oup.com/biometrics/article/79/4/3418/7587562},
	doi = {10.1111/biom.13859},
	language = {en},
	number = {4},
	journal = {Biometrics},
	author = {Sarvet, Aaron L. and Wanis, Kerollos N. and Young, Jessica G. and Hernandez-Alejandro, Roberto and Stensrud, Mats J.},
	month = dec,
	year = {2023},
	pages = {3418--3430},
}

@misc{wang_causal_2024,
	title = {Causal {Inference} with {Complex} {Treatments}: {A} {Survey}},
	shorttitle = {Causal {Inference} with {Complex} {Treatments}},
	url = {http://arxiv.org/abs/2407.14022},
	doi = {10.48550/arXiv.2407.14022},
	publisher = {arXiv},
	author = {Wang, Yingrong and Li, Haoxuan and Zhu, Minqin and Wu, Anpeng and Xiong, Ruoxuan and Wu, Fei and Kuang, Kun},
	month = jul,
	year = {2024},
	note = {arXiv:2407.14022 [stat]},
}

@article{kennedy2024semiparametric,
  title={Semiparametric doubly robust targeted double machine learning: a review},
  author={Kennedy, Edward H},
  journal={Handbook of statistical methods for precision medicine},
  pages={207--236},
  year={2024},
  publisher={Chapman and Hall/CRC}
}

@article{robins2000marginal,
  title = {Marginal structural models and causal inference in epidemiology},
  author = {Robins, James M. and Hernan, Miguel A. and Brumback, Babette},
  journal = {Epidemiology},
  volume = {11},
  number = {5},
  pages = {550--560},
  year = {2000},
  publisher = {LWW}
}

@article{rudolph2026,
  title={Everything all at once: On choosing an estimand for multi-component environmental exposures},
  author={Rudolph, Kara E and Inose, Shodai and Williams, Nicholas T and D{\'\i}az, Iv{\'a}n and Calderon, Lucia and Torres, Jacqueline M and Kioumourtzoglou, Marianthi-Anna},
  journal={Epidemiology},
  volume={37},
  number={6},
  pages={807--818},
  year={2026},
  doi={10.1097/EDE.0000000000002022}
}

@article{huang2026,
  title={Multivariate incremental effects for continuous treatments: Studying the health effects of environmental mixtures},
  author={Huang, Zhuochao and Dong, Kejin and Lin, Tuo and Antonelli, Joseph},
  journal={arXiv preprint arXiv:2604.23534},
  year={2026}
}

@article{levis2024stochastic,
  title={Stochastic interventions, sensitivity analysis, and optimal transport},
  author={Levis, Alexander W and Kennedy, Edward H and McClean, Alec and Balakrishnan, Sivaraman and Wasserman, Larry},
  journal={arXiv preprint arXiv:2411.14285},
  year={2024}
}

@article{haneuse2013estimation,
  title={Estimation of the effect of interventions that modify the received treatment},
  author={Haneuse, Sebastian and Rotnitzky, Andrea},
  journal={Statistics in medicine},
  volume={32},
  number={30},
  pages={5260--5277},
  year={2013},
  publisher={Wiley Online Library}
}

@article{munoz2012population,
  title={Population intervention causal effects based on stochastic interventions},
  author={Mu{\~n}oz, Iv{\'a}n D{\'\i}az and Van Der Laan, Mark},
  journal={Biometrics},
  volume={68},
  number={2},
  pages={541--549},
  year={2012},
  publisher={Oxford University Press}
}

@article{jiang2025mvtp,
  title={Exploring the effects of mechanical ventilator settings with modified vector-valued treatment policies},
  author={Jiang, Ziren and Crooke, Philip S and Marini, John J and Huling, Jared D},
  journal={arXiv preprint arXiv:2507.09809},
  year={2025}
}

@article{koo2026partial,
  title={Causal Effects of Modified Treatment Policies under Positivity Violations: A Partial Identification Approach},
  author={Koo, Taehyeon and Stuart, Elizabeth A and Rudolph, Kara E and Miles, Caleb H},
  journal={arXiv preprint arXiv:2608.23971},
  year={2026}
}

@article{diaz2023lmtp,
  title={Nonparametric causal effects based on longitudinal modified treatment policies},
  author={D{\'\i}az, Iv{\'a}n and Williams, Nicholas and Hoffman, Katherine L and Schenck, Edward J},
  journal={Journal of the American Statistical Association},
  volume={118},
  number={542},
  pages={846--857},
  year={2023}
}

@article{li2019multiple,
  title={Propensity score weighting for causal inference with multiple treatments},
  author={Li, Fan and Li, Fan},
  journal={The Annals of Applied Statistics},
  volume={13},
  number={4},
  pages={2389--2415},
  year={2019}
}

@article{pashley2023fractional,
  title={Causal inference for multiple treatments using fractional factorial designs},
  author={Pashley, Nicole E and Bind, Marie-Am{\'e}lie C},
  journal={Canadian Journal of Statistics},
  volume={51},
  number={2},
  pages={444--468},
  year={2023}
}
\bibliographystyle{plainnat}

\newpage 

\appendix

\section{Notation}\label{app:notation}

Table~\ref{tab:active-sets} collects the notation used throughout, grouped by topic, with the place where each symbol is introduced. The barred symbols $\bba_\cS$, $\bbA_\cS$ and $\bblambda_\cS$ are the reduced-coordinate, intercept-augmented versions (length $1+|\cS|$) of $\ba$, $\bA$ and $\blambda$ restricted to an active set $\cS$. By zero-padding, $\ba^\top\blambda_\cS=\bba_\cS^\top\bblambda_\cS$, so the two forms agree inside $h(\cdot)$. The barred forms are the ones contracted with $\bOmega$, $Z_\ba$ and $g$, and the main text drops the bar where the dimension is clear from context.

\begin{longtable}{>{\raggedright\arraybackslash}p{0.20\textwidth}>{\raggedright\arraybackslash}p{0.56\textwidth}>{\raggedright\arraybackslash}p{0.13\textwidth}}
\caption{Notation.}\label{tab:active-sets}\\
\toprule
Notation & Description & Introduced \\
\midrule
\endfirsthead
\toprule
Notation & Description & Introduced \\
\midrule
\endhead
\bottomrule
\endfoot
\multicolumn{3}{l}{\textit{Data, treatments and counterfactuals}}\\*
\midrule
$\bO=(\bX,\bA,Y)\sim\PP_0$ & Observed data: covariates in $\cX\subseteq\R^p$, binary treatment vector, outcome, where $\PP_0$ is the true law and we observe $n$ i.i.d.\ copies & Sec.~\ref{sec:setting} \\
$\cA=\{0,1\}^m$, $\ba$ & Treatment space with $m$ components, and a generic value in it & Sec.~\ref{sec:setting} \\
$\pi_\ba(\bx)$, $\bpi(\bx)$ & Propensity pmf $\PP_0(\bA=\ba\mid\bX=\bx)$ and its vector over $\cA$ & Sec.~\ref{sec:setting} \\
$m_k(\bx)$, $\bm(\bx)$ & Marginal $\PP_0(A_k=1\mid\bX=\bx)$ and its vector & Sec.~\ref{sec:setting} \\
$S(\bx)$ & Observed conditional support $\supp\{\PP_0(\cdot\mid\bx)\}$, distinct from the active-coordinate set $\cS$ & Sec.~\ref{sec:setting} \\
$\cP(\bx)$ & All conditional laws of $\bA$ given $\bX=\bx$ & Sec.~\ref{sec:setting} \\
$\mu_\ba(\bx)$, $\bmu(\bx)$ & $\E[Y\mid\bX=\bx,\bA=\ba]$ and the vector $(\mu_\ba(\bx))_{\ba}$ & Sec.~\ref{sec:setting} \\
$Y(\ba)$, $Y^{\Q}$ & Potential outcome under $\ba$ and outcome under treatment drawn from law $\Q$ & Sec.~\ref{sec:setting} \\
$\Q$, $q_\ba(\bx)$ & Generic counterfactual treatment law and its pmf, written $q_\ba(\bx;\bdelta)$ when the dependence on $\bdelta$ is shown & Sec.~\ref{sec:setting} \\
$\psi(\bdelta)$ & Estimand $\E[Y^{\Q_{\bdelta}}]$ & Sec.~\ref{sec:intro} \\
$\eta=(\bpi,\bmu)$, $\hat\eta$ & Nuisance functions and their estimates & Sec.~\ref{sec:estimation} \\
\midrule
\multicolumn{3}{l}{\textit{Odds scaling, divergence and tilt}}\\*
\midrule
$\bdelta$, $\delta_k$, $\delta$ & Odds-scaling factors, where $\delta$ is the common value when $\bdelta=\delta\mathbf 1$ (application) & Sec.~\ref{sec:setting} \\
$T_{\delta_k}$, $\zeta_{k,\bdelta}$, $\bzeta_{\bdelta}$ & Odds map $T_{\delta_k}(m_k)=\delta_km_k/\{1+(\delta_k-1)m_k\}$, and the target marginals $\zeta_{k,\bdelta}(\bx)=T_{\delta_k}\{m_k(\bx)\}$ & Sec.~\ref{sec:setting} \\
$\cP^S_{\bdelta}(\bx)$ & Laws supported on $S(\bx)$ that attain the target marginals $\bzeta_{\bdelta}(\bx)$ & Sec.~\ref{sec:setting} \\
$\Q_{\bdelta}(\bx)$ & Divergence projection of $\PP_0(\cdot\mid\bx)$ onto $\cP^S_{\bdelta}(\bx)$, equation \eqref{Q} & Sec.~\ref{sec:selecting-q} \\
$\D$ & User-specified separable divergence & Sec.~\ref{sec:selecting-q} \\
$\phi$, $\dot\phi$, $\ddot\phi$ & Convex generator of the divergence, and its derivatives (Condition~\ref{cond:f-div}) & Sec.~\ref{sec:computing-q} \\
$h$, $\dot h$, $\ddot h$ & Tilt function $h(t)=\argmax_{u\ge0}\{tu-\phi(u)\}$ and its derivatives & Sec.~\ref{sec:computing-q} \\
$\blambda\in\R^{m+1}$ & Lagrange multipliers with intercept $\lambda_0$, where $\ba^\top\blambda:=\lambda_0+\sum_ka_k\lambda_k$ & Sec.~\ref{sec:computing-q} \\
$\cM\{S(\bx)\}$ & Marginal polytope $\operatorname{conv}\{\ba:\ba\in S(\bx)\}$ & Sec.~\ref{sec:computing-q} \\
$r_\ba=q_\ba/\pi_\ba$ & Tilt ratio, $h(\ba^\top\blambda)$ & Sec.~\ref{sec:choosing-divergence-practice} \\
$s$, $c_s$ & Cressie--Read index and constant, $\ddot\phi(u)=c_su^{-s}$ & App.~\ref{sec:two-mechanisms} \\
\midrule
\multicolumn{3}{l}{\textit{Constraint system and regularity}}\\*
\midrule
$\bF=(F_0,F_k)$ & Normalization ($F_0$) and marginal-odds ($F_k$) constraints, $\bF(\bpi,\bdelta_\cK;\blambda_\cK)=\mathbf 0$ & Sec.~\ref{sec:eif} \\
$\bOmega$, $\omega_{\min}(\cdot)$ & Constraint Jacobian $\partial\bF/\partial\blambda_\cK$ and minimum eigenvalue of a symmetric matrix & Sec.~\ref{sec:eif} \\
$\Lambda_\cS(\bx)$ & Convex set of multipliers on which Conditions~\ref{cond:tilt-regularity}--\ref{cond:unif-bound} are imposed (the path from $\bpi$ to $\hat\bpi$ in the main text, enlarged in App.~\ref{app:lambda-def}) & App.~\ref{app:lambda-def} \\
$C_h$, $\ofl$ & Bound on $|h|,|\dot h|,|\ddot h|$ (Condition~\ref{cond:tilt-regularity}) and lower bound on $\omega_{\min}(\bOmega)$ (Condition~\ref{cond:unif-bound}) & Sec.~\ref{sec:eif} \\
$M$ & Bound on $|Y|$ and $|\mu_\ba|$ (Assumption~\ref{asmpt:outcome-bounded}) & Sec.~\ref{sec:eif} \\
$d(\bx)$ & $1+|\cK(\bx)|$, dimension of the reduced system (Condition~\ref{cond:nac}) & Sec.~\ref{sec:eif} \\
$Z_\ba$, $g$, $\bDelta^2_{\bm}$ & Terms of the EIF (Proposition~\ref{prop:eif}): $Z_\ba=\mu_\ba\pi_\ba\dot h\,\ba_\cK$, the centred score $g$, and the diagonal matrix with zero in the intercept slot and $\delta_k/\{\delta_km_k+1-m_k\}^2$ for $k\in\cK$ & Sec.~\ref{sec:eif} \\
$\beta$, $C$ & Margin exponent and constant (Condition~\ref{cond:margin}) & App.~\ref{sec:adaptive} \\
\midrule
\multicolumn{3}{l}{\textit{Active sets}}\\*
\midrule
$\cK(\bx)$ & Population active set $\{k: m_k(\bx) \in (0,1)\}=\{k:b_k(\bx)>0\}$, where $\cK(\bx)=\emptyset$ is allowed, with $b^\star(\bx)=\infty$ & Sec.~\ref{sec:identification} \\
$\cS(\bx)$ & A generic active-coordinate set, $\cS(\bx)\subseteq\{1,\ldots,m\}$ & Sec.~\ref{sec:identification} \\
$b_k(\bx)$, $b^\star(\bx)$ & Distance $\min\{m_k,1-m_k\}$ of $m_k(\bx)$ from $\{0,1\}$, and $\min_{k:b_k>0}b_k$. The quantity $b_k[\bpi^*](\bx)$ is the same distance computed from the propensity $\bpi^*$ & Sec.~\ref{sec:estimation} \\
$\cK(\cdot)$ & Known active set (as opposed to estimated): a measurable function of $\bx$ equal to $\cK(\bx)$ (Corollary~\ref{cor:onestep-fixed}) & Sec.~\ref{sec:step0} \\
$\en$, $\epsilon$ & Threshold shrinking with $n$ ($\en\in(0,1/2)$, $\en\to0$), and a fixed threshold & Sec.~\ref{sec:estimation}, App.~\ref{sec:adaptive} \\
$\cKen(\bx)$ & Population $\en$-thresholded active set $\{k: \en < m_k(\bx) < 1-\en\}$ & Sec.~\ref{sec:estimation} \\
$\hcKen(\bx)$ & Estimated $\en$-thresholded active set, using $\hat m_k(\bx)=\sum_{\ba:a_k=1}\hat\pi_\ba(\bx)$ & Sec.~\ref{sec:estimation} \\
$\cK_{\epsilon}(\bx)$, $\hat\cK_{\epsilon}(\bx)$ & Population and estimated fixed-$\epsilon$ active sets, $\{k:b_k(\bx)>\epsilon\}$ and its estimate & App.~\ref{sec:adaptive} \\
$\bdelta_{\cS}$ & Intervention restricted to an active set $\cS$ (any set, not only $\cS\subseteq\cK$): $\bdelta_\cS=(\delta_k)_{k\in\cS}$. The constraints for $k\notin\cS$ are deleted from the program, not reset to $\delta_k=1$ & Sec.~\ref{sec:identification} \\
$\psi(\bdelta_{\cS})$ & Estimand with the marginal constraints imposed on $\cS$ only, which equals $\psi(\bdelta)$ when $\cS=\cK$ & Sec.~\ref{sec:identification} \\
$\blambda_{\cS}$ & Multipliers $(\lambda_0,(\lambda_k)_{k\in\cS})$ in $\R^{m+1}$, with $\lambda_k=0$ for $k\notin\cS$ & Sec.~\ref{sec:identification} \\
$\bblambda_{\cS}$ & Reduced-coordinate multipliers: intercept plus $\cS$ & App.~\ref{app:notation} \\
$\bba_{\cS}$, $\bbA_{\cS}$ & Reduced treatment vectors: intercept plus the active coordinates & App.~\ref{app:notation} \\
$\varphi(\bO;\eta,\bdelta_\cS)$ & Uncentred influence function of Proposition~\ref{prop:eif}, extended to any fixed $\cS$ in Remark~\ref{rmk:oracle}, with nuisances $\eta=(\bpi,\bmu)$ & Sec.~\ref{sec:eif} \\
\midrule
\multicolumn{3}{l}{\textit{Estimation and inference}}\\*
\midrule
$\P f$, $\P_nf$ & $\E_{\PP_0}f(\bO)$ and $n^{-1}\sum_if(\bO_i)$. The symbol $\P$ denotes expectation and $\PP$ probability & Sec.~\ref{sec:estimation} \\
$\|f\|_2$, $\|f\|_\infty$, $\|\cdot\|_{op}$ & $L_2(\PP_0)$ norm, sup norm and operator norm & Sec.~\ref{sec:estimation} \\
$\lesssim$, $\cO_p$, $o_p$, $\rightsquigarrow$ & Inequality up to a constant, stochastic boundedness and convergence to zero in probability, and convergence in distribution & Sec.~\ref{sec:estimation} \\
$\bO_1^n$ & Sample on which $\hat\eta$ is fit (an independent sample or the complementary fold) & Sec.~\ref{sec:estimation} \\
$\hat\psi(\bdelta)$ & One-step estimator $\P_n\varphi(\bO;\hat\eta,\bdelta_{\hcKen})$, and the known-active-set version is $\hat\psi(\bdelta_{\cK(\cdot)})$ & Sec.~\ref{sec:estimation} \\
$\hat\sigma^2$, $\alpha$, $z_{1-\alpha/2}$ & Empirical variance of the plug-in score, nominal level of the Wald interval, and normal quantile & Sec.~\ref{sec:step0} \\
$\gamma$, $\tilde r_n$ & Rate exponent in $\rho_n=\cO_p(n^{-\gamma})$ (Corollary~\ref{cor:adaptive}), and a deterministic sequence with $\rho_n=\cO_p(\tilde r_n)$ & App.~\ref{sec:adaptive} \\
$\rho_n$ & Sup-norm error in the plug-in marginals, $\sup_{k,\bx}|\hat m_k(\bx)-m_k(\bx)|$, where $\rho_n=o_p(\en)$ gives $\hcKen=\cK$ with probability tending to one (Lemma~\ref{lem:exact-recovery}) & Sec.~\ref{sec:step0} \\
$R_n$ & Second-order product rate $\max_\ba\|\hat\pi_\ba-\pi_\ba\|_2\max_\ba\|\hat\mu_\ba-\mu_\ba\|_2+\max_\ba\|\hat\pi_\ba-\pi_\ba\|_2^2$ & Sec.~\ref{sec:step0} \\
$\cK_j$, $w_j$, $n_j$, $\hat w_j$, $\hat\psi_j$ & The $J$ values of $\cK(\cdot)$, their population weights and sample sizes, sample proportions, and within-stratum estimates & Sec.~\ref{sec:step0} \\
\midrule
\multicolumn{3}{l}{\textit{Sensitivity analysis}}\\*
\midrule
$\mu_\ba(\bx,\ba')$ & $\E[Y(\ba)\mid\bA=\ba',\bx]$, which equals $\mu_\ba(\bx)$ at $\ba'=\ba$ & Sec.~\ref{sec:sensitivity} \\
$\Gamma\ge1$ & Bound on the counterfactual risk ratio (Assumption~\ref{asmpt:bounded}), where $\Gamma=1$ is conditional mean exchangeability & Sec.~\ref{sec:sensitivity} \\
$\tilde\psi(\bdelta)$ & Functional identified when ignorability fails, $\E[\sum_\ba\mu_\ba\pi_\ba h(\ba^\top\blambda_\cK)]$ & Sec.~\ref{sec:sensitivity} \\
$\psi_{\ell,\Gamma}$, $\psi_{u,\Gamma}$ & Lower and upper bounds on $\psi(\bdelta_\cK)$ (Proposition~\ref{prop:bounds}) & Sec.~\ref{sec:sensitivity} \\
$\psi_{b,1}$, $\psi_{b,2}$ & Bias-bound components (Proposition~\ref{prop:eifbounds}) & App.~\ref{app:sensitivity-eif} \\
$\varphi_{1,\ba}$, $\varphi_{2,\ba}$, $\varphi_{u,\Gamma}$ & Scores for the bound components and for the upper bound & App.~\ref{app:sensitivity-eif} \\
$\varrho$, $\tau_n$, $V_n$ & Margin exponent at $\Gamma\mu_\ba=1$, $\max_\ba\|\hat\mu_\ba-\mu_\ba\|_\infty$, and extra upper-bound rate $\tau_n^{1+\varrho}$ & App.~\ref{app:asymptotic-theory} \\
\midrule
\multicolumn{3}{l}{\textit{Simulations}}\\*
\midrule
$\kappa$ & Overlap parameter: propensity coefficients are divided by $\kappa$ (Study 2) & Sec.~\ref{sec:study2} \\
$\psi_0$ & True parameter value, used to compute bias and coverage & Sec.~\ref{sec:sims} \\
$\xi$, $\sigma_\pi$, $\sigma_\mu$ & Standard normal noise in the outcome and in the perturbed nuisances, and noise scales in the consistent-nuisance scenarios & App.~\ref{app:study1} \\
\end{longtable}

For any $\cS(\bx)\subseteq\{1,\ldots,m\}$, $\bdelta_\cS=(\delta_k)_{k\in\cS}$ denotes the intervention imposing the marginal-odds constraint \eqref{marginal-odds} for $k\in\cS(\bx)$ only. The constraints for $k\notin\cS(\bx)$ are deleted from the program rather than reset to $\delta_k=1$, so $\blambda_\cS$ solves the reduced constraint system on $\cS$ with $\lambda_k=0$ for $k\notin\cS$, made explicit below. When $\cS(\bx)=\cK(\bx)$ the deleted constraints are redundant and $\psi(\bdelta_\cK)=\psi(\bdelta)$. For a thresholded or estimated $\cS(\bx)$ they need not be.

\begin{remark}[Constraint system]\label{rmk:constraints}
The Lagrange multipliers $\bblambda_\cK$ of Proposition~\ref{prop:kkt}, restricted to the active coordinates $\cK(\bx)$ plus intercept, are the root of the constraint system $\bF(\bpi(\bx),\bdelta_\cK;\bblambda_\cK)=\mathbf 0$ of Proposition~\ref{prop:pointwiseinvert}: $F_0$ is the normalization and $F_k$ equates the tilted mean of $a_k$ to $\zeta_{k,\bdelta}(\bx)$, matching the marginal-odds constraints of \eqref{feasible-set} restricted to $\cK(\bx)$. The sums there run over $\ba\in\cA$ rather than $\ba\in S(\bx)$ purely for notational uniformity. The two are numerically identical, since $\pi_\ba(\bx)=0$ off $S(\bx)$. The matrix $\bOmega(\bblambda_\cK) = \partial\bF/\partial\bblambda_\cK$ is the constraint Jacobian of Proposition~\ref{prop:pointwiseinvert}.
\end{remark}

\section{Regularity conditions and divergence family}\label{app:regularity}

This appendix collects the concrete implications of the regularity conditions for the divergence family used throughout the paper. Table~\ref{tab:divergences} records the tilt function and its first two derivatives for each divergence, as functions of the dual index $t=\bba^\top\bblambda$.

Condition~\ref{cond:tilt-existence} holds for every finite $\bdelta$ whenever $\PP_0(\cdot\mid\bx)$ has full support, i.e. $S(\bx)=\cA$: the marginal polytope of the full support is $[0,1]^m$, and $\bzeta_{\bdelta}(\bx)\in[0,1]^m$ for every finite $\bdelta$ by construction of $T_{\delta_k}$ (Section~\ref{sec:setting}). Feasibility can only fail once the observed support is restricted.

\begin{table}[h]
\centering
\caption{Tilt derivatives by dual index $t=\bba^\top\bblambda$.}
\label{tab:divergences}
\begin{tabular}{lccc}
\hline
Distance & $h(t)$ & $\dot h(t)$ & $\ddot h(t)$ \\
\hline
KL divergence & $\exp(t)$ & $\exp(t)$ & $\exp(t)$ \\
Reverse KL & $1/(1-t)$ & $1/(1-t)^2$ & $2/(1-t)^3$ \\
Pearson $\chi^2$ & $\max(0,1+t)$ & $1$ & $0$ \\
Squared Hellinger & $1/(1-t)^2$ & $2/(1-t)^3$ & $6/(1-t)^4$ \\
\hline
\end{tabular}
\end{table}

\subsection{Precise definition of $\Lambda_\cS(\bx)$}\label{app:lambda-def}

In the main text, Condition~\ref{cond:tilt-regularity} requires $\Lambda_\cS(\bx)$ to be a convex set containing $\bblambda_\cS(\bpi';\bdelta_\cS)$ for every $\bpi'$ in the convex hull of the propensities under consideration. This supplies the multiplier values visited by the second-order expansions in $\bpi$ of Lemma~\ref{lem:second-order}, which move the propensity across the hull and so traverse the corresponding solution path.

The results of Appendix~\ref{sec:adaptive} and Theorem~\ref{thm:master-expansion} compare tilted laws at different active sets, through Lemma~\ref{lem:tilted-law-stability}. For those results, Conditions~\ref{cond:tilt-regularity}--\ref{cond:unif-bound} are read with $\Lambda_\cS(\bx)$ enlarged to also contain $\bblambda_{\cS'',-k}(\bx)$ for every $\cS''\supseteq\cS$ and $k\in\cS''\setminus\cS$ arising in the nested chains of that lemma, where $\bblambda_{\cS'',-k}$ denotes $\bblambda_{\cS''}$ with its $k$th entry set to zero. These are the multiplier values visited by the active-set perturbation of Lemma~\ref{lem:tilted-law-stability}.

\begin{remark}[Checking Conditions~\ref{cond:tilt-regularity}--\ref{cond:unif-bound} on $\Lambda_\cS(\bx)$]\label{rmk:segment-conditions}
The path requirement of Condition~\ref{cond:tilt-regularity} comes at no additional cost in Condition~\ref{cond:unif-bound}. Fix $\bblambda'$ and write $\bpi_t=(1-t)\bpi+t\hat\bpi$. Then $\bOmega(\bblambda',\bpi_t)=(1-t)\bOmega(\bblambda',\bpi)+t\,\bOmega(\bblambda',\hat\bpi)$ because $\bOmega$ is linear in $\bpi$ at fixed multiplier, and $\omega_{\min}$ is concave on symmetric matrices, so $\omega_{\min}[\bOmega(\bblambda',\bpi_t)]\ge\ofl$ whenever the bound holds at $\bpi$ and $\hat\bpi$. Consequently $\|d\blambda_\cS(\bpi_t)/dt\|\lesssim\ofl^{-1}\|\hat\bpi(\bx)-\bpi(\bx)\|$, and a convex neighbourhood of $\blambda_\cS(\bpi(\bx))$ of that radius contains the path. For each divergence in Table~\ref{tab:divergences} the sup over such a neighbourhood reduces to its boundary. For KL, $h(t)=e^t$ and every derivative equals $h$. For reverse KL, $h(t)=1/(1-t)$ on $t<1$. For squared Hellinger, $h(t)=1/(1-t)^2$ on $t<1$. In each case $|h|,|\dot h|,|\ddot h|$ are non-decreasing in $t$, so a bound at the extreme points bounds the interior. For Pearson $\chi^2$ the tilt $h(t)=\max(0,1+t)$ is affine on $t>-1$, where $\dot h\equiv1$ and $\ddot h\equiv0$. Condition~\ref{assumption:interior} places the solution there, since $q_\ba(\bx;\bdelta)>0$ is exactly $\bba_\cS^\top\bblambda_\cS>-1$, so the kink at $t=-1$ is never attained and the differentiability requirement is met. The same interiority argument applies to any divergence whose tilt clips at the boundary.
\end{remark}

\subsection{What Condition~\ref{cond:unif-bound} implies at the true active set}\label{sec:margin-implications}

Applied at $\cS=\cK$, Condition~\ref{cond:unif-bound} is a restriction on $\PP_0$ that holds or fails at every fixed $n$. It is not an asymptotic statement. Combine it with the tilt-regularity bound $\dot h\le C_h$ (Condition~\ref{cond:tilt-regularity}) via the variational characterization of the minimum eigenvalue, $\omega_{\min}(M)\le v^\top Mv/\|v\|^2$ for any nonzero $v$: for any $k\in\cK(\bx)$, let $e_k\in\R^{d(\bx)}$ denote the standard basis vector for the coordinate-$k$ slot of the reduced system (Condition~\ref{cond:nac}), and $e_0$ the intercept slot, so that $\|e_k\|^2=1$ and $\|e_0-e_k\|^2=2$. Then
\begin{gather*}
e_k^\top\bOmega(\bblambda_\cK) e_k=\sum_{\ba:a_k=1}\pi_\ba(\bx)\dot h(\bba^\top\bblambda_\cK)\le C_hm_k(\bx),\\
\frac{(e_0-e_k)^\top\bOmega(\bblambda_\cK)(e_0-e_k)}{2}=\frac12\sum_{\ba:a_k=0}\pi_\ba(\bx)\dot h(\bba^\top\bblambda_\cK)\le \frac{C_h}{2}\{1-m_k(\bx)\}.
\end{gather*}
Bounding $\omega_{\min}(\bOmega(\bblambda_\cK))$ by each of these two quantities and minimizing over $k\in\cK(\bx)$ gives $\omega_{\min}(\bOmega(\bblambda_\cK))\le C_h\,b^\star(\bx)$ for every $\bx$. So Condition~\ref{cond:unif-bound} at $\cK$ forces
\[
b^\star(\bx)\ \ge\ \ofl/C_h\qquad\text{for every }\bx\in\cX:
\]
no active marginal may lie arbitrarily close to $\{0,1\}$ at any $n$. Consequently, $\cKen(\bx)=\cK(\bx)$ identically once $\epsilon_n<\ofl/C_h$, and the margin condition (Condition~\ref{cond:margin}) holds automatically, for every finite $\beta$.

This bound uses only the \emph{upper} half of the spectral relationship $\omega_{\min}(\bOmega)\lesssim b^\star$, and is one-directional: a hard margin does not imply Condition~\ref{cond:unif-bound}. Two or more active treatments assigned in near-lockstep can drive $\inf_{\bx\in\cX}\omega_{\min}(\bOmega)$ to zero through near-collinear constraint directions -- Condition~\ref{cond:nac} still holding at each individual $\bx$ (Proposition~\ref{prop:pointwiseinvert}) -- even when every marginal stays interior. This is a third mechanism, distinct from the two divergence-relevant mechanisms of Section~\ref{sec:two-mechanisms} below, which both concern a single coordinate or cell approaching a boundary rather than correlation across coordinates. Condition~\ref{cond:unif-bound} is therefore a joint margin-and-conditioning requirement, not a restatement of the margin condition under a different name.

\subsection{Two mechanisms of constraint-system degeneracy}\label{sec:two-mechanisms}

Two distinct degeneracies of the constraint Jacobian $\bOmega$ arise in this setting, governed by different features of the tilt derivative $\dot h$ in the tilt ratio $r_\ba(\bx):=q_\ba(\bx)/\pi_\ba(\bx)=h(\ba^\top\blambda)$ (Proposition~\ref{prop:kkt}).

\textbf{Marginal degeneracy} ($m_k(\bx)\to\{0,1\}$, a rare treatment component): here $\pi_\ba(\bx)\to0$ for the affected cells while $r_\ba(\bx)=\cO(1)$, so $\omega_{\min}(\bOmega)\to0$ at the same order, $\cO(m_k(\bx))$ (or $\cO(1-m_k(\bx))$), \emph{for every one of the four divergences of Table~\ref{tab:distributions_clean}}. Any apparent cross-divergence ranking under this mechanism is not comparable without first fixing a common normalization of $\phi$, since $\bOmega_{c\phi}=\bOmega_\phi/c$ for any $c>0$. We therefore do not rank the four divergences here.

\textbf{Target-boundary degeneracy} ($\bzeta_{\bdelta}\to\partial\cM\{S(\bx)\}$, an aggressive target near a structural zero): here $r_\ba(\bx)\to0$ in some cell for fixed $\pi_\ba(\bx)$, so $\dot h(\ba^\top\blambda)=1/\ddot\phi(r_\ba)\to0$ for the barrier divergences, at a rate governed by the Cressie--Read power-family index $s$ developed below. Unlike marginal degeneracy, this mechanism \emph{is} divergence-comparable without renormalization, because it is a property of $\ddot\phi$ (hence of $h$ alone) rather than a magnitude of $\bOmega$ that is being compared. Reverse KL degrades fastest and Pearson $\chi^2$ is unaffected while every cell stays positive (once a cell clips to zero, interiority, Condition~\ref{assumption:interior}, fails and the theory does not cover that regime): in a separate numerical illustration ($m=3$, a structural zero, $\bdelta=(1,\delta_2,1)$ as $\delta_2$ sweeps toward the polytope boundary), $\mathrm{cond}(\bOmega)$ for reverse KL reaches $\sim\!1.8\times10^6$ near the boundary while $\chi^2$ stays flat at $\sim\!46$, confirmed independently at a single interior point ($\mathrm{cond}(\bOmega)=45.6$ at $\bdelta=(1,2.8,1)$) -- the same direction and order of magnitude as the running example's own Table~\ref{tab:omega-sweep} above (forward-KL-to-reverse-KL and $\chi^2$-flat-vs-reverse-KL-degrading, though the two examples' absolute figures are not comparable to each other, since they differ in $m$, support, and target). The near-boundary digits in this illustration carry residual numerical error ($\sim\!10^{-5}$--$10^{-6}$) and should not be read as final to more than one significant figure.

The four divergences considered throughout this paper are members of the Cressie--Read power family, with $\ddot\phi(u)=c_s\,u^{-s}$ for $s\in\{0,1,\tfrac32,2\}$ (forward KL, Pearson $\chi^2$, squared Hellinger, reverse KL are $s=1,0,\tfrac32,2$ respectively) and constants $c_0=c_1=c_2=1$, $c_{3/2}=\tfrac12$ (squared Hellinger's constant is not $1$: $\phi(u)=(\sqrt u-1)^2$ gives $\ddot\phi(u)=\tfrac12u^{-3/2}$). Since $\dot h(\ba^\top\blambda)=1/\ddot\phi(r_\ba)=r_\ba^{\,s}/c_s$, writing $\bOmega_s(\bx) = \sum_{\ba\in S(\bx)}\pi_\ba(\bx)\,r_\ba(\bx)^{\,s}/c_s\,\bba\bba^\top$, the target-boundary mechanism above becomes a single rate rather than four separate facts: as the relevant cell's tilt ratio $r\to0$, $\omega_{\min}(\bOmega_s)\asymp r^{s}$ and $\mathrm{cond}(\bOmega_s)\asymp r^{-s}$, up to the divergence-specific constant $c_s$. Here $s=0$ ($\chi^2$) is unaffected by small but positive cells (an emptied cell violates interiority, Condition~\ref{assumption:interior}), $s\ge1$ (forward KL and above) is subject to the gradient barrier, and $s\ge2$ (reverse KL) is most severely zero-avoiding. This rate does \emph{not} extend to the marginal-degeneracy mechanism above, where $\omega_{\min}$ vanishes at order $\cO(m_k(\bx))$ for every $s$: the two mechanisms act through different features of $\bOmega$ -- the tilt weight $1/\ddot\phi(r)=r^s/c_s$ as $r\to0$, versus the mixing weight $\pi_\ba\to0$ at fixed $r$ -- and only the former reduces to a single $s$-indexed rate. Values $s\in(0,1)$, outside the four divergences considered in the main text, are barrier-free while still having $\dot h$ increasing in $r$, so no member of the family is simultaneously best under both mechanisms.

A divergence favorable under one mechanism need not be favorable under the other: Hellinger has a gradient barrier and so, like reverse KL, degrades under target-boundary degeneracy even though it is not being ranked under marginal degeneracy above. Neither mechanism is rare in the settings this paper targets: marginal degeneracy arises whenever a treatment component is uncommon. Target-boundary degeneracy arises routinely under structural zeros (Section~\ref{sec:example}'s eligibility-restricted indicators are one example of the underlying phenomenon, though there the known active set of Corollary~\ref{cor:onestep-fixed} removes the degenerate coordinates rather than requiring divergence-side robustness to them).

\section{Additional results: divergence additivity}\label{app:additivity}

Section~\ref{sec:choosing-divergence} noted that the choice of divergence $\phi$ can change $\psi(\bdelta)$. This appendix gives the exact identity governing that dependence and identifies when it vanishes.

\begin{proposition}[Divergence additivity]\label{prop:additivity}
Under Assumptions~\ref{consistency}--\ref{asmpt:outcome-bounded}, fix $\bdelta$ and let $\Q_{\bdelta,\phi},\Q_{\bdelta,\phi'}$ denote the divergence-closest laws \eqref{Q} under two divergences $\phi,\phi'$ satisfying Condition~\ref{cond:f-div}, each also satisfying Condition~\ref{cond:tilt-existence} (so both solutions exist), with pmfs $q^\phi_\ba(\bx),q^{\phi'}_\ba(\bx)$. Assume Condition~\ref{cond:nac} (no redundant active constraints). For each $\bx$, let $S(\bx)=\supp(\PP_0(\cdot\mid\bx))$, let $N(\bx)=\{\bv\in\R^{S(\bx)}: \textstyle\sum_{\ba\in S(\bx)}v_\ba=0,\ \sum_{\ba\in S(\bx)}a_kv_\ba=0\ \forall k\in\cK(\bx)\}$ be the null space of the marginal-constraint design restricted to $S(\bx)$, and let $\Pi_{N(\bx)}$ be orthogonal projection onto $N(\bx)$. Then
\begin{align}
    \psi_\phi(\bdelta) - \psi_{\phi'}(\bdelta) = \E\left\langle q^\phi(\bX)-q^{\phi'}(\bX),\ \Pi_{N(\bX)}\bmu(\bX)\right\rangle, \label{eqn:additivity}
\end{align}
where $\bmu(\bx)=(\mu_\ba(\bx))_{\ba\in S(\bx)}$, and $\dim N(\bx) = |S(\bx)| - 1 - |\cK(\bx)|$. The right-hand side vanishes whenever $\bmu(\bx)$ is affine on $S(\bx)$ (that is, $\Pi_{N(\bx)}\bmu(\bx)=0$), and otherwise depends on how far $\bmu(\bx)$ departs from affine (Remark~\ref{rmk:invariance-interactions} below).
\end{proposition}
\begin{proof}
Both $q^\phi(\bx)$ and $q^{\phi'}(\bx)$ satisfy the same $1+|\cK(\bx)|$ affine constraints defining $\cP_{\bdelta}^S(\bx)$ (normalization plus the $|\cK(\bx)|$ active marginal-odds constraints), so their difference $q^\phi(\bx)-q^{\phi'}(\bx)$ lies in the constraint null space, i.e. is orthogonal to the row space spanned by $\{\mathbf 1,(a_k)_{k\in\cK(\bx)}\}$ restricted to $S(\bx)$---equivalently, $q^\phi(\bx)-q^{\phi'}(\bx)\in N(\bx)$. Writing $\psi_\phi(\bdelta)=\E\langle q^\phi(\bX),\bmu(\bX)\rangle$ (Proposition~\ref{prop:identification}) and subtracting the two identifying expressions, $\psi_\phi(\bdelta)-\psi_{\phi'}(\bdelta) = \E\langle q^\phi(\bX)-q^{\phi'}(\bX),\bmu(\bX)\rangle$. Since $q^\phi(\bx)-q^{\phi'}(\bx)\in N(\bx)$, this inner product depends on $\bmu(\bx)$ only through its projection onto $N(\bx)$, giving \eqref{eqn:additivity}. The dimension count $\dim N(\bx)=|S(\bx)|-1-|\cK(\bx)|$ follows because the constraint row space has dimension exactly $1+|\cK(\bx)|$ under Condition~\ref{cond:nac} (no redundant active constraints).
\end{proof}

\begin{remark}[When invariance survives interactions]\label{rmk:invariance-interactions}
Proposition~\ref{prop:additivity} shows the sufficient condition for exact $\phi$-invariance of $\psi(\bdelta)$ is that $\bmu$ is \emph{affine on $S(\bx)$}, i.e. $\bmu(\bx)\in N(\bx)^\perp$. Absent structural zeros, this is essentially the additivity condition one would naively expect: $\mu$ has no interaction among treatment components (no nonzero coefficient on any product $a_ja_k$, $j\ne k$, on the full support), and the size of the gap in \eqref{eqn:additivity} scales with how far $\mu$ departs from additive, i.e. with the strength of these interactions. What is less obvious is that affine-on-$S(\bx)$ is strictly weaker than ``no interaction among treatment components'' once the support is not full: a structural zero can make an outcome model with genuine interaction terms affine on $S(\bx)$ anyway, and hence still $\phi$-invariant. For example, if a structural zero forces $a_j=1\Rightarrow a_k=1$ on $S(\bx)$, then the interaction term $a_ja_k$ coincides with $a_j$ on $S(\bx)$, so an outcome model with a nonzero $a_ja_k$ coefficient can still be affine on $S(\bx)$ and hence $\phi$-invariant.
\end{remark}

\section{Sensitivity analysis: influence functions}\label{app:sensitivity-eif}

\begin{proposition}\label{prop:bounds}
Under Assumption~\ref{consistency}, Condition~\ref{cond:tilt-existence} (applied to $\bdelta_\cK$), and Assumption~\ref{asmpt:bounded} for some $\Gamma\ge1$,
\[
\psi_{\ell,\Gamma}(\bdelta_\cK) \le \psi(\bdelta_\cK) \le \psi_{u,\Gamma}(\bdelta_\cK),
\]
where $\psi_{\ell,\Gamma}(\bdelta_\cK) = \tilde\psi(\bdelta) -(1-\Gamma^{-1})\E\left[\sum_{\ba\in \cA}\mu_\ba(\bX)(\pi_\ba(\bX)-\pi^2_{\ba}(\bX))h(\ba^\top\blambda_{\cK})\right]$ and
\[
\psi_{u,\Gamma}(\bdelta_\cK) = \tilde\psi(\bdelta) + \E\left[\sum_{\ba\in\cA}\{\pi_\ba(\bX)-\pi_\ba^2(\bX)\}h(\ba^\top\blambda_{\cK})\left[(\Gamma-1)\mu_\ba(\bX)\bI\{\Gamma\mu_\ba(\bX)\le1\} + \{1-\mu_\ba(\bX)\}\bI\{\Gamma\mu_\ba(\bX)>1\}\right]\right],
\]
with $\tilde\psi(\bdelta) = \E[\sum_{\ba\in\cA}\mu_\ba(\bX)\pi_\ba(\bX)h(\ba^\top\blambda_{\cK})]$.
\end{proposition}

\begin{proposition}\label{prop:eifbounds}
Under Assumption~\ref{asmpt:outcome-bounded} and Conditions~\ref{cond:f-div}--\ref{cond:unif-bound}, with Conditions~\ref{cond:tilt-regularity} and~\ref{cond:unif-bound} imposed at $(\cK,\bpi)$, recall that $\blambda_{\cK} = \blambda\{\bpi(\bX);\bdelta_{\cK(\bX)}\}$, and consider the following functionals:
\begin{align*}
\psi_{b,1}(\bdelta_{\cK}) &= \E\left[\sum_{\ba\in\cA}\mu_\ba(\bX)(\pi_\ba(\bX)-\pi_\ba^2(\bX))h(\ba^\top\blambda_{\cK})\right],\\
\psi_{b,2}(\bdelta_{\cK}) &= \E\left[\sum_{\ba\in\cA}(1-\mu_\ba(\bX))(\pi_\ba(\bX)-\pi_\ba^2(\bX))h(\ba^\top\blambda_{\cK})\right].
\end{align*}
In $\psi_{b,2}$ the constant $1$ is the outcome upper limit of Assumption~\ref{asmpt:bounded}, so $1-\mu_\ba(\bX)$ is the headroom between the conditional mean and that limit.
Then, treating $\cK(\bx)$ as known for all $\bx$, we have:
\begin{align*}
\eif\bigl[\psi_{b,1}(\bdelta_{\cK})\bigr] &= \sum_{\ba\in \cA}\left\{\bI\{\bA = \ba\}\{Y - \mu_\ba(\bX)\}(1-\pi_\ba(\bX))h(\ba^\top\blambda_{\cK}) \right.\\
&\left.+ \mu_\ba(\bX)\bigl(\bI\{\bA = \ba\} - \pi_\ba(\bX)\bigr)\bigl(1-2\pi_\ba(\bX)\bigr)h(\ba^\top\blambda_{\cK})\right. \\
&\left.- \mu_\ba(\bX)(\pi_\ba(\bX)-\pi_\ba^2(\bX))\dot h(\ba^\top\blambda_{\cK})\bba_{\cK}^\top \inv{\bOmega(\bblambda_{\cK})}g(\bbA_{\cK};\bblambda_{\cK},\bpi(\bX))\right. \\
&\left.+ \mu_\ba(\bX)(\pi_\ba(\bX)-\pi_\ba^2(\bX))h(\ba^\top\blambda_{\cK})\right\} - \psi_{b,1}(\bdelta_\cK),\\
\eif\bigl[\psi_{b,2}(\bdelta_{\cK})\bigr] &= \sum_{\ba\in\cA}\left\{\bI\{\bA = \ba\}\{Y-\mu_\ba(\bX)\}(\pi_\ba(\bX)-1)h(\ba^\top\blambda_{\cK})\right. \\
&\left.+(1-\mu_\ba(\bX))\bigl(\bI\{\bA = \ba\} - \pi_\ba(\bX)\bigr)\bigl(1-2\pi_\ba(\bX)\bigr)h(\ba^\top\blambda_{\cK}) \right.\\
&\left.- (1-\mu_\ba(\bX))(\pi_\ba(\bX)-\pi_\ba^2(\bX))\dot h(\ba^\top\blambda_{\cK})\bba_{\cK}^\top\inv{\bOmega(\bblambda_{\cK})}g(\bbA_{\cK};\bblambda_{\cK}, \bpi(\bX)) \right.\\
&\left. + (1-\mu_\ba(\bX))(\pi_\ba(\bX)-\pi_\ba^2(\bX))h(\ba^\top\blambda_{\cK})\right\} - \psi_{b,2}(\bdelta_\cK).
\end{align*}
\end{proposition}

\begin{remark}\label{rmk:boundestimator}
Proposition~\ref{prop:eifbounds}, in combination with Propositions~\ref{prop:eif} and~\ref{prop:bounds}, suggest the following estimator for the bounds $\psi_{\ell,\Gamma}(\bdelta_\cK),\psi_{u,\Gamma}(\bdelta_\cK)$, with the active set estimated by $\hcKen$ (which equals $\cK$ with probability tending to one, Lemma~\ref{lem:exact-recovery}):
\begin{align*}
\hat\psi_{\ell,\Gamma}(\bdelta)&=\P_n\left\{\hat\varphi(\bO;\bdelta_{\hcKen}) - (1-\Gamma^{-1})\sum_{\ba\in\cA}\hat\varphi_{1,\ba}(\bO;\bdelta_{\hcKen})\right\},\\
\hat\psi_{u,\Gamma}(\bdelta)&=\P_n\left\{\hat\varphi(\bO;\bdelta_{\hcKen}) + \sum_{\ba\in\cA}\left[(\Gamma-1)\hat\varphi_{1,\ba}(\bO;\bdelta_{\hcKen})\bI\{\Gamma\hat\mu_\ba(\bX)\le1\} + \hat\varphi_{2,\ba}(\bO;\bdelta_{\hcKen})\bI\{\Gamma\hat\mu_\ba(\bX)>1\}\right]\right\},
\end{align*}
where
\begin{align*}
\varphi_{1,\ba}(\bO;\bdelta_{\cK}) &= \bI\{\bA = \ba\}\{Y - \mu_\ba(\bX)\}(1-\pi_\ba(\bX))h(\ba^\top\blambda_{\cK}) \\
&+ \mu_\ba(\bX)\bigl(\bI\{\bA = \ba\} - \pi_\ba(\bX)\bigr)\bigl(1-2\pi_\ba(\bX)\bigr)h(\ba^\top\blambda_{\cK}) \\
&- \mu_\ba(\bX)(\pi_\ba(\bX)-\pi_\ba^2(\bX))\dot h(\ba^\top\blambda_{\cK})\bba_{\cK}^\top \inv{\bOmega(\bblambda_{\cK})}g(\bbA_{\cK};\bblambda_{\cK}, \bpi(\bX)) \\
&+ \mu_\ba(\bX)(\pi_\ba(\bX)-\pi_\ba^2(\bX))h(\ba^\top\blambda_{\cK}),\\
\varphi_{2,\ba}(\bO;\bdelta_{\cK})&= \bI\{\bA = \ba\}\{Y-\mu_\ba(\bX)\}(\pi_\ba(\bX)-1)h(\ba^\top\blambda_{\cK}) \\
&+(1-\mu_\ba(\bX))\bigl(\bI\{\bA = \ba\} - \pi_\ba(\bX)\bigr)\bigl(1-2\pi_\ba(\bX)\bigr)h(\ba^\top\blambda_{\cK}) \\
&- (1-\mu_\ba(\bX))(\pi_\ba(\bX)-\pi_\ba^2(\bX))\dot h(\ba^\top\blambda_{\cK})\bba_{\cK}^\top\inv{\bOmega(\bblambda_{\cK})}g(\bbA_{\cK};\bblambda_{\cK}, \bpi(\bX)) \\
&+ (1-\mu_\ba(\bX))(\pi_\ba(\bX)-\pi_\ba^2(\bX))h(\ba^\top\blambda_{\cK}),
\end{align*}
and where the set $\hcKen$ depends on the threshold parameter $\en$, as in Theorem~\ref{thm:onestep}.

At any active set $\cS$, the population estimating function behind $\hat\psi_{u,\Gamma}$ is
\begin{align}\label{eqn:phi-u-gamma}
\varphi_{u,\Gamma}(\bO;\eta,\bdelta_\cS)
&= \varphi(\bO;\eta,\bdelta_\cS)
+ \sum_{\ba\in\cA}\Bigl[(\Gamma-1)\varphi_{1,\ba}(\bO;\eta,\bdelta_\cS)\bI\{\Gamma\mu_\ba(\bX)\le1\}
+ \varphi_{2,\ba}(\bO;\eta,\bdelta_\cS)\bI\{\Gamma\mu_\ba(\bX)>1\}\Bigr],
\end{align}
so that $\psi_{u,\Gamma}(\bdelta_\cS)=\P[\varphi_{u,\Gamma}(\bO;\eta,\bdelta_\cS)]$ and $\hat\psi_{u,\Gamma}$ above is its plug-in empirical analogue at $\cS=\hcKen$. Unlike $\varphi$ and the lower-bound score, $\varphi_{u,\Gamma}$ is not pathwise differentiable in $\eta$, because of the indicators. The margin condition on $\Gamma\mu_\ba(\bX)$ in Lemma~\ref{lem:second-order} is what controls the resulting plug-in error.
\end{remark}

\section{Asymptotic theory: lemmas and master expansion}\label{app:asymptotic-theory}

\begin{lemma}[Jacobian and tilt stability]\label{lem:Omega-stability}
Fix $\bx$, a componentwise positive $\bdelta_\cS$, and an active set $\cS(\bx)$ with $b_k(\bx)>0$ for all $k\in\cS(\bx)$ (so Conditions~\ref{cond:tilt-existence} and~\ref{assumption:interior} apply to the reduced system on $\cS$). Write $\bbA=\bbA_\cS,\bblambda=\bblambda_\cS$ for the treatment and multiplier vectors restricted to $\cS$, together with an intercept, with $\bblambda=\bblambda\{\bpi(\bx);\bdelta_\cS\}$, $\hat\bblambda=\bblambda\{\hat\bpi(\bx);\bdelta_\cS\}$, and $\bOmega(\bblambda,\bpi)=\sum_{\ba\in\cA}\pi_\ba(\bx)\dot h(\bba_\cS^\top\bblambda)\bba_\cS\bba_\cS^\top$ the reduced constraint Jacobian. Under Conditions~\ref{cond:tilt-existence}, \ref{assumption:interior}, \ref{cond:tilt-regularity}, and~\ref{cond:unif-bound} applied to the reduced system on $\cS$ at $\bpi^*\in\{\bpi(\bx),\hat\bpi(\bx)\}$, $\|\hat\bblambda-\bblambda\|\lesssim\|\hat\bpi(\bx)-\bpi(\bx)\|$, $\|\bOmega(\hat\bblambda,\hat\bpi)^{-1}-\bOmega(\bblambda,\bpi)^{-1}\|_{op}\lesssim\|\hat\bpi(\bx)-\bpi(\bx)\|$, and $|h(\bbA^\top\hat\bblambda)-h(\bbA^\top\bblambda)|\lesssim\|\hat\bpi(\bx)-\bpi(\bx)\|$. If Conditions~\ref{cond:tilt-regularity} and~\ref{cond:unif-bound} hold on $\Lambda_\cS(\bx)$ uniformly in $\bx$ and over the active sets considered, the implicit constants above are free of $\bx$ and all three bounds hold in $L_2(\PP_0)$ norm with $\max_{\ba\in\cA}\|\hat\pi_\ba-\pi_\ba\|_2$ on the right-hand side.
\end{lemma}

\begin{lemma}[Smooth second-order and empirical-process remainder]\label{lem:second-order}
For any finite linear combination $\varphi^\star_\cS$ (bounded coefficients, possibly depending on $\bI\{\Gamma\mu_\ba(\bX)\le1\}$) of the smooth uncentered scores $\varphi(\bO;\eta,\bdelta_\cS)$, $\varphi_{1,\ba}(\bO;\eta,\bdelta_\cS)$, $\varphi_{2,\ba}(\bO;\eta,\bdelta_\cS)$ ($\ba\in\cA$) computed on an active set $\cS$ measurable w.r.t.\ the training sample (fixed, or ranging over $\{\cKen,\hcKen\}$), under sample splitting or cross-fitting, $L_2(\PP_0)$-consistency, uniform boundedness, a componentwise positive $\bdelta_\cS$, and Conditions~\ref{cond:tilt-existence}--\ref{cond:unif-bound} on $\cS$: $\|\hat\varphi^\star_\cS-\varphi^\star_\cS\|_2=o_p(1)$, $\P[\hat\varphi^\star_\cS-\varphi^\star_\cS]=\cO_p(R_n)$ and $(\P_n-\P)[\hat\varphi^\star_\cS-\varphi^\star_\cS]=o_p(n^{-1/2})$, where $R_n$ is the second-order product rate of Theorem~\ref{thm:onestep}. Replacing a true upper-bound indicator $\bI\{\Gamma\mu_\ba(\bX)\le1\}$ by its plug-in analogue $\bI\{\Gamma\hat\mu_\ba(\bX)\le1\}$ contributes an additional $\cO_p(\tau_n^{1+\varrho})$ to the drift and $o_p(n^{-1/2})$ to the empirical-process term, under a margin condition $\P[|\Gamma\mu_\ba(\bX)-1|\le t]\lesssim t^\varrho$ and with $\tau_n=\max_{\ba\in\cA}\|\hat\mu_\ba-\mu_\ba\|_\infty=o_p(1)$ (sup-norm consistency, required for the upper-bound score only). Thus for the upper-bound score, $\P[\hat\varphi^\star_\cS-\varphi^\star_\cS]=\cO_p(R_n+\tau_n^{1+\varrho})$ and $(\P_n-\P)[\hat\varphi^\star_\cS-\varphi^\star_\cS]=o_p(n^{-1/2})$.
\end{lemma}

\begin{lemma}[Tilted-law stability under removal of near-boundary coordinates]\label{lem:tilted-law-stability}
Fix $\bx$, and for an active set $\cS(\bx)\subseteq\cK(\bx)$ and $\bpi^*\in\{\bpi,\hat\bpi\}$ write $q_{\cS,\ba}(\bx)=\pi^*_\ba(\bx)h(\ba^\top\blambda_\cS(\bpi^*(\bx);\bdelta_\cS))$, so that $q_\ba(\bx;\bdelta_{\cK(\bx)})=q_{\cK(\bx),\ba}(\bx)$ at $\bpi^*=\bpi$ when $\cK(\bx)$ is the active set at $\bx$. Suppose $\bdelta_\cS$ is fixed componentwise positive, and Conditions~\ref{cond:tilt-existence}, \ref{assumption:interior}, \ref{cond:nac}, and~\ref{cond:tilt-regularity} hold uniformly for all nested active sets $\cS'(\bx)\subseteq\cS(\bx)\subseteq\cK(\bx)$ and $\bpi^*\in\{\bpi,\hat\bpi\}$. Conditions~\ref{cond:tilt-regularity}--\ref{cond:unif-bound} are read throughout with the enlarged $\Lambda_\cS(\bx)$ of Appendix~\ref{app:lambda-def}. For Condition~\ref{cond:unif-bound}, enumerate $\cS(\bx)\setminus\cS'(\bx)=\{k_1,\ldots,k_r\}$ in nondecreasing order of $b_{k_j}[\bpi^*](\bx):=\min\{\pi^*\text{-marginal},1-\pi^*\text{-marginal}\}$ at the coordinate in play (near-boundary coordinates dropped first) and set $\cS_0=\cS,\cS_j=\cS_{j-1}\setminus\{k_j\}$. Suppose Condition~\ref{cond:unif-bound} holds uniformly at every post-drop set $\cS_1(\bx),\ldots,\cS_r(\bx)=\cS'(\bx)$ and $\bpi^*\in\{\bpi,\hat\bpi\}$ -- \emph{not} at the starting set $\cS_0=\cS$ itself. For the thresholded active sets of the master expansion (Theorem~\ref{thm:master-expansion}) this means every set in the shrinking sequence strictly inside $\cK$, and for $\cS=\cK,\cS'=\cK_\epsilon$ (Lemma~\ref{lem:boundary-tapering}) it means every set strictly between $\cK$ and $\cK_\epsilon$. Uniform nonsingularity at the full set $\cK$ is never required. Condition~\ref{cond:unif-bound} at $\cS_1$ already forces the same bound $b^\star_{\cS_1}(\bx)\ge\ofl/C_h$ on the coordinates remaining in $\cS_1$ (by the argument of Appendix~\ref{sec:margin-implications} applied to $\cS_1$ in place of $\cK$). Consequently, whenever $\cS\setminus\cS'$ is required to contain a coordinate with $b_{k}(\bx)<\ofl/C_h$ -- the near-boundary case the lemma targets -- at most one such coordinate can be dropped at a given $\bx$ under this hypothesis, so the chain construction is stated for general $r$ but is applied in this paper only at $r=1$. Then there exists $C<\infty$, uniform over $\bx$, over $\bpi^*\in\{\bpi,\hat\bpi\}$, and over the relevant nested active sets, such that
\[
\sum_{\ba\in\cA}\left|q_{\cS,\ba}(\bx)-q_{\cS',\ba}(\bx)\right|\le C\sum_{k\in\cS(\bx)\setminus\cS'(\bx)}b_k[\bpi^*](\bx),
\]
where $b_k[\bpi^*](\bx):=\min\{m^*_k(\bx),1-m^*_k(\bx)\}$ is computed from the same $\bpi^*$ used to define $q_{\cS,\ba},q_{\cS',\ba}$ (so $b_k[\bpi^*]=b_k$ at $\bpi^*=\bpi$, and $b_k[\bpi^*]=\hat b_k$ at $\bpi^*=\hat\bpi$). The argument is identical in either case.
\end{lemma}

\begin{lemma}[Boundary tapering of active-set thresholding]\label{lem:boundary-tapering}
For $\epsilon\in(0,1/2)$ recall $\cK_\epsilon(\bx)=\{k:b_k(\bx)>\epsilon\}$ and $b^\star(\bx)=\min_{k:b_k(\bx)>0}b_k(\bx)$ ($=\infty$ if no such $k$ exists). Assume Assumption~\ref{asmpt:outcome-bounded}, the conditions of Lemma~\ref{lem:tilted-law-stability} hold across $\cK,\cK_\epsilon$, and the corresponding scores $\varphi$ are uniformly bounded over the active sets considered. Two consequences of these hypotheses are worth stating. First, by the disclosure in Lemma~\ref{lem:tilted-law-stability}, once $\epsilon<\ofl/C_h$ every coordinate of $\cK(\bx)\setminus\cK_\epsilon(\bx)$ is near-boundary, so the hypotheses permit at most one such coordinate at each $\bx$. The factor $m$ below is therefore not tight under them. Second, since $\varphi_\cS$ involves $\inv{\bOmega(\blambda_\cS)}$ and Condition~\ref{cond:unif-bound} is deliberately not imposed at $\cK$, uniform boundedness of $\varphi_\cK$ is itself a restriction on $\|\inv{\bOmega(\blambda_\cK)}\|_{op}$. It is used only for the second-moment bound, the first-moment bound requiring no control of that inverse. Then there exists $C<\infty$ such that, for all $\epsilon\in(0,1/2)$ and all $\bx$,
\[
\left|\P\left[\{\varphi_\cK(\bO;\eta,\bdelta_\cK)-\varphi_{\cK_\epsilon}(\bO;\eta,\bdelta_{\cK_\epsilon})\}\mid\bX=\bx\right]\right|\le C\sum_{k\in\cK(\bx)\setminus\cK_\epsilon(\bx)}b_k(\bx).
\]
If in addition the global margin condition (Condition~\ref{cond:margin}, exponent $\beta$) holds, then $|\P[\varphi_\cK-\varphi_{\cK_\epsilon}]|\lesssim\epsilon^{1+\beta}$ and $\P[(\varphi_\cK-\varphi_{\cK_\epsilon})^2]\lesssim\epsilon^\beta$. Equivalently, since $\P[\varphi_\cS]=\psi(\bdelta_\cS)$ for any active set $\cS$, $|\psi(\bdelta_{\cK_\epsilon})-\psi(\bdelta)|\lesssim\epsilon^{1+\beta}$. The lemma instantiates at $\epsilon=\epsilon_n$ shrinking (used in Theorem~\ref{thm:master-expansion}) and at fixed $\epsilon$ (used for the fixed-threshold estimand of Corollary~\ref{cor:adaptive}).
\end{lemma}

\begin{lemma}[Exact active-set recovery under Condition~\ref{cond:unif-bound} at $\cK$]\label{lem:exact-recovery}
Suppose Conditions~\ref{cond:tilt-regularity}--\ref{cond:unif-bound} hold at $(\cK,\bpi)$ and $(\cK,\hat\bpi)$ with constant $\ofl>0$ (Appendix~\ref{sec:margin-implications}, Appendix~\ref{app:regularity}), and let $\epsilon_n\to0$ satisfy $\rho_n=o_p(\epsilon_n)$. Then $\cKen(\bx)=\cK(\bx)$ identically for every $\bx$ once $\epsilon_n<\ofl/C_h$, and $\PP\{\hcKen(\bX)=\cK(\bX)\ \PP_0\text{-a.s.}\}\to1$. Consequently terms (i)--(ii) of \eqref{eqn:step0} vanish with probability tending to one, and $\hat\psi(\bdelta)$ has the same limiting distribution as the true-active-set estimator $\P_n\varphi(\bO;\hat\eta,\bdelta_\cK)$.
\end{lemma}
\begin{proof}
By Appendix~\ref{sec:margin-implications}, Conditions~\ref{cond:tilt-regularity}--\ref{cond:unif-bound} at $(\cK,\bpi)$ force $b^\star(\bx)\ge\ofl/C_h$ for every $\bx$, so $\cKen(\bx)=\{k:b_k(\bx)>\epsilon_n\}$ and $\cK(\bx)=\{k:b_k(\bx)>0\}$ coincide identically once $\epsilon_n<\ofl/C_h$: any $k\in\cK(\bx)$ has $b_k(\bx)\ge b^\star(\bx)\ge\ofl/C_h>\epsilon_n$, and the reverse inclusion is immediate. This step is deterministic.

For the estimated set, $t\mapsto\min\{t,1-t\}$ is $1$-Lipschitz, so $|\hat b_k(\bx)-b_k(\bx)|\le|\hat m_k(\bx)-m_k(\bx)|\le\rho_n$ for every $k$ and $\bx$. Define
\[
\cE_n=\left\{\rho_n<\min\!\left(\epsilon_n,\ \ofl/C_h-\epsilon_n\right)\right\}.
\]
Since $\epsilon_n\to0$, for all $n$ large enough that $\epsilon_n\le\ofl/(2C_h)$ the minimum equals $\epsilon_n$, so $\cE_n\supseteq\{\rho_n/\epsilon_n<1\}$ and $\PP(\cE_n)\to1$ by $\rho_n=o_p(\epsilon_n)$. On $\cE_n$: if $k\in\cK(\bx)$ then $b_k(\bx)\ge\ofl/C_h$, so $\hat b_k(\bx)\ge\ofl/C_h-\rho_n>\epsilon_n$ and $k\in\hcKen(\bx)$. If $k\notin\cK(\bx)$ then $b_k(\bx)=0$, so $\hat b_k(\bx)\le\rho_n<\epsilon_n$ and $k\notin\hcKen(\bx)$. Both inclusions hold for every $k$ and, since $\rho_n$ is a supremum over $k$ and $\bx$, uniformly in $\bx$. Hence $\hcKen(\bx)=\cK(\bx)$ on $\cE_n$, and on the same event $\hat\psi(\bdelta)=\P_n\varphi(\bO;\hat\eta,\bdelta_\cK)$ exactly.
\end{proof}

\begin{lemma}[Consistency of the plug-in variance estimator]\label{lem:variance-consistency}
Under the hypotheses of Theorem~\ref{thm:onestep}, with $\hat\sigma^2=\P_n[\{\varphi(\bO;\hat\eta,\bdelta_{\hcKen})-\hat\psi(\bdelta)\}^2]$,
\[
\hat\sigma^2\overset{\PP}{\to}\Var[\varphi(\bO;\eta,\bdelta)].
\]
The same conclusion holds for the known-active-set estimator under the hypotheses of Corollary~\ref{cor:onestep-fixed}, and for the fixed-threshold estimator under the hypotheses of Corollary~\ref{cor:adaptive}, with $\cK(\cdot)$ or $\cK_\epsilon$ and the corresponding target substituted throughout.
\end{lemma}
\begin{proof}
Write $\hat\varphi_\cK=\varphi(\bO;\hat\eta,\bdelta_\cK)$ and $\varphi_\cK=\varphi(\bO;\eta,\bdelta_\cK)$, and let $\cE_n$ be the event on which $\hcKen(\bx)=\cK(\bx)$ uniformly in $\bx$, so that $\PP(\cE_n)\to1$ by Lemma~\ref{lem:exact-recovery}. On $\cE_n$ the plug-in scores coincide with their true-active-set counterparts observation by observation, $\varphi(\bO_i;\hat\eta,\bdelta_{\hcKen})=\hat\varphi_\cK(\bO_i)$ for every $i$, whence $\hat\sigma^2=\P_n[\{\hat\varphi_\cK-\P_n\hat\varphi_\cK\}^2]$.

Lemma~\ref{lem:second-order} at the fixed set $\cS=\cK$ gives $\|\hat\varphi_\cK-\varphi_\cK\|_2=o_p(1)$. Under sample splitting the evaluation observations are independent of $\hat\varphi_\cK$ given the training sample, so $\P[(\hat\varphi_\cK-\varphi_\cK)^2\mid\bO_1^n]=\|\hat\varphi_\cK-\varphi_\cK\|_2^2=o_p(1)$ and Markov's inequality gives $\P_n[(\hat\varphi_\cK-\varphi_\cK)^2]=o_p(1)$. Under cross-fitting the same argument applies on each of the finitely many evaluation folds and the foldwise conclusions average. Assumption~\ref{asmpt:outcome-bounded}, Conditions~\ref{cond:tilt-regularity}--\ref{cond:unif-bound} at $\cK$, and $\bdelta>0$ componentwise (which bounds $\bDelta^2_{\bm}$) make $\varphi_\cK$ bounded, hence square-integrable, so $\P_n\varphi_\cK^2\overset{\PP}{\to}\P\varphi_\cK^2$ by the law of large numbers, and $\P_n(\hat\varphi_\cK+\varphi_\cK)^2\le2\P_n(\hat\varphi_\cK-\varphi_\cK)^2+8\P_n\varphi_\cK^2=\cO_p(1)$. Cauchy--Schwarz then gives
\[
\left|\P_n\hat\varphi_\cK^2-\P_n\varphi_\cK^2\right|
\le\left\{\P_n(\hat\varphi_\cK-\varphi_\cK)^2\right\}^{1/2}\left\{\P_n(\hat\varphi_\cK+\varphi_\cK)^2\right\}^{1/2}
=o_p(1),
\]
so $\P_n\hat\varphi_\cK^2\overset{\PP}{\to}\P\varphi_\cK^2$. For the mean, $\P_n\hat\varphi_\cK=\P\varphi_\cK+\P[\hat\varphi_\cK-\varphi_\cK]+(\P_n-\P)\hat\varphi_\cK$, where the second term is $\cO_p(R_n)=o_p(1)$ by Lemma~\ref{lem:second-order} and the third is $o_p(1)$ by the law of large numbers together with $\|\hat\varphi_\cK-\varphi_\cK\|_2=o_p(1)$. Moreover, $\P\varphi_\cK=\psi(\bdelta_\cK)=\psi(\bdelta)$ by Propositions~\ref{prop:identification} and~\ref{prop:eif}, since $\varphi_\cS$ is an uncentered efficient influence function for $\psi(\bdelta_\cS)$ at every active set $\cS$. Hence $\P_n\hat\varphi_\cK\overset{\PP}{\to}\psi(\bdelta)$. Hence, on $\cE_n$,
\[
\hat\sigma^2=\P_n\hat\varphi_\cK^2-(\P_n\hat\varphi_\cK)^2
\overset{\PP}{\to}\P\varphi_\cK^2-(\P\varphi_\cK)^2=\Var[\varphi(\bO;\eta,\bdelta)],
\]
and since $\PP(\cE_n)\to1$ the convergence holds unconditionally.

For the known-active-set estimator the argument is identical with $\cE_n$ omitted, since $\cK(\cdot)$ is known and no recovery step arises. For the fixed-threshold estimator, Condition~\ref{cond:localized-margin} together with the active-set sup-norm rate assumed in Corollary~\ref{cor:adaptive} gives $\PP_0\{\hat\cK_\epsilon(\bX)\ne\cK_\epsilon(\bX)\}=o_p(1)$. Uniform boundedness of the scores converts this into $\|\varphi(\bO;\hat\eta,\bdelta_{\hat\cK_\epsilon})-\varphi(\bO;\eta,\bdelta_{\cK_\epsilon})\|_2=o_p(1)$ once the fixed-active-set nuisance difference is added, controlled by Lemma~\ref{lem:second-order} at $\cS=\cK_\epsilon$, and the first- and second-moment argument above applies verbatim with $\psi(\bdelta_{\cK_\epsilon})$ in place of $\psi(\bdelta)$ and $\Var[\varphi(\bO;\eta,\bdelta_{\cK_\epsilon})]$ in place of $\Var[\varphi(\bO;\eta,\bdelta)]$.
\end{proof}

\begin{theorem}[Master expansion for estimated-active-set one-step estimators]\label{thm:master-expansion}
Let $\varphi^\star(\bO;\eta,\bdelta_\cK)$ denote the uncentered EIF for $\psi(\bdelta_\cK)$, for $\psi_{\ell,\Gamma}(\bdelta_\cK)$, or the upper-bound score $\varphi_{u,\Gamma}$. Write $\psi^\star(\bdelta_\cK)$ for $\psi(\bdelta_\cK)$, $\psi_{\ell,\Gamma}(\bdelta_\cK)$ or $\psi_{u,\Gamma}(\bdelta_\cK)$ respectively, and $\hat\psi^\star(\bdelta_\cK)=\P_n\varphi^\star(\bO;\hat\eta,\bdelta_{\hcKen})$ for the corresponding one-step estimator. Suppose $\varphi^\star(\bO;\eta,\bdelta_\cS)$ is uniformly bounded over every active set $\cS(\bx)$ with $\cKen(\bx)\cap\hcKen(\bx)\subseteq\cS(\bx)\subseteq\cK(\bx)$, with $\Var[\varphi^\star(\bO;\eta,\bdelta_\cK)]\in(0,\infty)$. Under Conditions~\ref{cond:f-div}--\ref{cond:nac} and Assumption~\ref{asmpt:outcome-bounded}, and, when $\varphi^\star$ is the EIF of $\psi(\bdelta_\cK)$, Assumptions~\ref{consistency}--\ref{ignorability} (the two bounds are functionals of the observed-data law and need neither), $L_2$-consistent nuisances, the margin condition (Condition~\ref{cond:margin}, exponent $\beta$), $\epsilon_n^{1+\beta}=o(n^{-1/2})$ with $\rho_n=o_p(\epsilon_n)$, and the following regularity conditions: (a) Condition~\ref{cond:tilt-regularity} holding uniformly over every active set $\cS(\bx)$ with $\cKen(\bx)\cap\hcKen(\bx)\subseteq\cS(\bx)\subseteq\cK(\bx)$ and $\bpi^*\in\{\bpi,\hat\bpi\}$, (b) Condition~\ref{cond:unif-bound} holding uniformly at every post-drop set in the nested chain from $\cK(\bx)$ to $\cKen(\bx)$ specified in Lemma~\ref{lem:tilted-law-stability} (dropping coordinates in nondecreasing order of $b_k(\bx)$) and $\bpi^*\in\{\bpi,\hat\bpi\}$, with a single constant $\ofl>0$ not depending on $n$, but \emph{not} at $\cK(\bx)$ itself (for $n$ large enough that $\epsilon_n<\ofl/C_h$ this forces $|\cK(\bx)\setminus\cKen(\bx)|\le1$ at every $\bx$, Lemma~\ref{lem:tilted-law-stability}), and (c) on the event $\cG_n=\{\rho_n\le\epsilon_n/2\}$, Conditions~\ref{cond:tilt-regularity}--\ref{cond:unif-bound} holding uniformly over $\cKen(\bx)\cap\hcKen(\bx)\subseteq\cS(\bx)\subseteq\cKen(\bx)\cup\hcKen(\bx)$ and $\bpi^*\in\{\bpi,\hat\bpi\}$ (as for the chain, Condition~\ref{cond:unif-bound} at $\cKen(\bx)$ itself requires that no coordinate have $b_k(\bx)$ strictly between $\epsilon_n$ and $\ofl/C_h$, Appendix~\ref{sec:margin-implications}): with $R_n$ as in Theorem~\ref{thm:onestep}, $\tau_n=\max_{\ba\in\cA}\|\hat\mu_\ba-\mu_\ba\|_\infty$, and, for the upper-bound score only, the additional margin condition $\P[|\Gamma\mu_\ba(\bX)-1|\le t]\lesssim t^\varrho$ for every $\ba\in\cA$ and $t>0$, let $V_n=\tau_n^{1+\varrho}$ for the upper bound and $V_n=0$ otherwise. Then
\[
\hat\psi^\star(\bdelta_\cK)-\psi^\star(\bdelta_\cK) = (\P_n-\P)\varphi^\star(\bO;\eta,\bdelta_\cK) + \cO_p(R_n+V_n) + o_p(n^{-1/2}).
\]
Consequently, if $R_n+V_n=o_p(n^{-1/2})$, $\sqrt n\{\hat\psi^\star(\bdelta_\cK)-\psi^\star(\bdelta_\cK)\}\rightsquigarrow\cN(0,\Var[\varphi^\star(\bO;\eta,\bdelta_\cK)])$.
\end{theorem}

\begin{remark}[Scope of Theorem~\ref{thm:master-expansion}]\label{rmk:thm-scope-master}
The chain hypothesis asks for Condition~\ref{cond:unif-bound} at the post-drop sets from $\cK$ to $\cKen$ but not at $\cK$ itself. Since $\ofl$ is a single constant not depending on $n$, any coordinate with $0<b_k(\bx)<\ofl/C_h$ would eventually fall in the excluded gap, so $b^\star(\bx)\ge\ofl/C_h$ for every $\bx$, as under Condition~\ref{cond:unif-bound} at $\cK$ (Remark~\ref{rmk:thm-scope}). Then $\cKen=\cK$ identically for large $n$ and $\hcKen=\cK$ with probability tending to one (Lemma~\ref{lem:exact-recovery}), so the margin condition and the rate $\epsilon_n^{1+\beta}=o(n^{-1/2})$ are not binding. The theorem's content is the extension of Theorem~\ref{thm:onestep} to the three scores $\varphi^\star$. Near-boundary tolerance comes instead from Corollary~\ref{cor:adaptive}.
\end{remark}

\subsection{Fixed-threshold estimand}\label{sec:adaptive}

Theorem~\ref{thm:onestep} and Corollary~\ref{cor:onestep-fixed} both require Condition~\ref{cond:unif-bound} at the true active set, which forces every genuinely stochastic marginal to stay a fixed distance from $\{0,1\}$ (Appendix~\ref{sec:margin-implications}). When that cannot be assumed -- some marginals may lie arbitrarily close to the boundary -- a different estimand is needed. The threshold sequence $\epsilon_n\to0$ in Theorem~\ref{thm:onestep} is in any case a device for asymptotic normality, not a practical prescription, since in practice one fixes $\epsilon>0$ for a given $n$. Both considerations point to the same fix, a \emph{fixed-threshold estimand} $\psi(\bdelta_{\cK_\epsilon})$, intervening only on coordinates bounded away from the boundary by $\epsilon$, using $\cK_\epsilon(\bx):=\{k:b_k(\bx)>\epsilon\}$:
\begin{equation}\label{eq:adaptive-estimand}
\psi(\bdelta_{\cK_{\epsilon}}) = \PP_0\left[\sum_{\ba \in \cA} \mu_\ba(\bX) \pi_\ba(\bX)h\left\{\ba^\top\bblambda(\bpi(\bX);\bdelta_{\cK_\epsilon})\right\}\right].
\end{equation}

Because $\bdelta_{\cK_\epsilon}$ differs from $\bdelta_\cK$ only by dropping the coordinates $\cK(\bx)\setminus\cK_\epsilon(\bx)$ that $\cK_\epsilon$ excludes, the fixed-threshold estimand itself -- before any estimation -- already differs from the target $\psi(\bdelta)$ by an amount governed by how much probability mass sits near the true boundary $\{0,1\}$ that $\cK_\epsilon$ excludes. The following proposition characterizes the bias of the fixed-threshold estimand with respect to $\psi(\bdelta)$ under an assumed margin condition that restricts the density of $b^\star(\bx)$. However, it is a fixed bias that does not vanish with $n$.

\begin{condition}[Margin condition]\label{cond:margin}
There exists $\beta>0, C<\infty$ such that $\PP_0\bigl(0<b^\star(\bX)\le t\bigr)\le Ct^\beta$ for all $t>0$.
\end{condition}

\begin{proposition}[Bias of the fixed threshold estimand]\label{prop:bias}
Under Assumption~\ref{asmpt:outcome-bounded}, Condition~\ref{cond:margin} with exponent $\beta$, and the hypotheses of Lemma~\ref{lem:boundary-tapering} (at $\bpi^*=\bpi$: Conditions~\ref{cond:tilt-existence}--\ref{cond:tilt-regularity} for all active sets between $\cK_\epsilon$ and $\cK$, Condition~\ref{cond:unif-bound} at every post-drop set of the chain from $\cK$ to $\cK_\epsilon$, and uniformly bounded scores), for every $\epsilon\in(0,1/2)$,
\[
|\psi(\bdelta)-\psi(\bdelta_{\cK_{\epsilon}})| \lesssim \epsilon^{1+\beta}.
\]
\end{proposition}
\begin{proof}
Lemma~\ref{lem:boundary-tapering}, applied at the fixed threshold $\epsilon$ under Condition~\ref{cond:margin}, gives
\[
\bigl|\P[\varphi_\cK-\varphi_{\cK_\epsilon}]\bigr|\lesssim\epsilon^{1+\beta}.
\]
By Propositions~\ref{prop:identification} and~\ref{prop:eif}, $\P[\varphi_\cS(\bO;\eta,\bdelta_\cS)]=\psi(\bdelta_\cS)$ at every active set $\cS$. Taking $\cS=\cK$ and $\cS=\cK_\epsilon$, and using $\psi(\bdelta_\cK)=\psi(\bdelta)$ by redundancy of the constraints for $k\notin\cK(\bX)$, the display becomes $|\psi(\bdelta)-\psi(\bdelta_{\cK_\epsilon})|\lesssim\epsilon^{1+\beta}$. Lemma~\ref{lem:boundary-tapering} supplies this for all sufficiently small $\epsilon$. For $\epsilon$ bounded away from zero the bound holds trivially after enlarging the implicit constant, since $|\psi(\bdelta)-\psi(\bdelta_{\cK_\epsilon})|\le2\max_\ba\|\mu_\ba\|_\infty$ by boundedness.
\end{proof}

Proposition~\ref{prop:bias} suggests that the bias of the fixed-threshold estimand is relatively small under the assumed margin condition. Moreover, this condition is relatively mild: for example, when $\beta = 1$ this simply requires that the density of $b^\star(\bx)$ is bounded.

On the other hand, establishing a CLT and associated inference for the corresponding estimator $\hat\psi(\bdelta_{\hat\cK_\epsilon})$ is a separate question from the approximation bias above: it requires recovering $\cK_\epsilon$ from estimated marginals, which is controlled not by mass near the true boundary, as in Proposition~\ref{prop:bias}, but by how much mass sits near the \emph{fixed} cutoffs $\{\epsilon,1-\epsilon\}$ themselves:

\begin{condition}[Localized margin condition]\label{cond:localized-margin}
Fix $\epsilon\in(0,1/2)$. There exists $\beta>0$ such that $\PP_0[b^\star_\epsilon(\bX)\le t]\lesssim t^\beta$ for all $t\in(0,\epsilon/2)$, where $b^\star_\epsilon(\bx)=\min_k\min\{|m_k(\bx)-\epsilon|,|m_k(\bx)-(1-\epsilon)|\}$. This controls the density of $m_k(\bX)$ only near the fixed cutoffs $\{\epsilon,1-\epsilon\}$: the relevant bands $m_k(\bX)\in[\epsilon-t,\epsilon+t]$ lie strictly within $[\epsilon/2,3\epsilon/2]$ and never reach the boundaries $\{0,1\}$, so it places no restriction on the density of $m_k(\bX)$ near $\{0,1\}$ itself. When there is concern about non-negligible probability of very small but nonzero marginal probabilities that could violate Condition~\ref{cond:margin}, Condition~\ref{cond:localized-margin} may be the preferable assumption.
\end{condition}

Under this condition we may then establish the root-$n$ consistency and asymptotic normality of the fixed-threshold estimator.

\begin{corollary}[Asymptotic normality for the fixed-threshold estimand]\label{cor:adaptive}
Fix $\epsilon\in(0,1/2)$, $\bdelta>0$. Under Condition~\ref{cond:f-div}, Assumptions~\ref{consistency}--\ref{ignorability} and~\ref{asmpt:outcome-bounded}, Conditions~\ref{cond:tilt-existence}--\ref{cond:unif-bound} applied to the reduced system on $\cS(\bx)\in\{\cK_\epsilon(\bx),\hat\cK_\epsilon(\bx)\}$ at $\bpi^*\in\{\bpi,\hat\bpi\}$ (and not at $\cK$ itself), sample splitting or cross-fitting, $\Var[\varphi(\bO;\eta,\bdelta_{\cK_\epsilon})]>0$, $L_2(\PP_0)$-consistency of $\hat\mu_\ba$ for every $\ba$, $\hat\bpi(\bx)$ in the simplex over $\cA$, Condition~\ref{cond:localized-margin} with parameter $\beta$, an active-set sup-norm rate $\rho_n=\cO_p(n^{-\gamma})$ with $\gamma>\frac1{2\beta}$, and the product nuisance rate $R_n=o_p(n^{-1/2})$ of Theorem~\ref{thm:onestep}, the estimator $\hat\psi(\bdelta_{\hat\cK_\epsilon})=\P_n\varphi(\bO;\hat\eta,\bdelta_{\hat\cK_\epsilon})$ satisfies $\sqrt n\{\hat\psi(\bdelta_{\hat\cK_\epsilon})-\psi(\bdelta_{\cK_\epsilon})\}\rightsquigarrow\cN(0,\Var[\varphi(\bO;\eta,\bdelta_{\cK_\epsilon})])$. The corresponding empirical variance estimator is consistent by Lemma~\ref{lem:variance-consistency}, so for each fixed $\alpha\in(0,1)$ its Wald intervals attain asymptotic coverage $1-\alpha$.
\end{corollary}

The rate condition stated above, $\rho_n=\cO_p(n^{-\gamma})$ with $\gamma>\frac1{2\beta}$, is the special case $\tilde r_n=n^{-\gamma}$ of a more general, deterministic-sequence condition: it suffices that $\rho_n=\cO_p(\tilde r_n)$ for some deterministic $\tilde r_n\to0$ with $\tilde r_n^\beta=o(n^{-1/2})$. The nuisance-rate condition invoked as ``the product nuisance rate of Theorem~\ref{thm:onestep}'' is, explicitly, $\max_{\ba\in\cA}\|\hat\pi_\ba-\pi_\ba\|_2\cdot\max_{\ba\in\cA}\|\hat\mu_\ba-\mu_\ba\|_2+\max_{\ba\in\cA}\|\hat\pi_\ba-\pi_\ba\|_2^2=o_p(n^{-1/2})$.

\paragraph{Comparing Theorem~\ref{thm:onestep} and Corollary~\ref{cor:adaptive}.} The estimated-active-set estimator's rate requirement is comparatively unconstrained (Section~\ref{sec:step0}) because Condition~\ref{cond:unif-bound} at the true active set $\cK$ already forces $b^\star(\bx)\ge\ofl/C_h$, so $\cKen(\bx)=\cK(\bx)$ exactly once $\epsilon_n<\ofl/C_h$ (Lemma~\ref{lem:exact-recovery}) and no active-set-thresholding error remains to control. No analogous separation is available at the fixed interior cutoffs $\{\epsilon,1-\epsilon\}$: crossing them changes the intervention by an order-one amount regardless of $n$, so the fixed-threshold estimand instead needs a genuine active-set convergence rate ($\gamma>\frac12$ at $\beta=1$). Separately, Conditions~\ref{cond:tilt-regularity} and~\ref{cond:unif-bound} are required only over the pair $\{\cK_\epsilon(\bx),\hat\cK_\epsilon(\bx)\}$ in Corollary~\ref{cor:adaptive}, versus at the true active set $\cK$ itself in Theorem~\ref{thm:onestep} -- a weaker, more easily verified requirement, since the fixed-$\epsilon$ active sets stay bounded away from $\cK(\bx)$'s own possibly-boundary coordinates, at the cost of the estimand shift $\psi(\bdelta_{\cK_\epsilon})$ against $\psi(\bdelta_\cK)$ quantified in Lemma~\ref{lem:boundary-tapering}.

\begin{remark}[Outcome boundedness relaxation]
Assumption~\ref{asmpt:outcome-bounded}'s uniform boundedness of $Y,\mu_\ba$ could be relaxed to sub-Gaussian tails or finite $(2+\delta)$-moments at the cost of additional technical complexity. Since outcomes in most causal applications are naturally bounded, we retain uniform boundedness as the simpler and sufficient assumption.
\end{remark}

\section{Proofs}\label{app:proofs}

\subsection{Identification and bounds}

\subsection{Proof of Proposition~\ref{prop:kkt}}
Fix $\bx$. Write $u_\ba:=q_\ba(\bx)/\pi_\ba(\bx)$ for $\ba\in S(\bx)$, where $\pi_\ba(\bx)>0$ by definition of the support (so $q_\ba(\bx)=0$ for $\ba\notin S(\bx)$, per \eqref{feasible-set}). The objective $\sum_{\ba\in S(\bx)}\pi_\ba(\bx)\phi(u_\ba)$ is convex in $\bu$ by Condition~\ref{cond:f-div}, subject to the $m+1$ affine constraints $\sum_{\ba\in S(\bx)}\pi_\ba(\bx)u_\ba=1$ and $\sum_{\ba\in S(\bx)}a_k\pi_\ba(\bx)u_\ba=\zeta_{k,\bdelta}(\bx)$, $k=1,\ldots,m$, and $\bu\ge\mathbf 0$. Since the constraints are affine, a feasible point satisfying the KKT conditions is optimal, and the hypothesis supplies finite multipliers for which the candidate below is feasible. Introduce multipliers $\blambda=(\lambda_0,\ldots,\lambda_m)^\top$ for the $m+1$ affine constraints and write $t_\ba:=\ba^\top\blambda=\lambda_0+\sum_{k=1}^m a_k\lambda_k$. The Lagrangian is
\[
\cL(\bu,\blambda) = \sum_{\ba\in S(\bx)}\pi_\ba(\bx)\{\phi(u_\ba)-t_\ba u_\ba\} + \lambda_0 + \sum_{k=1}^m\lambda_k\zeta_{k,\bdelta}(\bx),
\]
where the last two terms do not depend on $\bu$. Minimizing $\cL$ over $\bu\ge\mathbf 0$, for fixed $\blambda$, therefore separates coordinate by coordinate: for each $\ba\in S(\bx)$,
\[
u_\ba^\star = \argmin_{u\ge0}\{\phi(u)-t_\ba u\} = \argmax_{u\ge0}\{t_\ba u-\phi(u)\} = h(t_\ba),
\]
the middle equality because minimizing $\phi(u)-t_\ba u$ is the same as maximizing its negative, and the last by the definition of $h$ in Proposition~\ref{prop:kkt}'s statement. This argmax is taken directly over $u\ge0$, so the non-negativity constraint is built into $h$ itself rather than assumed away. Because $\phi$ is strictly convex (Condition~\ref{cond:f-div}), $u_\ba^\star$ is the unique minimizer whenever it exists, so $\bu^\star=(h(\ba^\top\blambda))_{\ba\in S(\bx)}$ is the unique minimizer of $\cL(\cdot,\blambda^\star)$. At any feasible $\bu$ the multiplier terms vanish, so $\sum_\ba\pi_\ba\phi(u_\ba)=\cL(\bu,\blambda^\star)\ge\cL(\bu^\star,\blambda^\star)=\sum_\ba\pi_\ba\phi(u^\star_\ba)$, and $\bu^\star$ is feasible by hypothesis. Hence $\bu^\star$ is primal optimal, and it is the unique primal optimum because the objective is strictly convex on a convex feasible set. Substituting $q_\ba(\bx)=\pi_\ba(\bx)u_\ba^\star$ gives $q_\ba(\bx;\bdelta)=\pi_\ba(\bx)h(\ba^\top\blambda)$, with $\blambda$ fixed by the constraints defining \eqref{feasible-set} together with normalization. When $h(t)>0$, $u^\star=h(t)$ is interior to $[0,\infty)$, so first-order optimality of $\max_{u\ge0}\{tu-\phi(u)\}$ gives $\dot\phi(u^\star)=t$. Since $\ddot\phi>0$ makes $\dot\phi$ injective on $(0,\infty)$ (Condition~\ref{cond:f-div}), this agrees with the more familiar form $h(t)=(\dot\phi)^{-1}(t)$. The argmax formulation additionally covers the boundary case $h(t)=0$, needed for Pearson $\chi^2$'s clipped tilt (Section~\ref{sec:computing-q}).\qed

\subsection{Proof of Proposition~\ref{prop:identification}}
By definition of the counterfactual mean and the law of total expectation,
\[
\psi(\bdelta) = \E[Y^{\Q_{\bdelta}}] = \int_\cX \sum_{\ba\in\cA} \E[Y(\ba)\mid\bX=\bx]\,q_\ba(\bx;\bdelta)\,d\PP_0(\bx).
\]
By ignorability (Assumption~\ref{ignorability}), $\E[Y(\ba)\mid\bX=\bx]=\E[Y\mid\bX=\bx,\bA=\ba]=\mu_\ba(\bx)$, and Proposition~\ref{prop:kkt} together with the active-set reduction above gives $q_\ba(\bx;\bdelta)=\pi_\ba(\bx)h(\ba^\top\blambda_\cK)$, with $\blambda_\cK$ existing by Condition~\ref{cond:tilt-existence} applied to $\bdelta_{\cK(\bx)}$. Substituting,
\[
\psi(\bdelta) = \int_\cX \sum_{\ba\in\cA}\mu_\ba(\bx)\pi_\ba(\bx)h(\ba^\top\blambda_\cK)\,d\PP_0(\bx) = \E\left[\sum_{\ba\in\cA}\mu_\ba(\bX)\pi_\ba(\bX)h(\ba^\top\blambda_\cK)\right],
\]
the first identifying expression. For the second, by consistency (Assumption~\ref{consistency}) and then iterating expectations,
\[
\E\left[Y\sum_{\ba\in\cA}\bI\{\bA=\ba\}h(\ba^\top\blambda_\cK)\right] = \E\left[\sum_{\ba\in\cA}Y(\ba)\bI\{\bA=\ba\}h(\ba^\top\blambda_\cK)\right] = \E\left[\sum_{\ba\in\cA}\mu_\ba(\bX)\pi_\ba(\bX)h(\ba^\top\blambda_\cK)\right] = \psi(\bdelta),
\]
where the final equality applies ignorability as above.\qed

\subsection{Proof of Proposition~\ref{prop:bounds}}
Recall $\mu_\ba(\bx,\ba')=\E[Y(\ba)\mid\bA=\ba',\bx]$ (Assumption~\ref{asmpt:bounded}). By consistency (Assumption~\ref{consistency}) and the law of iterated expectations,
\[
\E[Y(\ba)\mid\bx] = \mu_\ba(\bx,\ba)\Big(1-\sum_{\ba'\ne\ba}\pi_{\ba'}(\bx)\Big) + \sum_{\ba'\ne\ba}\mu_\ba(\bx,\ba')\pi_{\ba'}(\bx) = \mu_\ba(\bx,\ba) + \underbrace{\sum_{\ba'\ne\ba}\big\{\mu_\ba(\bx,\ba')-\mu_\ba(\bx,\ba)\big\}\pi_{\ba'}(\bx)}_{B_\ba(\bx)},
\]
Writing $\mu_\ba(\bx):=\mu_\ba(\bx,\ba)$, Assumption~\ref{asmpt:bounded} gives $\Gamma^{-1}\mu_\ba(\bx)\le\mu_\ba(\bx,\ba')\le\min\{\Gamma\mu_\ba(\bx),1\}$ for every $\ba'\ne\ba$ (so $\mu_\ba(\bx,\ba')=0$ wherever $\mu_\ba(\bx)=0$), hence
\[
(\Gamma^{-1}-1)\mu_\ba(\bx) \;\le\; \mu_\ba(\bx,\ba')-\mu_\ba(\bx) \;\le\; \begin{cases}(\Gamma-1)\mu_\ba(\bx), & \Gamma\mu_\ba(\bx)\le1,\\ 1-\mu_\ba(\bx), & \Gamma\mu_\ba(\bx)>1,\end{cases}
\]
the upper case split coming from $\min\{\Gamma\mu_\ba(\bx),1\}-\mu_\ba(\bx)$, with $\mu_\ba(\bx,\ba')\le1$ when the cap binds. This case split depends only on $\bx$ and $\ba$, not on $\ba'$, so it may be pulled outside $\sum_{\ba'\ne\ba}$ together with the rest of the bracket. Multiplying by $\pi_{\ba'}(\bx)\ge0$ and summing over $\ba'\ne\ba$,
\[
(\Gamma^{-1}-1)\mu_\ba(\bx)\sum_{\ba'\ne\ba}\pi_{\ba'}(\bx) \;\le\; B_\ba(\bx) \;\le\; \sum_{\ba'\ne\ba}\pi_{\ba'}(\bx)\Big[(\Gamma-1)\mu_\ba(\bx)\bI\{\Gamma\mu_\ba(\bx)\le1\} + \big(1-\mu_\ba(\bx)\big)\bI\{\Gamma\mu_\ba(\bx)>1\}\Big],
\]
and since $\sum_{\ba'\ne\ba}\pi_{\ba'}(\bx)=1-\pi_\ba(\bx)$ and the bracket does not depend on $\ba'$,
\[
(\Gamma^{-1}-1)(1-\pi_\ba(\bx))\mu_\ba(\bx) \;\le\; B_\ba(\bx) \;\le\; (1-\pi_\ba(\bx))\Big[(\Gamma-1)\mu_\ba(\bx)\bI\{\Gamma\mu_\ba(\bx)\le1\} + \big(1-\mu_\ba(\bx)\big)\bI\{\Gamma\mu_\ba(\bx)>1\}\Big].
\]
Finally, by Proposition~\ref{prop:identification}'s derivation,
\begin{align*}
\psi(\bdelta)&=\E\Big[\sum_{\ba\in\cA}\E[Y(\ba)\mid\bX]\pi_\ba(\bX)h(\ba^\top\blambda_\cK)\Big]\\
&=\tilde\psi(\bdelta)+\E\Big[\sum_{\ba\in\cA}B_\ba(\bX)\pi_\ba(\bX)h(\ba^\top\blambda_\cK)\Big].
\end{align*}
Since $\pi_\ba(\bx)\ge0$ and $h(\ba^\top\blambda_\cK)\ge0$ (Proposition~\ref{prop:kkt}), multiplying the bounds on $B_\ba(\bx)$ above by $\pi_\ba(\bx)h(\ba^\top\blambda_\cK)$ preserves both inequalities, and $\pi_\ba(\bx)(1-\pi_\ba(\bx))=\pi_\ba(\bx)-\pi_\ba^2(\bx)$, so taking expectations and summing over $\ba\in\cA$ gives $\psi_{\ell,\Gamma}(\bdelta_\cK)\le\psi(\bdelta_\cK)\le\psi_{u,\Gamma}(\bdelta_\cK)$ with $\psi_{\ell,\Gamma},\psi_{u,\Gamma}$ exactly as stated.\qed

\subsection{Efficient influence functions}

\subsection{Proof of Proposition~\ref{prop:pointwiseinvert}}
Fix $\bx$. Interiority (Condition~\ref{assumption:interior}) gives $q_\ba(\bx)=\pi_\ba(\bx)h(\ba^\top\blambda_\cK)>0$ on the support, hence $h(\ba^\top\blambda_\cK)>0$ for every active $\ba$, so $u^\star:=h(\ba^\top\blambda_\cK)$ is interior to $[0,\infty)$. By the same first-order argument used in the proof of Proposition~\ref{prop:kkt}, $\dot\phi(u^\star)=\ba^\top\blambda_\cK$, and differentiating the inverse relationship $h=(\dot\phi)^{-1}$ gives $\dot h(\ba^\top\blambda_\cK)=1/\ddot\phi(u^\star)$. Since Condition~\ref{cond:f-div} gives $0<\ddot\phi(u)<\infty$ for every $u>0$, this is strictly positive for every divergence satisfying Condition~\ref{cond:f-div} -- no separate hypothesis on $\ddot h$ is needed, and in particular this covers Pearson $\chi^2$ ($\ddot\phi\equiv1$), which a bound on $\ddot h$ alone would exclude since $\ddot h\equiv0$ there. Combined with the span condition (Condition~\ref{cond:nac}), $\bOmega(\bblambda_\cK)=\sum_{\ba\in\cA}\pi_\ba(\bx)\dot h(\bba_\cK^\top\bblambda_{\cK})\bba_\cK\bba_\cK^\top$ is therefore a sum of positive-semidefinite rank-one terms whose active directions span $\R^{d(\bx)}$, so $\omega_{\min}(\bOmega)>0$ at this $\bx$.\qed

\subsection{Proof of Proposition~\ref{prop:eif}}
Fix $\bx\in\cX$ and treat $\bX$ as discrete. The general case follows by the same argument with $\eif[\E[Y\mid\bX=\bx]]=\bI\{\bX=\bx\}(Y-\E[Y\mid\bX=\bx])/\PP_0(\bX=\bx)$ interpreted as a Gateaux derivative along the usual nonparametric submodel \citep{kennedy2024semiparametric}. Write $\phi_\ba(\bO):=\bI\{\bA=\ba\}(Y-\mu_\ba(\bX))/\pi_\ba(\bX)$, a proof-local device used only through the product $\phi_\ba(\bO)\pi_\ba(\bX)=\bI\{\bA=\ba\}(Y-\mu_\ba(\bX))$, so that no inverse propensity survives in the resulting influence function. By the product rule applied to $\psi(\bdelta)=\sum_\bx\sum_{\ba\in\cA}\mu_\ba(\bx)\pi_\ba(\bx)h(\bba_\cK^\top\bblambda_\cK(\bpi(\bx);\bdelta_\cK))\PP_0(\bX=\bx)$ (Proposition~\ref{prop:identification}), differentiating each of the four factors in turn and using $\eif[\mu_\ba(\bx)]=\bI\{\bX=\bx\}\bI\{\bA=\ba\}(Y-\mu_\ba(\bx))/\{\PP_0(\bx)\pi_\ba(\bx)\}$, $\eif[\pi_\ba(\bx)]=\bI\{\bX=\bx\}(\bI\{\bA=\ba\}-\pi_\ba(\bx))/\PP_0(\bx)$, and $\eif[\PP_0(\bX=\bx)]=\bI\{\bX=\bx\}-\PP_0(\bx)$:
\begin{align*}
\eif[\psi(\bdelta)] &= \sum_{\bx}\sum_{\ba\in\cA}\Big\{\eif[\mu_\ba(\bx)]\,\pi_\ba(\bx)h(\bba_\cK^\top\bblambda_\cK)\PP_0(\bx) + \mu_\ba(\bx)\,\eif[\pi_\ba(\bx)]\,h(\bba_\cK^\top\bblambda_\cK)\PP_0(\bx) \\
&\qquad+ \mu_\ba(\bx)\pi_\ba(\bx)\,\eif[h(\bba_\cK^\top\bblambda_\cK)]\,\PP_0(\bx) + \mu_\ba(\bx)\pi_\ba(\bx)h(\bba_\cK^\top\bblambda_\cK)\,\eif[\PP_0(\bx)]\Big\}.
\end{align*}
Summing the first two terms over $\bx$ (each collapses the outer $\bI\{\bX=\bx\}$ against the observed $\bX$) and over $\ba$ (each $\bI\{\bA=\ba\}$ selects the observed $\bA$):
\[
\sum_{\ba\in\cA}\bI\{\bA=\ba\}\big(Y-\mu_\ba(\bX)\big)h(\bba_\cK^\top\bblambda_\cK) + \mu_\ba(\bX)\big(\bI\{\bA=\ba\}-\pi_\ba(\bX)\big)h(\bba_\cK^\top\bblambda_\cK) = Yh(\bbA_\cK^\top\bblambda_\cK) - \sum_{\ba\in\cA}\mu_\ba(\bX)\pi_\ba(\bX)h(\bba_\cK^\top\bblambda_\cK),
\]
using $\phi_\ba(\bO)\pi_\ba(\bX)=\bI\{\bA=\ba\}(Y-\mu_\ba(\bX))$ and that the $\mu_\ba(\bX)\bI\{\bA=\ba\}$ and $-\mu_\ba(\bX)\bI\{\bA=\ba\}(\ldots)$-type cross terms telescope. Summing the fourth term over $\bx,\ba$ and applying Proposition~\ref{prop:identification}'s first identifying expression, $\sum_\bx\sum_\ba\mu_\ba(\bx)\pi_\ba(\bx)h(\bba_\cK^\top\bblambda_\cK)\{\bI\{\bX=\bx\}-\PP_0(\bx)\} = \sum_\ba\mu_\ba(\bX)\pi_\ba(\bX)h(\bba_\cK^\top\bblambda_\cK) - \psi(\bdelta)$. Adding this to the display above cancels the $\sum_\ba\mu_\ba(\bX)\pi_\ba(\bX)h(\bba_\cK^\top\bblambda_\cK)$ term exactly, leaving $Yh(\bbA_\cK^\top\bblambda_\cK) - \psi(\bdelta)$ from terms one, two, and four combined.

It remains to evaluate the third term, $\sum_\ba\mu_\ba(\bX)\pi_\ba(\bX)\eif[h(\bba_\cK^\top\bblambda_\cK)]$. By Remark~\ref{rmk:constraints}, $\bblambda_\cK$ solves $\bF(\bpi(\bx),\bdelta_\cK;\bblambda_\cK)=\mathbf0$. Since $\bF$ is continuously differentiable and $\partial\bF/\partial\bblambda_\cK=\bOmega(\bblambda_\cK)$ is invertible (Proposition~\ref{prop:pointwiseinvert}), the implicit function theorem gives $\eif[\bblambda_\cK]=-\bOmega(\bblambda_\cK)^{-1}\,\partial\bF/\partial\bpi\cdot\eif[\bpi(\bx)]$. Differentiating Remark~\ref{rmk:constraints}'s two displays in $\bpi$ (using $\zeta_{k,\bdelta}(\bx)=\delta_km_k(\bx)/\{\delta_km_k(\bx)+1-m_k(\bx)\}$ and $\partial\zeta_{k,\bdelta}/\partial m_k=\delta_k/\{\delta_km_k(\bx)+1-m_k(\bx)\}^2$, the $(k,k)$ entry of $\bbDelta^2_{\bm}(\bX)$) and contracting with $\eif[\pi_{\ba'}(\bx)]=\bI\{\bX=\bx\}(\bI\{\bA=\ba'\}-\pi_{\ba'}(\bx))/\PP_0(\bx)$, summed over $\ba'\in\cA$, gives
\begin{align*}
\partial\bF/\partial\bpi\cdot\eif[\bpi(\bx)] &= \Big\{h(\bbA_\cK^\top\bblambda_\cK)\bbA_\cK - \E_{\Q_{\bdelta}}[\bbA_\cK\mid\bX] - \bbDelta^2_{\bm}(\bX)\big(\bbA_\cK-\E[\bbA_\cK\mid\bX]\big)\Big\}\bI\{\bX=\bx\}/\PP_0(\bx) \\
&= g(\bbA_\cK;\bblambda_\cK,\bpi(\bX))\,\bI\{\bX=\bx\}/\PP_0(\bx),
\end{align*}
using $\sum_{\ba'}\bar\ba'_\cK\pi_{\ba'}(\bx)h(\bar\ba'_\cK{}^\top\bblambda_\cK)=\E_{\Q_{\bdelta}}[\bbA_\cK\mid\bx]$ (the $\Q_{\bdelta}$-tilted first moment, since $q_{\ba'}(\bx)=\pi_{\ba'}(\bx)h(\bar\ba'_\cK{}^\top\bblambda_\cK)$ by Proposition~\ref{prop:kkt}) and, for the marginal rows, $\sum_{\ba':a'_k=1}(\bI\{\bA=\ba'\}-\pi_{\ba'}(\bx))=A_k-m_k(\bx)=\bar A_{\cK,k}-\E[\bar A_{\cK,k}\mid\bx]$. Hence $\eif[\bblambda_\cK]=-\bOmega(\bblambda_\cK)^{-1}g(\bbA_\cK;\bblambda_\cK,\bpi(\bX))\,\bI\{\bX=\bx\}/\PP_0(\bx)$, and by the chain rule,
\[
\eif[h(\bba_\cK^\top\bblambda_\cK)] = \dot h(\bba_\cK^\top\bblambda_\cK)\,\bba_\cK^\top\eif[\bblambda_\cK] = -\dot h(\bba_\cK^\top\bblambda_\cK)\,\bba_\cK^\top\bOmega(\bblambda_\cK)^{-1}g(\bbA_\cK;\bblambda_\cK,\bpi(\bX))\,\bI\{\bX=\bx\}/\PP_0(\bx).
\]
Multiplying by $\mu_\ba(\bx)\pi_\ba(\bx)\PP_0(\bx)$, summing over $\bx$ (collapsing $\bI\{\bX=\bx\}$ against $\bX$), and over $\ba\in\cA$ gives
\[
-\sum_{\ba\in\cA}Z_{\bba}(\bX)^\top\bOmega(\bblambda_\cK)^{-1}g(\bbA_\cK;\bblambda_\cK,\bpi(\bX)),
\]
with $Z_{\bba}(\bX)=\mu_\ba(\bX)\pi_\ba(\bX)\dot h(\bba_\cK^\top\bblambda_\cK)\bba_\cK$ as in the statement. Combining with the terms-one/two/four result above,
\[
\eif[\psi(\bdelta)] = Yh(\bbA_\cK^\top\bblambda_\cK) - \sum_{\ba\in\cA}Z_{\bba}(\bX)^\top\bOmega(\bblambda_\cK)^{-1}g(\bbA_\cK;\bblambda_\cK,\bpi(\bX)) - \psi(\bdelta),
\]
and dropping the $-\psi(\bdelta)$ centering term recovers the stated uncentered EIF $\varphi(\bO;\bdelta)$. Since the model is nonparametric, the tangent space is all of $L^2_0(\PP_0)$ and the influence function is unique, hence efficient.\qed

\subsection{Proof of Proposition~\ref{prop:eifbounds}}
Fix $\bx\in\cX$, treat $\bX$ as discrete as in the proof of Proposition~\ref{prop:eif}, and reuse that proof's four building blocks $\eif[\mu_\ba(\bx)]$, $\eif[\pi_\ba(\bx)]$, $\eif[h(\bba_\cK^\top\bblambda_\cK)]$, and $\eif[\PP_0(\bx)]$ directly. Since $\pi_\ba(\bx)-\pi_\ba^2(\bx)=\pi_\ba(\bx)\{1-\pi_\ba(\bx)\}$, the product rule gives $\eif[\pi_\ba(\bx)-\pi_\ba^2(\bx)]=\{1-2\pi_\ba(\bx)\}\eif[\pi_\ba(\bx)]$.

By the product rule applied to $\psi_{b,1}(\bdelta_\cK)=\sum_\bx\sum_{\ba\in\cA}\mu_\ba(\bx)\{\pi_\ba(\bx)-\pi_\ba^2(\bx)\}h(\bba_\cK^\top\bblambda_\cK)\PP_0(\bx)$, differentiating each of the four factors:
\begin{itemize}
\item $\eif[\mu_\ba(\bx)]\{\pi_\ba(\bx)-\pi_\ba^2(\bx)\}h(\bba_\cK^\top\bblambda_\cK)\PP_0(\bx)$, summed over $\bx$, uses $\eif[\mu_\ba(\bx)]\pi_\ba(\bx)=\bI\{\bX=\bx\}\bI\{\bA=\ba\}(Y-\mu_\ba(\bx))/\PP_0(\bx)$ to give $\bI\{\bA=\ba\}(Y-\mu_\ba(\bX))\{1-\pi_\ba(\bX)\}h(\bba_\cK^\top\bblambda_\cK)$.
\item $\mu_\ba(\bx)\eif[\pi_\ba(\bx)-\pi_\ba^2(\bx)]h(\bba_\cK^\top\bblambda_\cK)\PP_0(\bx)$, summed over $\bx$, gives
\[
\mu_\ba(\bX)\{\bI\{\bA=\ba\}-\pi_\ba(\bX)\}\{1-2\pi_\ba(\bX)\}h(\bba_\cK^\top\bblambda_\cK).
\]
\item $\mu_\ba(\bx)\{\pi_\ba(\bx)-\pi_\ba^2(\bx)\}\eif[h(\bba_\cK^\top\bblambda_\cK)]\PP_0(\bx)$, summed over $\bx$, uses the proof of Proposition~\ref{prop:eif}'s
\begin{align*}
\eif[h(\bba_\cK^\top\bblambda_\cK)]&=-\dot h(\bba_\cK^\top\bblambda_\cK)\bba_\cK^\top\bOmega(\bblambda_\cK)^{-1}\\
&\qquad\times g(\bbA_\cK;\bblambda_\cK,\bpi(\bX))\bI\{\bX=\bx\}/\PP_0(\bx)
\end{align*}
to give
\begin{align*}
-\mu_\ba(\bX)\{\pi_\ba(\bX)-\pi_\ba^2(\bX)\}\dot h(\bba_\cK^\top\bblambda_\cK)\bba_\cK^\top\bOmega(\bblambda_\cK)^{-1}g(\bbA_\cK;\bblambda_\cK,\bpi(\bX));
\end{align*}
\item $\mu_\ba(\bx)\{\pi_\ba(\bx)-\pi_\ba^2(\bx)\}h(\bba_\cK^\top\bblambda_\cK)\eif[\PP_0(\bx)]$, summed over $\bx$ and $\ba$, gives $\sum_\ba\mu_\ba(\bX)\{\pi_\ba(\bX)-\pi_\ba^2(\bX)\}h(\bba_\cK^\top\bblambda_\cK) - \psi_{b,1}(\bdelta_\cK)$ by definition of $\psi_{b,1}(\bdelta_\cK)$.
\end{itemize}
Summing the first three bullets over $\ba\in\cA$ and adding the fourth bullet's non-centering part inside the same sum gives $\eif[\psi_{b,1}(\bdelta_\cK)]$ exactly as stated, with $-\psi_{b,1}(\bdelta_\cK)$ contributed once, outside the sum, by the fourth bullet's centering part.

The derivation for $\psi_{b,2}(\bdelta_\cK)=\sum_\bx\sum_{\ba\in\cA}\{1-\mu_\ba(\bx)\}\{\pi_\ba(\bx)-\pi_\ba^2(\bx)\}h(\bba_\cK^\top\bblambda_\cK)\PP_0(\bx)$ is identical, replacing $\mu_\ba(\bx)$ by $1-\mu_\ba(\bx)$ throughout: since $\eif[1-\mu_\ba(\bx)]=-\eif[\mu_\ba(\bx)]$, the first bullet's factor $\{1-\pi_\ba(\bX)\}$ becomes $\{\pi_\ba(\bX)-1\}$ (the sign flip), while the remaining three bullets carry over with $\mu_\ba$ replaced by $1-\mu_\ba$ unchanged, giving $\eif[\psi_{b,2}(\bdelta_\cK)]$ exactly as stated.\qed

\subsection{Estimation}
Recall $\P f$, $\P_n f$, and the $L_2(\PP_0)$ norm $\|f\|_2$ from Section~\ref{sec:estimation}, and write $\bO_1^n$ for the independent sample (or, under cross-fitting, the complementary fold) on which $\hat\bpi,\hat\bmu$ are fit, as in Theorem~\ref{thm:onestep}. Let $\P[\hat f(\bX)]=\E[\hat f(\bX)\mid\bO_1^n]$, treating $\hat f$ as fixed conditional on $\bO_1^n$.

\subsection{Proof of Lemma~\ref{lem:Omega-stability}}
Fix $\bx$ and suppress dependence on $\bx,\cS(\bx)$ throughout. By Condition~\ref{cond:tilt-existence} applied to $\bpi(\bx)$ and $\hat\bpi(\bx)$, $\bblambda,\hat\bblambda$ exist. Let $\bF(\bpi,\bdelta_\cS;\bblambda)=\mathbf 0$ denote the reduced constraint system on $\cS$ (Remark~\ref{rmk:constraints}, restricted to $\cS$ in place of $\cK$), so that $\partial\bF/\partial\bblambda=\bOmega(\bblambda,\bpi)=\sum_{\ba\in\cA}\pi_\ba(\bx)\dot h(\bba_\cS^\top\bblambda)\bba_\cS\bba_\cS^\top$. By Condition~\ref{cond:unif-bound}, $\bOmega(\bblambda,\bpi)^{-1}$ exists with uniformly bounded operator norm, $\|\bOmega(\bblambda,\bpi)^{-1}\|_{op}+\|\bOmega(\hat\bblambda,\hat\bpi)^{-1}\|_{op}\lesssim1$. Since $\bF$ is continuously differentiable in $(\bblambda,\bpi)$, and $\partial\bF/\partial\bpi$ is uniformly bounded -- $\partial F_0/\partial\pi_{\ba'}=h(\bba'_\cS{}^\top\bblambda)$ is bounded by Condition~\ref{cond:tilt-regularity}'s bound on $|h|$, and $\partial F_k/\partial\pi_{\ba'}=a'_k h(\bba'_\cS{}^\top\bblambda)-\partial\zeta_{k,\bdelta}/\partial\pi_{\ba'}$'s second piece is the $\bbDelta^2_{\bm}(\bX)$-diagonal term of Proposition~\ref{prop:eif}'s proof, also bounded since $\delta_km_k(\bx)+1-m_k(\bx)\ge\min(1,\delta_k)>0$, and $\bba_\cS$ itself is bounded since $\ba$ is binary and $m$ is fixed -- the implicit function theorem gives that $\bpi\mapsto\bblambda_\cS(\bpi;\bdelta_\cS)$ is continuously differentiable with $\partial\bblambda/\partial\bpi=-\inv{\bOmega_\cS}(\partial\bF/\partial\bpi)$ wherever Conditions~\ref{cond:tilt-existence} and~\ref{cond:unif-bound} hold. Because Conditions~\ref{cond:tilt-existence}, \ref{cond:tilt-regularity} and~\ref{cond:unif-bound} are imposed on the convex set $\Lambda_\cS(\bx)$, which contains the solutions at $\bpi$ and at $\hat\bpi$ and hence the segment joining them, that derivative is bounded in operator norm by $C_h/\ofl$ along the whole segment, and the mean value inequality gives
\[
\|\hat\bblambda-\bblambda\|\lesssim\|\hat\bpi-\bpi\|.
\]
By the inverse-difference identity $\bOmega(\hat\bblambda,\hat\bpi)^{-1}-\bOmega(\bblambda,\bpi)^{-1}=\bOmega(\bblambda,\bpi)^{-1}\{\bOmega(\bblambda,\bpi)-\bOmega(\hat\bblambda,\hat\bpi)\}\bOmega(\hat\bblambda,\hat\bpi)^{-1}$ and the uniform operator-norm bound above,
\[
\|\bOmega(\hat\bblambda,\hat\bpi)^{-1}-\bOmega(\bblambda,\bpi)^{-1}\|_{op}\lesssim\|\bOmega(\hat\bblambda,\hat\bpi)-\bOmega(\bblambda,\bpi)\|_{op}\le\|\bOmega(\hat\bblambda,\hat\bpi)-\bOmega(\bblambda,\hat\bpi)\|_{op}+\|\bOmega(\bblambda,\hat\bpi)-\bOmega(\bblambda,\bpi)\|_{op}.
\]
For the first term, holding $\hat\bpi$ fixed, $\bOmega(\hat\bblambda,\hat\bpi)-\bOmega(\bblambda,\hat\bpi)=\sum_{\ba\in\cA}\hat\pi_\ba(\bx)\{\dot h(\bba_\cS^\top\hat\bblambda)-\dot h(\bba_\cS^\top\bblambda)\}\bba_\cS\bba_\cS^\top$. By the mean value theorem, $|\dot h(\bba_\cS^\top\hat\bblambda)-\dot h(\bba_\cS^\top\bblambda)|\le\sup_{u\text{ on the segment joining }\bba_\cS^\top\hat\bblambda,\,\bba_\cS^\top\bblambda}|\ddot h(u)|\,|\bba_\cS^\top(\hat\bblambda-\bblambda)|\lesssim\|\hat\bblambda-\bblambda\|$, where the supremum is at most $C_h$ because the segment joining $\bblambda$ and $\hat\bblambda$ lies in $\Lambda_\cS(\bx)$, on which Condition~\ref{cond:tilt-regularity} bounds $|\ddot h|$, and using that $\bba_\cS$ is bounded. Also $\|\bba_\cS\bba_\cS^\top\|_{op}=\|\bba_\cS\|^2\lesssim1$. Since $\cA$ is finite and $\sum_{\ba\in\cA}\hat\pi_\ba(\bx)\lesssim1$,
\[
\|\bOmega(\hat\bblambda,\hat\bpi)-\bOmega(\bblambda,\hat\bpi)\|_{op}\le\sum_{\ba\in\cA}\hat\pi_\ba(\bx)|\dot h(\bba_\cS^\top\hat\bblambda)-\dot h(\bba_\cS^\top\bblambda)|\,\|\bba_\cS\bba_\cS^\top\|_{op}\lesssim\|\hat\bblambda-\bblambda\|.
\]
For the second term, holding $\bblambda$ fixed, $\bOmega(\bblambda,\hat\bpi)-\bOmega(\bblambda,\bpi)=\sum_{\ba\in\cA}\{\hat\pi_\ba(\bx)-\pi_\ba(\bx)\}\dot h(\bba_\cS^\top\bblambda)\bba_\cS\bba_\cS^\top$, so by Condition~\ref{cond:tilt-regularity}'s boundedness of $\dot h$ and $\bba_\cS\bba_\cS^\top$, and finiteness of $\cA$,
\[
\|\bOmega(\bblambda,\hat\bpi)-\bOmega(\bblambda,\bpi)\|_{op}\le\sum_{\ba\in\cA}|\hat\pi_\ba(\bx)-\pi_\ba(\bx)|\,|\dot h(\bba_\cS^\top\bblambda)|\,\|\bba_\cS\bba_\cS^\top\|_{op}\lesssim\sum_{\ba\in\cA}|\hat\pi_\ba(\bx)-\pi_\ba(\bx)|\lesssim\|\hat\bpi(\bx)-\bpi(\bx)\|.
\]
Combining, $\|\bOmega(\hat\bblambda,\hat\bpi)^{-1}-\bOmega(\bblambda,\bpi)^{-1}\|_{op}\lesssim\|\hat\bblambda-\bblambda\|+\|\hat\bpi-\bpi\|\lesssim\|\hat\bpi-\bpi\|$, the second claim. For the tilt bound, the mean value theorem and Condition~\ref{cond:tilt-regularity}'s bound on $|\dot h|$ over $\Lambda_\cS(\bx)$ give $|h(\bbA^\top\hat\bblambda)-h(\bbA^\top\bblambda)|\lesssim\|\hat\bblambda-\bblambda\|\lesssim\|\hat\bpi-\bpi\|$, the third claim. Finally, if Conditions~\ref{cond:tilt-regularity} and~\ref{cond:unif-bound} hold on $\Lambda_\cS(\bx)$ uniformly over $\bx$ and over the family of active sets $\cS(\bx)$ at which the lemma is later invoked (e.g.\ $\{\cK(\bx),\cKen(\bx),\hcKen(\bx)\}$), then the constants $C_h$ and $\ofl$ above are free of $\bx$, so every implicit constant in the three claims is as well. The maps $\bx\mapsto\bblambda_\cS(\bpi(\bx);\bdelta_\cS)$ and $\bx\mapsto\bblambda_\cS(\hat\bpi(\bx);\bdelta_\cS)$ are measurable, being the unique solutions of the reduced system at measurable $\bpi,\hat\bpi$ under Condition~\ref{cond:unif-bound}. Taking $L_2(\PP_0)$ norms of both sides of all three claims above, evaluated at $\bX$, yields the three stated $L_2$ bounds with $\max_{\ba\in\cA}\|\hat\pi_\ba-\pi_\ba\|_2$ on the right-hand side, since $\|\hat\bpi(\bX)-\bpi(\bX)\|\le\sum_{\ba\in\cA}|\hat\pi_\ba(\bX)-\pi_\ba(\bX)|$ and $\cA$ is finite.\qed

\subsection{Proof of Lemma~\ref{lem:second-order}}
Throughout, the active set $\cS(\bX)$ is fixed and common to $\hat\varphi^\star_\cS$ and $\varphi^\star_\cS$, so all reduced-coordinate objects have the same dimension. Conditions~\ref{cond:tilt-existence} and~\ref{assumption:interior}, within the lemma's stated range, supply Lemma~\ref{lem:Omega-stability}'s own hypotheses. By Condition~\ref{cond:tilt-regularity} and Lemma~\ref{lem:Omega-stability} (uniform tilt and Jacobian stability under Condition~\ref{cond:unif-bound}), the maps $\bpi\mapsto\bblambda(\bpi;\bdelta_\cS)$, $\bpi\mapsto h(\bba_\cS^\top\bblambda(\bpi;\bdelta_\cS))$, and $\bpi\mapsto\bOmega_\cS(\bblambda_\cS;\bpi)^{-1}$ are uniformly Lipschitz (Lemma~\ref{lem:Omega-stability} directly) and twice continuously differentiable on $\Lambda_\cS(\bx)$ with uniformly bounded first and second derivatives, by differentiating $\partial\bblambda/\partial\bpi=-\inv{\bOmega_\cS}(\partial\bF/\partial\bpi)$ a second time. The three maps consume different derivatives of $h$, so we account for them separately. For $\bpi\mapsto\bblambda$ and $\bpi\mapsto h(\bba_\cS^\top\bblambda)$, writing $J=\partial\bblambda/\partial\bpi$, the total derivative $\partial\bOmega_\cS/\partial\pi_{\ba'}=\dot h(\bba'_\cS{}^\top\bblambda)\bba'_\cS\bba'_\cS{}^\top+\sum_{\ba\in\cA}\pi_\ba\ddot h(\bba_\cS^\top\bblambda)(\bba_\cS^\top J_{\cdot\ba'})\bba_\cS\bba_\cS^\top$ stops at $\ddot h$, so their second derivatives involve only $h,\dot h,\ddot h$, the derivatives of $\zeta_{k,\bdelta}$ --- bounded because $\delta_km_k(\bx)+1-m_k(\bx)=1+(\delta_k-1)m_k(\bx)\ge\min(1,\delta_k)>0$ for $m_k(\bx)\in[0,1]$ once $\bdelta_\cS$ is fixed and componentwise positive, which is where that hypothesis is consumed --- and $\inv{\bOmega_\cS}$. For $\bpi\mapsto\inv{\bOmega_\cS}$ only the first derivative is required below, $\partial\inv{\bOmega_\cS}/\partial\bpi=-\inv{\bOmega_\cS}(\partial\bOmega_\cS/\partial\bpi)\inv{\bOmega_\cS}$, which again stops at $\ddot h$: it enters only through $\ddot W_{\ba,\cS}$, and no step takes a third derivative of $W_{\ba,\cS}$. All bounds are then uniform by Conditions~\ref{cond:tilt-regularity} and~\ref{cond:unif-bound} on $\Lambda_\cS(\bx)$, whose path-of-multipliers requirement supplies the multiplier values along the interpolated-propensity path that the expansion below visits.

Define $e_A=(\bI\{\bA=\bb\})_{\bb\in\cA}$, $\Delta_\pi(\bX)=\hat\bpi(\bX)-\bpi(\bX)$, $\Delta_{\mu,\ba}(\bX)=\hat\mu_\ba(\bX)-\mu_\ba(\bX)$, and $\|\Delta_\pi\|_2=\max_{\bb\in\cA}\|\hat\pi_\bb-\pi_\bb\|_2$, $\|\Delta_\mu\|_2=\max_{\bb\in\cA}\|\hat\mu_\bb-\mu_\bb\|_2$.

Let $W_{\ba,\cS}(\bpi)$ be one of the smooth propensity weights appearing in the scores below, with uniformly bounded first and second derivatives in the full propensity vector $\bpi=(\pi_\ba)_{\ba\in\cA}$. Consider $\Psi_{\nu,\ba,\cS}(\PP_0)=\E[\nu_\ba(\bX)W_{\ba,\cS}(\bpi(\bX))]$, where $\nu_\ba(\bX)=\mu_\ba(\bX)$ or $\nu_\ba(\bX)=1-\mu_\ba(\bX)$. Let $s_\nu=1$ in the first case and $s_\nu=-1$ in the second, so $\hat\nu_\ba-\nu_\ba=s_\nu(\hat\mu_\ba-\mu_\ba)$. The estimating function at estimated nuisances is
\[
\hat\varphi_{\nu,\ba,\cS}(\bO) = s_\nu\bI\{\bA=\ba\}\{Y-\hat\mu_\ba(\bX)\}\frac{W_{\ba,\cS}(\hat\bpi(\bX))}{\hat\pi_\ba(\bX)} + \hat\nu_\ba(\bX)W_{\ba,\cS}(\hat\bpi(\bX)) + \hat\nu_\ba(\bX)\dot W_{\ba,\cS}(\hat\bpi(\bX))^\top\{e_A-\hat\bpi(\bX)\},
\]
where $\dot W_{\ba,\cS}$ is the gradient with respect to the full propensity vector. The ratio in the first term is notation for the bounded closed-form term appearing in the estimating function. No positivity of $\pi_\ba$ is required, since the $\hat\pi_\ba(\bX)$ denominator cancels against every weight $W_{\ba,\cS}$ considered below.

Taking conditional expectations under the true law,
\[
\P[\hat\varphi_{\nu,\ba,\cS}(\bO)\mid\bX] = s_\nu\pi_\ba(\bX)\{\mu_\ba(\bX)-\hat\mu_\ba(\bX)\}\frac{W_{\ba,\cS}(\hat\bpi(\bX))}{\hat\pi_\ba(\bX)} + \hat\nu_\ba(\bX)W_{\ba,\cS}(\hat\bpi(\bX)) + \hat\nu_\ba(\bX)\dot W_{\ba,\cS}(\hat\bpi(\bX))^\top\{\bpi(\bX)-\hat\bpi(\bX)\}.
\]
Subtracting $\nu_\ba(\bX)W_{\ba,\cS}(\bpi(\bX))$ and using $\hat\nu_\ba-\nu_\ba=s_\nu(\hat\mu_\ba-\mu_\ba)$,
\begin{align}
\nonumber&\P[\hat\varphi_{\nu,\ba,\cS}(\bO)\mid\bX] - \nu_\ba(\bX)W_{\ba,\cS}(\bpi(\bX)) \\
\nonumber&\quad= \nu_\ba(\bX)\Big[W_{\ba,\cS}(\hat\bpi) - W_{\ba,\cS}(\bpi) + \dot W_{\ba,\cS}(\hat\bpi)^\top(\bpi-\hat\bpi)\Big] \\
&\quad\quad+ \{\hat\nu_\ba(\bX)-\nu_\ba(\bX)\}\Big[W_{\ba,\cS}(\hat\bpi) - \pi_\ba\frac{W_{\ba,\cS}(\hat\bpi)}{\hat\pi_\ba} + \dot W_{\ba,\cS}(\hat\bpi)^\top(\bpi-\hat\bpi)\Big]. \label{eqn:so-R1}
\end{align}
By Taylor's theorem and the smoothness established above (Lemma~\ref{lem:Omega-stability}),
\[
\Big|W_{\ba,\cS}(\hat\bpi) - W_{\ba,\cS}(\bpi) + \dot W_{\ba,\cS}(\hat\bpi)^\top(\bpi-\hat\bpi)\Big| \lesssim \|\hat\bpi-\bpi\|^2,
\]
and, for the weights used below, again invoking Lemma~\ref{lem:Omega-stability},
\[
\Big|W_{\ba,\cS}(\hat\bpi) - \pi_\ba\frac{W_{\ba,\cS}(\hat\bpi)}{\hat\pi_\ba} + \dot W_{\ba,\cS}(\hat\bpi)^\top(\bpi-\hat\bpi)\Big| \lesssim \|\hat\bpi-\bpi\|.
\]
Combining both bounds with \eqref{eqn:so-R1}, integrating over $\bX$, and applying Cauchy--Schwarz,
\[
\big|\P\hat\varphi_{\nu,\ba,\cS} - \P[\nu_\ba W_{\ba,\cS}(\bpi)]\big| \lesssim \P\|\hat\bpi(\bX)-\bpi(\bX)\|^2 + \P\big[|\hat\nu_\ba(\bX)-\nu_\ba(\bX)|\,\|\hat\bpi(\bX)-\bpi(\bX)\|\big] \lesssim \|\Delta_\pi\|_2^2 + \|\Delta_{\mu,\ba}\|_2\|\Delta_\pi\|_2.
\]
For the original estimand, take $W_{0,\ba,\cS}(\bpi)=\pi_\ba h(\bba_\cS^\top\bblambda_\cS(\bpi;\bdelta_\cS))$ and $\nu_\ba=\mu_\ba$. Write $e_\ba=(\bI\{\bb=\ba\})_{\bb\in\cA}$, $\bD_\cS=\partial\bF/\partial\bpi$, and $h_\ba=h(\bba_\cS^\top\bblambda_\cS)$. The product and implicit-function chain rules give
\begin{align*}
\dot W_{0,\ba,\cS}(\bpi)
&=h_\ba e_\ba+\pi_\ba\dot h(\bba_\cS^\top\bblambda_\cS)\left(\frac{\partial\bblambda_\cS}{\partial\bpi}\right)^\top\bba_\cS
=h_\ba e_\ba-\pi_\ba\dot h(\bba_\cS^\top\bblambda_\cS)\bD_\cS^\top\inv{\bOmega_\cS}\bba_\cS,
\end{align*}
using $\partial\bblambda_\cS/\partial\bpi=-\inv{\bOmega_\cS}\bD_\cS$ and the symmetry of $\bOmega_\cS$. Let $\bDelta^2_{\bm,\cS}(\bX)$ be diagonal with zero in the intercept slot and entries $\delta_k/\{\delta_km_k(\bX)+1-m_k(\bX)\}^2$ for $k\in\cS$. Since $m_k(\bx)=\sum_{\ba:a_k=1}\pi_\ba(\bx)$ and
\[
\frac{\partial}{\partial\pi_{\ba'}}\frac{\delta_km_k}{\delta_km_k+1-m_k}=\frac{\delta_k}{\{\delta_km_k+1-m_k\}^2}a'_k,
\]
the target marginals contribute to $\bD_\cS$ as well as the explicit $\pi_\ba$ in $\bF$, and the columns of $\bD_\cS$ at fixed $\bblambda_\cS$ are $\partial\bF/\partial\pi_{\ba'}=h_{\ba'}\bba'_\cS-\bDelta^2_{\bm,\cS}(\bX)\bba'_\cS$. Contracting against $e_A-\bpi(\bX)$ and using $q_{\ba'}(\bx)=\pi_{\ba'}(\bx)h_{\ba'}$,
\begin{align*}
\bD_\cS\{e_A-\bpi(\bX)\}
&=h(\bbA_\cS^\top\bblambda_\cS)\bbA_\cS-\sum_{\ba'\in\cA}\pi_{\ba'}h_{\ba'}\bba'_\cS-\bDelta^2_{\bm,\cS}(\bX)\Big\{\bbA_\cS-\sum_{\ba'\in\cA}\pi_{\ba'}\bba'_\cS\Big\}\\
&=h(\bbA_\cS^\top\bblambda_\cS)\bbA_\cS-\E_{\Q_{\bdelta_\cS}}[\bbA_\cS\mid\bX]-\bDelta^2_{\bm,\cS}(\bX)\{\bbA_\cS-\E[\bbA_\cS\mid\bX]\}
=:g_\cS(\bbA_\cS;\bblambda_\cS,\bpi(\bX)).
\end{align*}
Summing the generic score over $\ba\in\cA$, the first term contributes $(Y-\mu_{\bA})h(\bbA_\cS^\top\bblambda_\cS)$, the second $\sum_{\ba}\mu_\ba\pi_\ba h_\ba$, and the third splits along $\dot W_{0,\ba,\cS}$ into a direct part $\mu_{\bA}h(\bbA_\cS^\top\bblambda_\cS)-\sum_{\ba}\mu_\ba\pi_\ba h_\ba$ and an indirect part $-\sum_{\ba}Z_{\ba,\cS}(\bX)^\top\inv{\bOmega_\cS}g_\cS$, with $Z_{\ba,\cS}(\bX)=\mu_\ba(\bX)\pi_\ba(\bX)\dot h(\bba_\cS^\top\bblambda_\cS)\bba_\cS$. The $\mu$-level terms cancel in pairs, leaving
\[
\sum_{\ba\in\cA}\varphi_{\mu,\ba,\cS}(\bO)=h(\bbA_\cS^\top\bblambda_\cS)Y-\sum_{\ba\in\cA}Z_{\ba,\cS}(\bX)^\top\inv{\bOmega_\cS}g_\cS(\bbA_\cS;\bblambda_\cS,\bpi(\bX))=\varphi(\bO;\eta,\bdelta_\cS),
\]
which at $\cS=\cK$ is exactly \eqref{eqn:eifsimple}, with $Z_{\ba,\cK}=Z_\ba$, $g_\cK=g$, and $\bOmega_\cK=\bOmega(\bblambda_\cK)$. The same algebra with every nuisance-dependent quantity replaced by its estimated counterpart gives the observation-wise identity $\sum_{\ba\in\cA}\hat\varphi_{\mu,\ba,\cS}(\bO)=\varphi(\bO;\hat\eta,\bdelta_\cS)$.

For the bounding components, take $W_{1,\ba,\cS}(\bpi)=\pi_\ba(1-\pi_\ba)h_\ba$, whose gradient is $\dot W_{1,\ba,\cS}(\bpi)=(1-2\pi_\ba)h_\ba e_\ba-\pi_\ba(1-\pi_\ba)\dot h(\bba_\cS^\top\bblambda_\cS)\bD_\cS^\top\inv{\bOmega_\cS}\bba_\cS$. The contraction against $e_A-\bpi(\bX)$ produces the same $g_\cS$, so substitution into the generic score gives $\varphi_{1,\ba}(\bO;\eta,\bdelta_\cS)$ when $(\nu_\ba,s_\nu)=(\mu_\ba,1)$ and $\varphi_{2,\ba}(\bO;\eta,\bdelta_\cS)$ when $(\nu_\ba,s_\nu)=(1-\mu_\ba,-1)$, at the true nuisances and, with hats throughout, at the estimated ones. The rate bound above, summed over the finite set $\cA$ and over the bounded linear combinations defining $\varphi^\star_\cS$, gives $\P[\hat\varphi^\star_\cS-\varphi^\star_\cS]=\cO_p(R_n)$.

Turning to the empirical-process term: by sample splitting, $\hat\varphi^\star_\cS-\varphi^\star_\cS$ is fixed conditional on the training sample $\bO_1^n$. Under cross-fitting the same argument applies on each of the finitely many evaluation folds and the foldwise conclusions average. By $L_2(\PP_0)$-consistency of the nuisance estimators and continuity of the maps established above, $\|\hat\varphi^\star_\cS-\varphi^\star_\cS\|_2=o_p(1)$. Applying \citet{kennedy2020sharp}'s Lemma 2 (an independent-sample empirical-process bound: for $f$ estimated on a sample independent of the evaluation sample, $(\P_n-\P)(\hat f-f)=\cO_p(\|\hat f-f\|_2/\sqrt n)$), $(\P_n-\P)[\hat\varphi^\star_\cS-\varphi^\star_\cS]=o_p(n^{-1/2})$.

Finally, consider the estimated upper-bound indicator. Decompose the upper-bound score's full difference $\hat\varphi^\star_\cS(\hat\eta,\hat\bI)-\varphi^\star_\cS(\eta,\bI)$ into a smooth part at fixed estimated indicator coefficients, $\hat\varphi^\star_\cS(\hat\eta,\hat\bI)-\varphi^\star_\cS(\eta,\hat\bI)$, and an indicator-replacement part, $\varphi^\star_\cS(\eta,\hat\bI)-\varphi^\star_\cS(\eta,\bI)$. The rate bound just established covers the former: its implied constant depends on the linear-combination coefficients defining $\varphi^\star_\cS$ only through their uniform bound, and $\hat\bI_\ba(\bX)\in\{0,1\}$ is fixed conditional on $\bO_1^n$ (the training sample on which $\hat\mu_\ba$ is fit), so the bound applies verbatim with $\hat\bI_\ba$ in place of the true $\bI_\ba$ -- exactly what the statement's parenthetical ``possibly depending on $\bI\{\Gamma\mu_\ba(\bX)\le1\}$'' licenses. It remains to control the latter, the indicator-replacement part. Let $\bI_\ba(\bX)=\bI\{\Gamma\mu_\ba(\bX)\le1\}$, $\hat\bI_\ba(\bX)=\bI\{\Gamma\hat\mu_\ba(\bX)\le1\}$. For each $\ba\in\cA$ define $\Xi_\ba(\bO;\eta)=(\Gamma-1)\varphi_{1,\ba}(\bO;\eta,\bdelta_\cS) - \varphi_{2,\ba}(\bO;\eta,\bdelta_\cS)$. Then $\P[\Xi_\ba(\bO;\eta)\mid\bX]=\{\Gamma\mu_\ba(\bX)-1\}\{\pi_\ba(\bX)-\pi_\ba^2(\bX)\}h(\ba_\cS^\top\bblambda_\cS)$, so the leading non-smooth drift from replacing $\bI_\ba$ by $\hat\bI_\ba$ is $B_{\bI,\ba}=\P[\Xi_\ba(\bO;\eta)\{\hat\bI_\ba(\bX)-\bI_\ba(\bX)\}]$, which by boundedness satisfies $|B_{\bI,\ba}|\lesssim\P[|\Gamma\mu_\ba(\bX)-1|\,|\hat\bI_\ba(\bX)-\bI_\ba(\bX)|]$. The two indicators differ only if $\Gamma\mu_\ba(\bX)-1$ and $\Gamma\hat\mu_\ba(\bX)-1$ lie on opposite sides of zero, which forces $|\Gamma\mu_\ba(\bX)-1|\le\Gamma|\hat\mu_\ba(\bX)-\mu_\ba(\bX)|$. Hence $|\hat\bI_\ba(\bX)-\bI_\ba(\bX)|\le\bI\{|\Gamma\mu_\ba(\bX)-1|\le\Gamma|\hat\mu_\ba(\bX)-\mu_\ba(\bX)|\}$, and with $\tau_n=\max_{\ba\in\cA}\|\hat\mu_\ba-\mu_\ba\|_\infty$ (as in the statement above) and the margin condition $\P[|\Gamma\mu_\ba(\bX)-1|\le t]\lesssim t^\varrho$,
\[
|B_{\bI,\ba}| \lesssim \P\big[|\Gamma\mu_\ba(\bX)-1|\,\bI\{|\Gamma\mu_\ba(\bX)-1|\le\Gamma \tau_n\}\big] \le \Gamma \tau_n\,\P[|\Gamma\mu_\ba(\bX)-1|\le\Gamma \tau_n] \lesssim \tau_n^{1+\varrho}.
\]
Summing over the finite set $\cA$ gives the additional drift remainder $\cO_p(\tau_n^{1+\varrho})$. The empirical-process contribution of the indicator replacement is also negligible: the difference $\hat\bI_\ba(\bX)-\bI_\ba(\bX)$ is uniformly bounded and supported on $\bigcup_{\ba\in\cA}\{|\Gamma\mu_\ba(\bX)-1|\le\Gamma \tau_n\}$, whose probability is $\cO_p(\tau_n^\varrho)=o_p(1)$ by the margin condition. Conditional on the training sample, its variance is $\cO_p(\tau_n^\varrho/n)=o_p(n^{-1})$, so its empirical-process contribution is $o_p(n^{-1/2})$. Combining the drift and empirical-process bounds for the upper-bound score gives $\P[\hat\varphi^\star_\cS-\varphi^\star_\cS]=\cO_p(R_n+\tau_n^{1+\varrho})$ and $(\P_n-\P)[\hat\varphi^\star_\cS-\varphi^\star_\cS]=o_p(n^{-1/2})$, as claimed.\qed

\subsection{Proof of Lemma~\ref{lem:tilted-law-stability}}
It suffices to prove the result when $\cS'$ is obtained from $\cS$ by dropping a single coordinate $k$. The general result follows by dropping finitely many coordinates one at a time (see the end of this proof). We write the proof for $\bpi^*=\bpi$. The case $\bpi^*=\hat\bpi$ is identical throughout with $\pi_\ba$ replaced by $\hat\pi_\ba$, $m_k$ by $\hat m_k$, and $b_k$ by $\hat b_k$, since no step below uses any property of $\bpi$ beyond $\sum_{\ba\in\cA}\pi_\ba(\bx)=1$ and $\pi_\ba(\bx)\ge0$, both of which $\hat\bpi$ shares. Write $\cS'=\cS\setminus\{k\}$, and fix $\bx$, suppressing the dependence throughout. Consider first $m_k(\bx)\le1/2$, so $b_k(\bx)=m_k(\bx)$. The case $m_k(\bx)>1/2$ is symmetric after replacing $A_k$ by $1-A_k$.

First (Step 1), we show that the common multiplier coordinates are close. Let $\blambda_{\cS,-k}$ denote $\blambda_\cS$ with its $k$-th entry additionally set to zero, so $\blambda_{\cS,-k}$ and $\blambda_{\cS'}$ both lie in $\R^{m+1}$, zero outside $\cS'$. Let $\bba_{\cS'}\in\R^{1+|\cS'|}$ denote $\bar\ba$ restricted to the intercept and $\cS'$ coordinates, and let $\bblambda_{\cS,-k},\bblambda_{\cS'}\in\R^{1+|\cS'|}$ denote the corresponding restrictions of $\blambda_{\cS,-k},\blambda_{\cS'}$. Since both are zero outside $\{0\}\cup\cS'$, this restriction is lossless ($\|\blambda_{\cS,-k}-\blambda_{\cS'}\|=\|\bblambda_{\cS,-k}-\bblambda_{\cS'}\|$) and $\ba^\top\blambda_{\cS,-k}=\bba_{\cS'}^\top\bblambda_{\cS,-k}$, $\ba^\top\blambda_{\cS'}=\bba_{\cS'}^\top\bblambda_{\cS'}$ for every $\ba\in\cA$. We show
\[
\|\bblambda_{\cS,-k}-\bblambda_{\cS'}\|\lesssim m_k.
\]
Let $\bF_{\cS'}(\bblambda):=\sum_{\ba:a_k=0}\pi_\ba(\bx)h(\bba_{\cS'}^\top\bblambda)\bba_{\cS'}$ for $\bblambda\in\R^{1+|\cS'|}$ -- a square map $\R^{1+|\cS'|}\to\R^{1+|\cS'|}$ (the constraint-generating map of Remark~\ref{rmk:constraints}, restricted to the reduced $\cS'$-system and to the common slice $\{A_k=0\}$) -- and let $t_{\cS'}$ denote the corresponding target vector (normalization plus target marginals for coordinates in $\cS'$). Since $\bblambda_{\cS'}$ solves the reduced constraint system, $\sum_{\ba\in\cA}\pi_\ba(\bx)h(\bba_{\cS'}^\top\bblambda_{\cS'})\bba_{\cS'}=t_{\cS'}$. The rare-slice ($\{A_k=1\}$) contribution is $\sum_{\ba:a_k=1}\pi_\ba(\bx)h(\bba_{\cS'}^\top\bblambda_{\cS'})\bba_{\cS'}=\cO(m_k)$ by boundedness of $h$ (Condition~\ref{cond:tilt-regularity}) and $\sum_{\ba:a_k=1}\pi_\ba(\bx)=m_k$, so $\bF_{\cS'}(\bblambda_{\cS'})=t_{\cS'}+\cO(m_k)$. Similarly, since $\blambda_\cS$ solves the larger constraint system, its common coordinates satisfy $\sum_{\ba\in\cA}\pi_\ba(\bx)h(\ba^\top\blambda_\cS)\bar\ba=t_{\cS'}$ restricted to $\{0\}\cup\cS'$. On the common slice $\{A_k=0\}$, $h(\ba^\top\blambda_\cS)=h(\bba_{\cS'}^\top\bblambda_{\cS,-k})$ because on $\{a_k=0\}$ the $k$th entry of $\blambda_\cS$ is multiplied by $a_k=0$, so zeroing it changes nothing, and the rare-slice contribution is again $\cO(m_k)$, giving $\bF_{\cS'}(\bblambda_{\cS,-k})=t_{\cS'}+\cO(m_k)$. Combining, $\bF_{\cS'}(\bblambda_{\cS,-k})-\bF_{\cS'}(\bblambda_{\cS'})=\cO(m_k)$.

To invert this via the mean value theorem, the common-slice Jacobian $\partial\bF_{\cS'}/\partial\bblambda=\sum_{\ba:a_k=0}\pi_\ba(\bx)\dot h(\bba_{\cS'}^\top\bblambda)\bba_{\cS'}\bba_{\cS'}^\top$ must be bounded below (in $\omega_{\min}$) uniformly along the segment joining $\bblambda_{\cS,-k}$ and $\bblambda_{\cS'}$, not merely at the solution $\bblambda_{\cS'}$ itself. This Jacobian equals $\bOmega(\bblambda_{\cS'})$ (Condition~\ref{cond:unif-bound}'s object, summed over the full $\cS'$-system) minus the rare-slice contribution $\sum_{\ba:a_k=1}\pi_\ba(\bx)\dot h(\bba_{\cS'}^\top\bblambda_{\cS'})\bba_{\cS'}\bba_{\cS'}^\top$, whose operator norm is $\lesssim C_hm_k$ by Condition~\ref{cond:tilt-regularity}'s bound on $\dot h$ and $\sum_{\ba:a_k=1}\pi_\ba(\bx)=m_k$. By Weyl's inequality, $\omega_{\min}(\partial\bF_{\cS'}/\partial\bblambda)\ge\omega_{\min}[\bOmega(\bblambda')]-C_h'm_k\ge\ofl-C_h'm_k$ at every $\bblambda'\in\Lambda_{\cS'}(\bx)$ (Condition~\ref{cond:unif-bound}). The enlarged $\Lambda_{\cS'}(\bx)$ of Appendix~\ref{app:lambda-def} contains $\bblambda_{\cS,-k}$ and $\bblambda_{\cS'}$ and is convex, so the segment joining them lies in $\Lambda_{\cS'}(\bx)$ and the bound holds along it. Fix $c:=\ofl/(2C_h')$: for $m_k\le c$, the common-slice Jacobian is bounded below by $\ofl/2$ throughout a neighborhood containing the segment, so the mean value theorem gives $\|\bblambda_{\cS,-k}-\bblambda_{\cS'}\|\lesssim\|\bF_{\cS'}(\bblambda_{\cS,-k})-\bF_{\cS'}(\bblambda_{\cS'})\|\lesssim m_k$, and Step 2 below converts this into the target total-variation bound. For $m_k>c$, we bypass the multiplier estimate and bound the total-variation distance directly: since $q_\cS(\bx)$ and $q_{\cS'}(\bx)$ are both probability mass functions on $\cA$, $\sum_{\ba\in\cA}|q_{\cS,\ba}-q_{\cS',\ba}|\le2\le(2/c)m_k$. Either way, $\sum_{\ba\in\cA}|q_{\cS,\ba}-q_{\cS',\ba}|\lesssim m_k$.

Second (Step 2), we establish the total-variation bound for the remaining case $m_k\le c$ (the case $m_k>c$ was resolved directly above). Split $\sum_{\ba\in\cA}|q_{\cS,\ba}-q_{\cS',\ba}|$ into the rare slice $\{A_k=1\}$ and common slice $\{A_k=0\}$. On the rare slice, boundedness of $h$ (Condition~\ref{cond:tilt-regularity}) gives $\sum_{\ba:a_k=1}q_{\cS',\ba}=\sum_{\ba:a_k=1}\pi_\ba(\bx)h(\ba^\top\blambda_{\cS'})\lesssim\sum_{\ba:a_k=1}\pi_\ba(\bx)=m_k$. Writing $\zeta_{k,\bdelta}(\bx)=\delta_km_k(\bx)/\{\delta_km_k(\bx)+1-m_k(\bx)\}$ for the shifted marginal under the larger active set $\cS$, we also have $\sum_{\ba:a_k=1}q_{\cS,\ba}=\zeta_{k,\bdelta}(\bx)\lesssim m_k$. Hence
\[
\sum_{\ba:a_k=1}|q_{\cS,\ba}-q_{\cS',\ba}|\le\sum_{\ba:a_k=1}q_{\cS,\ba}+\sum_{\ba:a_k=1}q_{\cS',\ba}\lesssim m_k.
\]
On the common slice, $q_{\cS,\ba}=\pi_\ba(\bx)h(\bba_{\cS'}^\top\bblambda_{\cS,-k})$ and $q_{\cS',\ba}=\pi_\ba(\bx)h(\bba_{\cS'}^\top\bblambda_{\cS'})$, so by Lipschitz continuity of $h$ (Condition~\ref{cond:tilt-regularity}) and Step 1,
\[
\sum_{\ba:a_k=0}|q_{\cS,\ba}-q_{\cS',\ba}| = \sum_{\ba:a_k=0}\pi_\ba(\bx)\big|h(\bba_{\cS'}^\top\bblambda_{\cS,-k})-h(\bba_{\cS'}^\top\bblambda_{\cS'})\big| \lesssim \|\bblambda_{\cS,-k}-\bblambda_{\cS'}\|\sum_{\ba:a_k=0}\pi_\ba(\bx) \lesssim m_k.
\]
Combining the two slice bounds, $\sum_{\ba\in\cA}|q_{\cS,\ba}-q_{\cS',\ba}|\lesssim m_k=b_k(\bx)$. If instead $m_k(\bx)>1/2$, the same argument applies with rare slice $\{A_k=0\}$, probability $1-m_k(\bx)$, giving $\sum_{\ba\in\cA}|q_{\cS,\ba}-q_{\cS',\ba}|\lesssim1-m_k(\bx)=b_k(\bx)$.

Third (Step 3), we handle multiple coordinates. If $\cS\setminus\cS'$ has more than one coordinate, enumerate them $k_1,\ldots,k_r$ in nondecreasing order of $b_{k_j}[\bpi^*](\bx)$ (for some such enumeration, since distinct tie-breaks give formally distinct but equally sufficient hypotheses) and define the nested sequence $\cS=\cS_0\supset\cS_1\supset\cdots\supset\cS_r=\cS'$, $\cS_j=\cS_{j-1}\setminus\{k_j\}$. By hypothesis, Conditions~\ref{cond:tilt-existence}, \ref{assumption:interior}, \ref{cond:nac}, and~\ref{cond:tilt-regularity} hold at every $\cS_j$ (including $\cS_0$), while Condition~\ref{cond:unif-bound} holds at every post-drop set $\cS_1,\ldots,\cS_r$ -- Steps 1--2 above are applied at each consecutive pair $(\cS_{j-1},\cS_j)$ using Condition~\ref{cond:unif-bound} at $\cS_j$ only, never at $\cS_{j-1}$, so no nonsingularity assumption is used at $\cS_0=\cS$ itself. By the triangle inequality and Steps 1--2 applied at each $j$,
\[
\sum_{\ba\in\cA}|q_{\cS,\ba}-q_{\cS',\ba}| \le \sum_{j=1}^r\sum_{\ba\in\cA}|q_{\cS_{j-1},\ba}-q_{\cS_j,\ba}| \lesssim \sum_{j=1}^r b_{k_j}(\bx) = \sum_{k\in\cS\setminus\cS'}b_k(\bx).
\]
\qed

\subsection{Proof of Lemma~\ref{lem:boundary-tapering}}
Fix $\bx$ and suppress its dependence where this causes no confusion. For an active set $\cS$, write $\varphi_\cS(\bO;\eta,\bdelta)=Yh(\bA^\top\blambda_\cS)-C_\cS(\bO;\eta,\bdelta)$, where $C_\cS(\bO;\eta,\bdelta)=\sum_{\ba\in\cA}Z_{\bba}(\bX)^\top\bOmega(\bblambda_\cS)^{-1}g(\bbA_\cS;\bblambda_\cS,\bpi(\bX))$ is Proposition~\ref{prop:eif}'s correction term with $\cK$ generalized to $\cS$.

The correction term has conditional mean zero for each active set: conditional on $\bX=\bx$, $Z_{\bba}(\bx)$ and $\bOmega(\bblambda_\cS)^{-1}$ are fixed, and $g(\bbA_\cS;\bblambda_\cS,\bpi(\bx))$ is linear in $\{\bI\{\bA=\ba'\}-\pi_{\ba'}(\bx)\}_{\ba'\in\cA}$ (Proposition~\ref{prop:eif}'s proof), each term of which has conditional mean zero given $\bX=\bx$. Hence $\P[g(\bbA_\cS;\bblambda_\cS,\bpi(\bx))\mid\bX=\bx]=0$ and $\P[C_\cS(\bO;\eta,\bdelta)\mid\bX=\bx]=0$. It follows that
\begin{align*}
\P\big[\{\varphi_\cK(\bO;\eta,\bdelta_\cK)-\varphi_{\cK_\epsilon}(\bO;\eta,\bdelta_{\cK_\epsilon})\}\mid\bX=\bx\big] &= \P\big[Y\{h(\bA^\top\blambda_\cK)-h(\bA^\top\blambda_{\cK_\epsilon})\}\mid\bX=\bx\big] \\
&= \sum_{\ba\in\cA}\mu_\ba(\bx)\pi_\ba(\bx)\{h(\ba^\top\blambda_\cK)-h(\ba^\top\blambda_{\cK_\epsilon})\} \\
&= \sum_{\ba\in\cA}\mu_\ba(\bx)\{q_{\cK,\ba}(\bx)-q_{\cK_\epsilon,\ba}(\bx)\}.
\end{align*}
By uniform boundedness of $\mu_\ba$ and Lemma~\ref{lem:tilted-law-stability},
\[
\big|\P[\{\varphi_\cK-\varphi_{\cK_\epsilon}\}\mid\bX=\bx]\big| \le \max_{\ba\in\cA}\|\mu_\ba\|_\infty\sum_{\ba\in\cA}|q_{\cK,\ba}(\bx)-q_{\cK_\epsilon,\ba}(\bx)| \lesssim \sum_{k\in\cK(\bx)\setminus\cK_\epsilon(\bx)}b_k(\bx),
\]
proving the conditional mean bound.

Now, $\cK_\epsilon(\bx)\ne\cK(\bx)$ only when $0<b^\star(\bx)<\epsilon$, and whenever $k\in\cK(\bx)\setminus\cK_\epsilon(\bx)$, $b_k(\bx)\le\epsilon$. Since $m$ is fixed, $\sum_{k\in\cK(\bx)\setminus\cK_\epsilon(\bx)}b_k(\bx)\le m\epsilon\,\bI\{0<b^\star(\bx)\le\epsilon\}$. Therefore, by the margin condition (Condition~\ref{cond:margin}, $\PP_0[0<b^\star(\bX)\le t]\le Ct^\beta$),
\[
\big|\P[\varphi_\cK-\varphi_{\cK_\epsilon}]\big| \le \P\big[\big|\P[\varphi_\cK-\varphi_{\cK_\epsilon}\mid\bX]\big|\big] \lesssim \epsilon\,\P[0<b^\star(\bX)\le\epsilon] \lesssim \epsilon^{1+\beta}.
\]

For the second-moment bound, a boundary-tapering bound on the correction term is not needed: by uniform boundedness of the estimating functions, $|\varphi_\cK(\bO;\eta,\bdelta_\cK)-\varphi_{\cK_\epsilon}(\bO;\eta,\bdelta_{\cK_\epsilon})|\le C$, and the two functions agree whenever $\cK_\epsilon(\bX)=\cK(\bX)$. Hence
\[
\P\big[\{\varphi_\cK-\varphi_{\cK_\epsilon}\}^2\big] \le C^2\,\P[\cK_\epsilon(\bX)\ne\cK(\bX)] \le C^2\,\P[0<b^\star(\bX)\le\epsilon] \lesssim \epsilon^\beta.
\]
Finally, $\P[\varphi_\cS(\bO;\eta,\bdelta_\cS)]=\psi(\bdelta_\cS)$ for any active set $\cS$ (Proposition~\ref{prop:identification}, since $\varphi_\cS$ is an uncentered EIF for $\psi(\bdelta_\cS)$), so the bound on $|\P[\varphi_\cK-\varphi_{\cK_\epsilon}]|$ gives $|\psi(\bdelta_{\cK_\epsilon})-\psi(\bdelta)|\lesssim\epsilon^{1+\beta}$ directly.\qed

\subsection{Proof of Theorem~\ref{thm:onestep}}
Since $m$ is fixed, $\rho_n=\sup_{k,\bx}|\hat m_k(\bx)-m_k(\bx)|\le2^{m-1}\max_{\ba\in\cA}\|\hat\pi_\ba-\pi_\ba\|_\infty=o_p(\epsilon_n)$ by hypothesis. The theorem's Conditions~\ref{cond:tilt-regularity}--\ref{cond:unif-bound} hold at $(\cK,\bpi)$ and $(\cK,\hat\bpi)$, so Lemma~\ref{lem:exact-recovery} applies and gives that $\hat\psi(\bdelta)$ has the same limiting distribution as the true-active-set estimator $\P_n\varphi(\bO;\hat\eta,\bdelta_\cK)$. For the latter, write $\hat\varphi_\cK=\varphi(\bO;\hat\eta,\bdelta_\cK)$ and $\varphi_\cK=\varphi(\bO;\eta,\bdelta_\cK)$, and decompose
\[
\P_n\hat\varphi_\cK-\psi(\bdelta)
=(\P_n-\P)\varphi_\cK
+\P[\hat\varphi_\cK-\varphi_\cK]
+(\P_n-\P)[\hat\varphi_\cK-\varphi_\cK]
+\{\P\varphi_\cK-\psi(\bdelta)\}.
\]
The last bracket vanishes: $\varphi_\cK=\varphi(\bO;\eta,\bdelta)$ because the constraints for $k\notin\cK(\bX)$ are redundant, and $\P\varphi_\cS=\psi(\bdelta_\cS)$ at every active set $\cS$ by Propositions~\ref{prop:identification} and~\ref{prop:eif}, the latter giving the correction term conditional mean zero. Hence $\P\varphi_\cK=\psi(\bdelta_\cK)=\psi(\bdelta)$. The two middle terms are term (iii) of \eqref{eqn:step0} after exact active-set recovery, and Lemma~\ref{lem:second-order} applied at the fixed set $\cS=\cK$ bounds them by $\cO_p(R_n)$ and $o_p(n^{-1/2})$ respectively, giving
\[
\P_n\varphi(\bO;\hat\eta,\bdelta_\cK)-\psi(\bdelta) = (\P_n-\P)\varphi(\bO;\eta,\bdelta) + \cO_p(R_n) + o_p(n^{-1/2}),
\]
and the stated product nuisance rate gives $R_n=o_p(n^{-1/2})$. Since $\Var[\varphi(\bO;\eta,\bdelta)]>0$ by hypothesis, the central limit theorem applied to $(\P_n-\P)\varphi(\bO;\eta,\bdelta)$, together with Slutsky's theorem, gives $\sqrt n\{\hat\psi(\bdelta)-\psi(\bdelta)\}\rightsquigarrow\cN(0,\Var[\varphi(\bO;\eta,\bdelta)])$. Lemma~\ref{lem:variance-consistency} gives $\hat\sigma^2\overset{\PP}{\to}\Var[\varphi(\bO;\eta,\bdelta)]$, and a further application of Slutsky's theorem yields the stated Wald coverage.\qed

\subsection{Proof of Theorem~\ref{thm:master-expansion}}
Throughout, $\varphi^\star$ denotes any of the three objects the theorem's statement allows (the uncentered EIF for $\psi(\bdelta_\cK)$, for $\psi_{\ell,\Gamma}(\bdelta_\cK)$, or the upper-bound estimating function $\varphi_{u,\Gamma}$). The argument is identical for each, so we suppress the star. Terms (I)--(III) below are the $\varphi^\star$-generalizations of terms (i)--(iii) of \eqref{eqn:step0}, which is stated for $\varphi$ and $\bdelta$. The uppercase labels are used here to mark that wider scope. By \eqref{eqn:step0},
\begin{align*}
\hat\psi(\bdelta_\cK)-\psi(\bdelta_\cK) &= (\P_n-\P)\varphi_\cK \\
&\quad + \underbrace{\P[\varphi_{\cKen}-\varphi_\cK] + (\P_n-\P)[\varphi_{\cKen}-\varphi_\cK]}_{\text{(I)}} \\
&\quad + \underbrace{\P[\hat\varphi_{\hcKen}-\hat\varphi_{\cKen}] + (\P_n-\P)[\hat\varphi_{\hcKen}-\hat\varphi_{\cKen}]}_{\text{(II)}} \\
&\quad + \underbrace{\P[\hat\varphi_{\cKen}-\varphi_{\cKen}] + (\P_n-\P)[\hat\varphi_{\cKen}-\varphi_{\cKen}]}_{\text{(III)}}.
\end{align*}

For the upper-bound score, the plug-in indicator $\bI\{\Gamma\hat\mu_\ba(\bX)\le1\}$ is handled inside Lemma~\ref{lem:second-order}, whose second conclusion contributes the $V_n=\tau_n^{1+\varrho}$ term under the margin condition assumed above. It is carried through the display below without further comment. For term (I), threshold crossing: by Lemma~\ref{lem:boundary-tapering} applied at $\epsilon=\epsilon_n$, $|\P[\varphi_{\cKen}-\varphi_\cK]|\lesssim\epsilon_n^{1+\beta}=o(n^{-1/2})$ by hypothesis, and $\P[(\varphi_{\cKen}-\varphi_\cK)^2]\lesssim\epsilon_n^\beta$, so by Chebyshev's inequality $(\P_n-\P)[\varphi_{\cKen}-\varphi_\cK]=\cO_p(\sqrt{\epsilon_n^\beta/n})=\cO_p(n^{-1/2}\epsilon_n^{\beta/2})=o_p(n^{-1/2})$ since $\epsilon_n\to0$ (see also \citet{kennedy2020sharp}'s Lemma 2). Hence term (I) is $o_p(n^{-1/2})$.

For term (II), active-set estimation: let $\cG_n=\{\rho_n\le\epsilon_n/2\}$. By hypothesis $\rho_n=o_p(\epsilon_n)$, so $\P[\cG_n^c]=\P[\rho_n/\epsilon_n>1/2]\to0$. Note $\cG_n$ is fixed conditional on $\bO_1^n$ (the sample fitting $\hat\bpi$, hence $\hat m_k$). On $\cG_n$: if $m_k(\bX)=0$ then $\hat m_k(\bX)\le\epsilon_n/2<\epsilon_n$, and if $m_k(\bX)=1$ then $\hat m_k(\bX)\ge1-\epsilon_n/2>1-\epsilon_n$, so exact boundary coordinates are never falsely (de)activated. Hence $\hcKen(\bX)\subseteq\cK(\bX)$ on $\cG_n$ (the same argument as Lemma~\ref{lem:exact-recovery}'s estimated-set case). Moreover, on $\cG_n$, every $k\in\cKen(\bX)\triangle\hcKen(\bX)$ satisfies $\epsilon_n/2<b_k(\bX)\le3\epsilon_n/2$: if $k\in\cKen(\bX)\setminus\hcKen(\bX)$ then $b_k(\bX)>\epsilon_n$ and $\hat b_k(\bX)\le\epsilon_n$, so $b_k(\bX)\le\hat b_k(\bX)+\rho_n\le3\epsilon_n/2$. If $k\in\hcKen(\bX)\setminus\cKen(\bX)$ then $\hat b_k(\bX)>\epsilon_n$ and $b_k(\bX)\le\epsilon_n$, so $b_k(\bX)\ge\hat b_k(\bX)-\rho_n>\epsilon_n/2$. Since $m$ is fixed, $\sum_{k\in\cKen(\bX)\triangle\hcKen(\bX)}b_k(\bX)\le\tfrac32m\epsilon_n\bI\{0<b^\star(\bX)\le3\epsilon_n/2\}$ on $\cG_n$, and in particular $\{\hcKen(\bX)\ne\cKen(\bX)\}\cap\cG_n\subseteq\{0<b^\star(\bX)\le3\epsilon_n/2\}$.

Split $\hat\varphi_{\hcKen}-\hat\varphi_{\cKen}=(\varphi_{\hcKen}-\varphi_{\cKen})+\{(\hat\varphi_{\hcKen}-\varphi_{\hcKen})-(\hat\varphi_{\cKen}-\varphi_{\cKen})\}$, and control the two pieces separately.

\emph{(II-a): the true-$\eta$ active-set-replacement term $\varphi_{\hcKen}-\varphi_{\cKen}$.} Since $\hcKen(\bX),\cKen(\bX)$ are both fixed conditional on $\bO_1^n$, this is exactly Lemma~\ref{lem:boundary-tapering}'s proof mechanism applied with the pair $(\hcKen,\cKen)$ in place of $(\cK_\epsilon,\cK)$, conditional on $\bO_1^n$: the correction term $C_\cS(\bO;\eta,\bdelta)$ has conditional mean zero for every active set (Proposition~\ref{prop:eif}'s proof, at the true $\eta$), so $\P[\{\varphi_{\hcKen}-\varphi_{\cKen}\}\mid\bX,\bO_1^n]=\sum_\ba\mu_\ba(\bX)\{q_{\hcKen,\ba}(\bX)-q_{\cKen,\ba}(\bX)\}$, with $q$ evaluated at the true $\bpi$ throughout (Lemma~\ref{lem:tilted-law-stability}'s default $\bpi^*=\bpi$). On $\cG_n$, both $\hcKen(\bX)\subseteq\cK(\bX)$ and $\cKen(\bX)\subseteq\cK(\bX)$, so the common refinement $\cKen(\bX)\cap\hcKen(\bX)$ is nested inside both, and Lemma~\ref{lem:tilted-law-stability}'s total-variation bound applies to each of the two nested comparisons $\cKen(\bX)\to\cKen(\bX)\cap\hcKen(\bX)$ and $\hcKen(\bX)\to\cKen(\bX)\cap\hcKen(\bX)$. By the triangle inequality, $\sum_\ba|q_{\hcKen,\ba}(\bX)-q_{\cKen,\ba}(\bX)|\lesssim\sum_{k\in\cKen(\bX)\triangle\hcKen(\bX)}b_k(\bX)\lesssim\epsilon_n\bI\{0<b^\star(\bX)\le3\epsilon_n/2\}$ on $\cG_n$, using the every-$k$ bound just established. Hence, by boundedness of $\mu_\ba$,
\[
\big|\P\big[\{\varphi_{\hcKen}-\varphi_{\cKen}\}\bI\{\cG_n\}\mid\bO_1^n\big]\big| \lesssim \epsilon_n\,\P[0<b^\star(\bX)\le3\epsilon_n/2] \lesssim \epsilon_n^{1+\beta}=o(n^{-1/2}),
\]
and similarly, by boundedness, $\P[\{\varphi_{\hcKen}-\varphi_{\cKen}\}^2\bI\{\cG_n\}\mid\bO_1^n]\lesssim\P[0<b^\star(\bX)\le3\epsilon_n/2]\lesssim\epsilon_n^\beta$, so by Chebyshev's inequality $(\P_n-\P)[\{\varphi_{\hcKen}-\varphi_{\cKen}\}\bI\{\cG_n\}\mid\bO_1^n]=\cO_p(n^{-1/2}\epsilon_n^{\beta/2})=o_p(n^{-1/2})$. On $\cG_n^c$: since $\varphi_{\hcKen}-\varphi_{\cKen}$ is fixed conditional on $\bO_1^n$ and uniformly bounded, $|\P[\{\varphi_{\hcKen}-\varphi_{\cKen}\}\bI\{\cG_n^c\}\mid\bO_1^n]|\le C\bI\{\cG_n^c\}$, so for every $t>0$, $\P[\sqrt n\,|\P[\{\varphi_{\hcKen}-\varphi_{\cKen}\}\bI\{\cG_n^c\}\mid\bO_1^n]|>t]\le\P[\sqrt nC\bI\{\cG_n^c\}>t]\le\P[\cG_n^c]\to0$, so this drift contribution is $o_p(n^{-1/2})$. Similarly the empirical-process contribution on $\cG_n^c$ is $o_p(n^{-1/2})$. Combining, (II-a) is $o_p(n^{-1/2})$.

\emph{(II-b): the nuisance-drift difference $(\hat\varphi_{\hcKen}-\varphi_{\hcKen})-(\hat\varphi_{\cKen}-\varphi_{\cKen})$.} Split this difference over $\cG_n$ and $\cG_n^c$, exactly as in (II-a). On $\cG_n$, both $\hcKen$ and $\cKen$ lie in the family $\cKen\cap\hcKen\subseteq\cS\subseteq\cKen\cup\hcKen$ on which the theorem's hypothesis grants Conditions~\ref{cond:tilt-regularity}--\ref{cond:unif-bound}, and both are measurable with respect to the training sample $\bO_1^n$, so Lemma~\ref{lem:second-order} (widened to admit $\hcKen$) applies at $\cS=\hcKen$ and, as in Term (III) below, at $\cS=\cKen$: each gives $\P[\{\hat\varphi_\cS-\varphi_\cS\}\bI\{\cG_n\}]=\cO_p(R_n+V_n)$ and $(\P_n-\P)[\{\hat\varphi_\cS-\varphi_\cS\}\bI\{\cG_n\}]=o_p(n^{-1/2})$. Taking the difference, the $\cG_n$ contribution to (II-b) is $\cO_p(R_n+V_n)+o_p(n^{-1/2})$. On $\cG_n^c$, both $\hat\varphi_{\hcKen}-\varphi_{\hcKen}$ and $\hat\varphi_{\cKen}-\varphi_{\cKen}$ are uniformly bounded and fixed conditional on $\bO_1^n$, so exactly as in (II-a)'s $\cG_n^c$ treatment, $\P[\sqrt n\,|\P[\{(\hat\varphi_{\hcKen}-\varphi_{\hcKen})-(\hat\varphi_{\cKen}-\varphi_{\cKen})\}\bI\{\cG_n^c\}\mid\bO_1^n]|>t]\le\P[\cG_n^c]\to0$ for every $t>0$, and similarly for the empirical-process piece, so the $\cG_n^c$ contribution is $o_p(n^{-1/2})$. Combining, (II-b) is $\cO_p(R_n+V_n)+o_p(n^{-1/2})$.

Combining (II-a) and (II-b), term (II) is $\cO_p(R_n+V_n)+o_p(n^{-1/2})$.

For term (III), smooth nuisance drift: since $\cKen(\bX)$ is deterministic conditional on the observed law (a fixed, non-random function of $\bX$) and all reduced-coordinate objects are computed on this single, fixed active set, Lemma~\ref{lem:second-order} applies directly at $\cS=\cKen$, giving $\P[\hat\varphi_{\cKen}-\varphi_{\cKen}]=\cO_p(R_n+V_n)$ and $(\P_n-\P)[\hat\varphi_{\cKen}-\varphi_{\cKen}]=o_p(n^{-1/2})$, i.e.\ term (III) is $\cO_p(R_n+V_n)+o_p(n^{-1/2})$.

Combining terms (I)--(III) with \eqref{eqn:step0},
\[
\hat\psi^\star(\bdelta_\cK)-\psi^\star(\bdelta_\cK) = (\P_n-\P)\varphi(\bO;\eta,\bdelta_\cK) + \cO_p(R_n+V_n) + o_p(n^{-1/2}).
\]
When $R_n+V_n=o_p(n^{-1/2})$, the remainder is $o_p(n^{-1/2})$ and the central limit theorem applied to $(\P_n-\P)\varphi(\bO;\eta,\bdelta_\cK)$ gives $\sqrt n\{\hat\psi^\star(\bdelta_\cK)-\psi^\star(\bdelta_\cK)\}\rightsquigarrow\cN(0,\Var[\varphi^\star(\bO;\eta,\bdelta_\cK)])$.\qed

\subsection{Proof of Corollary~\ref{cor:adaptive}}
The proof follows the structure of Theorem~\ref{thm:master-expansion}'s proof, with three changes. First, since the target estimand is $\psi(\bdelta_{\cK_\epsilon})$ itself (not $\psi(\bdelta)$), there is no oracle-thresholding term of the form $\P[\varphi_{\cK_{\epsilon_n}}-\varphi_\cK]$ to control.

Second, the active-set-estimation term is controlled by the localized margin condition (Condition~\ref{cond:localized-margin}) around the fixed cutoffs $\{\epsilon,1-\epsilon\}$, rather than the global margin condition around $\{0,1\}$. Since $\rho_n=\cO_p(n^{-\gamma})$, let $\tilde r_n\to0$ satisfy $\rho_n=o_p(\tilde r_n)$ (e.g.\ $\tilde r_n=n^{-\gamma}\log n$), and let $\cG_n=\{\rho_n\le\tilde r_n\}$, so $\P[\cG_n^c]\to0$. On $\cG_n$, $\hat\cK_\epsilon(\bX)\ne\cK_\epsilon(\bX)$ only if some $\hat m_k(\bX)$ falls within $\tilde r_n$ of $\epsilon$ or $1-\epsilon$, which (up to a constant depending on $\rho_n\le\tilde r_n$) implies $b^\star_\epsilon(\bX)\lesssim\tilde r_n$. By the localized margin condition, $\P[\hat\cK_\epsilon(\bX)\ne\cK_\epsilon(\bX)\mid\bO_1^n]\lesssim\P[b^\star_\epsilon(\bX)\lesssim\tilde r_n]\lesssim\tilde r_n^\beta$ on $\cG_n$. By uniform boundedness of the estimating functions (no finer tapering argument is invoked here), $|\hat\varphi_{\hat\cK_\epsilon}-\hat\varphi_{\cK_\epsilon}|\le C\bI\{\hat\cK_\epsilon(\bX)\ne\cK_\epsilon(\bX)\}$, so
\[
\big|\P\big[\{\hat\varphi_{\hat\cK_\epsilon}-\hat\varphi_{\cK_\epsilon}\}\bI\{\cG_n\}\mid\bO_1^n\big]\big| \lesssim \tilde r_n^\beta = o(n^{-1/2})
\]
since $\gamma>\frac1{2\beta}$ gives $\tilde r_n^\beta\asymp n^{-\gamma\beta}\log^\beta n=o(n^{-1/2})$, and similarly $\P[\{\hat\varphi_{\hat\cK_\epsilon}-\hat\varphi_{\cK_\epsilon}\}^2\bI\{\cG_n\}\mid\bO_1^n]\lesssim\tilde r_n^\beta=o(1)$, so $(\P_n-\P)[\{\hat\varphi_{\hat\cK_\epsilon}-\hat\varphi_{\cK_\epsilon}\}\bI\{\cG_n\}\mid\bO_1^n]=\cO_p(\sqrt{\tilde r_n^\beta/n})=o_p(n^{-1/2})$. The $\cG_n^c$ contribution is handled exactly as in Theorem~\ref{thm:master-expansion}'s proof (uniform boundedness plus $\P[\cG_n^c]\to0$), giving $o_p(n^{-1/2})$ for both pieces there too. Combining, the active-set-estimation term is $o_p(n^{-1/2})$.

Third, since Conditions~\ref{cond:tilt-regularity}--\ref{cond:unif-bound} are assumed only for $\cS(\bx)\in\{\cK_\epsilon(\bx),\hat\cK_\epsilon(\bx)\}$ and $\bpi^*\in\{\bpi,\hat\bpi\}$ (this corollary's own hypothesis), Lemma~\ref{lem:second-order} applies at the fixed, non-random active set $\cS=\cK_\epsilon$ exactly as in Theorem~\ref{thm:master-expansion}'s proof of the nuisance-drift term, giving $\P[\hat\varphi_{\cK_\epsilon}-\varphi_{\cK_\epsilon}]=\cO_p(R_n)$ and $(\P_n-\P)[\hat\varphi_{\cK_\epsilon}-\varphi_{\cK_\epsilon}]=o_p(n^{-1/2})$.

Combining the three pieces, $\hat\psi(\bdelta_{\hat\cK_\epsilon})-\psi(\bdelta_{\cK_\epsilon})=(\P_n-\P)\varphi(\bO;\eta,\bdelta_{\cK_\epsilon})+\cO_p(R_n)+o_p(n^{-1/2})$, and the central limit theorem gives $\sqrt n\{\hat\psi(\bdelta_{\hat\cK_\epsilon})-\psi(\bdelta_{\cK_\epsilon})\}\rightsquigarrow\cN(0,\Var[\varphi(\bO;\eta,\bdelta_{\cK_\epsilon})])$ under the product nuisance rate of Theorem~\ref{thm:onestep}.\qed

\subsection{Missing outcomes}
If $Y$ is observed only when $R=1$ and $Y\perp R\mid\bX,\bA$, then $\E[Y\mid\bx,\ba,R=1]=\mu_\ba(\bx)$, so Proposition~\ref{prop:identification} identifies $\psi(\bdelta)$ with this regression in place of $\mu_\ba$. We do not derive the corresponding influence function or estimator here.

\subsection{Proof of Corollary~\ref{cor:onestep-fixed}}
When $\cK(\cdot)$ is a known measurable function of $\bx$, the estimator $\hat\psi(\bdelta_{\cK(\cdot)})=\P_n\varphi(\bO;\hat\eta,\bdelta_{\cK(\cdot)})$ plugs in $\cK(\cdot)$ directly: neither a thresholded set $\cKen(\cdot)$ nor an estimated set $\hcKen(\cdot)$ is ever formed, so terms (i)--(ii) of the decomposition \eqref{eqn:step0} do not arise at all (not merely vanish asymptotically). The decomposition therefore collapses to $\hat\psi(\bdelta_{\cK(\cdot)})-\psi(\bdelta_{\cK(\cdot)}) = (\P_n-\P)\varphi_{\cK(\cdot)} + \text{term (III)}$, and Lemma~\ref{lem:second-order} applied with $\cS=\cK(\cdot)$ gives term (III) $=\cO_p(R_n)+o_p(n^{-1/2})$ under the stated product nuisance rate. Since $\cK(\cdot)=\cK$ by hypothesis, the constraints for $k\notin\cK(\bx)$ are redundant, so $\varphi_{\cK(\cdot)}=\varphi(\bO;\eta,\bdelta)$ and $\psi(\bdelta_{\cK(\cdot)})=\psi(\bdelta)$. The centering above is therefore the $\psi(\bdelta)$ of the statement, and the limiting variance is $\Var[\varphi(\bO;\eta,\bdelta)]$. The claim follows from the CLT applied to $(\P_n-\P)\varphi_{\cK(\cdot)}$, exactly as in the proof of Theorem~\ref{thm:onestep} with the exact-recovery and active-set-estimation steps unnecessary.\qed

\subsection{Proof of the pooled-estimator representation}
When $\cK(\bx)$ takes finitely many values $\cK_1,\ldots,\cK_J$ with population weights $w_j=\PP_0[\cK(\bX)=\cK_j]$, write $n_j=\#\{i:\cK(\bX_i)=\cK_j\}$, $\hat w_j=n_j/n$, and $\hat\psi_j(\bdelta_{\cK_j})=n_j^{-1}\sum_{i:\cK(\bX_i)=\cK_j}\varphi(\bO_i;\hat\eta,\bdelta_{\cK_j})$ for the within-stratum one-step estimator (using the same fitted nuisances $\hat\eta$ throughout, not stratum-specific refits), with the convention $\hat w_j\hat\psi_j(\bdelta_{\cK_j}):=0$ if $n_j=0$. Since the active set plugged in for observation $i$ is $\cK(\bX_i)$, and $\cK(\bX_i)=\cK_j$ for every $i$ in stratum $j$ by definition of the strata, $\varphi(\bO_i;\hat\eta,\bdelta_{\cK(\bX_i)})=\varphi(\bO_i;\hat\eta,\bdelta_{\cK_j})$ there. Partitioning the sample average defining $\hat\psi(\bdelta_{\cK(\cdot)})$ by stratum gives
\[
\hat\psi(\bdelta_{\cK(\cdot)}) = \P_n\varphi(\bO;\hat\eta,\bdelta_{\cK(\cdot)}) = \sum_{j=1}^J \frac{n_j}{n}\cdot\frac{1}{n_j}\sum_{i:\cK(\bX_i)=\cK_j}\varphi(\bO_i;\hat\eta,\bdelta_{\cK_j}) = \sum_{j=1}^J \hat w_j\,\hat\psi_j(\bdelta_{\cK_j}),
\]
an exact algebraic identity, not merely an asymptotic equivalence: the pooled average on the right-hand side and $\hat\psi(\bdelta_{\cK(\cdot)})$ on the left are the same number for every $n$, not just the same limit. Corollary~\ref{cor:onestep-fixed}, applied to $\hat\psi(\bdelta_{\cK(\cdot)})$, therefore already gives $\sqrt n\{\hat\psi(\bdelta_{\cK(\cdot)})-\psi(\bdelta_{\cK(\cdot)})\}\rightsquigarrow\cN(0,\Var[\varphi(\bO;\eta,\bdelta_{\cK(\cdot)})])$. By the identity above this is simultaneously a CLT for the pooled average $\sum_j\hat w_j\hat\psi_j(\bdelta_{\cK_j})$, with no separate weighted-average argument needed beyond recognizing that the two expressions coincide exactly.\qed

\section{Simulation details}\label{app:simdetails}

This appendix provides complete specifications for the simulation study.

\subsection{Study 1 details: implementation validation and nuisance robustness}\label{app:study1}

\paragraph{Data generating process.} We generated $n=1000$ observations with covariates $(X_1,X_2)\stackrel{\text{iid}}{\sim}N(0,1)$ and $m=4$ binary treatments, assigned sequentially so that each depends on the covariates and all previous treatments:
\[
\logit\PP(A_1\mid\bX)=0.5X_1+0.3X_2,\qquad
\logit\PP(A_k\mid\bX,A_{<k})=0.5X_1+0.3X_2+0.4\sum_{j<k}A_j,\quad k=2,3,4,
\]
creating positive dependence among treatments (receiving earlier treatments increases the probability of receiving later ones, conditional on covariates). The outcome model included main effects and all pairwise interactions,
\[
Y=1.0+0.3X_1+0.4X_2+\sum_{k=1}^4 0.5\,A_k+\sum_{k<j}0.2\,A_kA_j+\xi,\qquad\xi\sim N(0,1).
\]
The target intervention was $\bdelta=(1.5,1.5,1.5,1.5)$ (a uniform $50\%$ increase in treatment odds), evaluated across all four divergences (KL, reverse KL, $\chi^2$, Hellinger). True parameters were computed by Monte Carlo integration ($10^6$ observations), giving $\psi_0=2.872$ (KL), $2.879$ (reverse KL), $2.854$ ($\chi^2$), and $2.876$ (Hellinger) -- the variation across divergences reflects the different counterfactual distributions $\Q_{\bdelta}$ satisfying the same marginal constraints.

\paragraph{Nuisance estimation scenarios.}
\begin{itemize}
\item \textbf{Oracle:} true sequential propensities and conditional outcome means from the known DGP, benchmarking the irreducible finite-sample and tilting error.
\item \textbf{Parametric (correct):} correctly specified logistic propensity models (including dependence on prior treatments) and a linear outcome model with all interactions, which achieves the $n^{-1/2}$ rate.
\item \textbf{XGBoost:} gradient-boosted trees for both propensity and outcome models, default tuning, no knowledge of the true functional forms.
\item \textbf{Consistent (fast rate):} independent mean-zero Gaussian noise added to the oracle nuisances, $\hat\pi_\ba(\bX)=\pi_\ba(\bX)+\sigma_\pi\cdot n^{-\gamma}\xi$ and $\hat\mu_\ba(\bX)=\mu_\ba(\bX)+\sigma_\mu\cdot n^{-\gamma}\xi$ with $\xi\sim N(0,1)$, $\gamma=0.3$, $\sigma_\pi,\sigma_\mu$ calibrated to the oracle functions' empirical standard deviations (propensities clipped to $[0,1]$ and renormalized after adding noise). Because the noise is mean-zero this is a \emph{consistent}, not misspecified, estimator with $\|\hat\pi_\ba-\pi_\ba\|_2=\|\hat\mu_\ba-\mu_\ba\|_2=\cO_p(n^{-0.3})$, so the product condition in Theorem~\ref{thm:onestep} gives $R_n=\cO_p(n^{-0.6})=o_p(n^{-1/2})$, satisfying the rate requirement. At $\gamma=1/2$ this construction would reproduce the sampling behavior of a correctly specified parametric estimator. We choose slower rates $\gamma<1/2$ to stress the estimator.
\item \textbf{Consistent (slow rate):} same construction with $\gamma=0.1$, giving $\|\hat\pi_\ba-\pi_\ba\|_2=\|\hat\mu_\ba-\mu_\ba\|_2=\cO_p(n^{-0.1})$ and $R_n=\cO_p(n^{-0.2})$, which \emph{violates} $R_n=o_p(n^{-1/2})$ -- consistent but outside the rate condition's stated regime.
\end{itemize}

\paragraph{Metrics.} For each of 1000 replications per method-divergence combination, we computed the one-step estimator with all treatments active, $\hat\psi(\bdelta)=\P_n[\varphi(\bO;\hat\eta,\bdelta_\cK)]$ with $\cK(\bx)=\{1,2,3,4\}$ (Corollary~\ref{cor:onestep-fixed}), and reported bias, empirical standard error, and Wald coverage (Section~\ref{sec:sims}). Every marginal propensity lies strictly inside $(0,1)$. Because the covariates are Gaussian the population infimum of the distance to $\{0,1\}$ is zero, but over $2\times10^6$ draws of $\bX$ the smallest distance is $0.014$ and $0.06\%$ of draws lie within $0.05$ of the boundary, so $\cK(\bx)=\{1,2,3,4\}$ for every $\bx$.

\subsection{Study 2 details: scaling, dimensionality, and overlap}\label{app:study2}

\paragraph{Data generating process.} We generated covariates $(X_1,X_2)\stackrel{\text{iid}}{\sim}N(0,1)$ and $m$ sequentially-assigned binary treatments, with an overlap parameter $\kappa$ scaling the linear predictor,
\[
\logit\PP(A_k=1\mid\bX,A_{<k})=\frac1\kappa\Big(0.5X_1+0.3X_2+0.4\sum_{j<k}A_j\Big),\quad k=1,\ldots,m.
\]
Larger $\kappa$ shrinks all coefficients toward zero, pushing marginal propensities toward $0.5$ (good overlap). The value $\kappa=1.2$ produces the strongest imbalance. The outcome combined main effects and adjacent-pair interactions, each normalized so the total treatment signal stays stable as $m$ grows,
\[
Y=1.0+0.3X_1+0.4X_2+\frac{0.5}{m}\sum_{k=1}^m A_k+\frac{0.3}{m-1}\sum_{k=1}^{m-1}A_kA_{k+1}+\xi,\qquad\xi\sim N(0,1).
\]
We crossed $n\in\{200,500,1000,2000,5000\}$, $m\in\{2,3,4,5,6\}$, and $\kappa\in\{1.2,1.5,2,3,5\}$, target $\bdelta=(1.5,\ldots,1.5)$ under KL divergence, 1000 replications per cell for each of the two estimators below ($125$ design cells -- one per $(n,m,\kappa)$ triple -- $250{,}000$ fits). Because the outcome contains treatment interactions, $\psi_0(\bdelta)=\E[Y^{\Q_{\bdelta}}]$ depends on the joint KL-tilted law and cannot be obtained from a marginal plug-in. We computed it by Gauss--Hermite tensor quadrature over $(X_1,X_2)\sim N(0,1)^2$, evaluating $\sum_\ba\pi_\ba(\bx)h(\ba^\top\lambda(\bx))\mu_\ba(\bx)$ at each node, agreeing with an independent Monte Carlo evaluation to within $3\times10^{-3}$. True parameters ranged from $\psi_0=1.413$ ($m=2,\kappa=5$) to $\psi_0=1.493$ ($m=6,\kappa=1.2$).

\paragraph{Estimators.} Both estimators used XGBoost for the sequential propensities and the outcome regression, differing only in the active set supplied to the one-step correction (Section~\ref{sec:estimation}).
\begin{itemize}
\item \textbf{Active set:} the one-step estimator with a vanishing active-set threshold $\epsilon_n=n^{-0.4}$, which recovers the active set from the estimated marginals, and a minimum active-set size of one (Theorem~\ref{thm:onestep}).
\item \textbf{Standard:} the known-active-set estimator with $\cK(\bx)=\{1,\ldots,m\}$ (no thresholding, Corollary~\ref{cor:onestep-fixed}).
\end{itemize}
An additional run of the fixed-threshold estimand of Appendix~\ref{sec:adaptive} with $\epsilon=0.1$ tracked both estimators closely (per-cell signed relative bias within $0.6$ percentage points of theirs). It targets a different quantity and is not reported further.

\paragraph{Metrics.} We report bias, empirical standard error, and Wald coverage (Section~\ref{sec:sims}) at the fully-crossed $(m,\kappa,\text{method})$ level. Summaries below describe the distribution across cells and never average a reported quantity over settings. Every fit produced a valid estimate. Empirical standard error followed the theoretical $n^{-1/2}$ rate: the log--log slope within each of the $50$ cells had median $-0.504$ (interquartile range $[-0.511,-0.501]$, full range $[-0.523,-0.485]$, as shown in Figure~\ref{fig:study2_scaling}), and was essentially flat in $m$ within each $\kappa$ once the outcome signal is normalized by $m$ (Figure~\ref{fig:study2_m_beta}). Bias was small throughout and driven more by sample size than dimension or method: median absolute relative bias fell with sample size (from $1.6\%$ at $n=200$ to $0.2\%$ at $n=5000$, with worst cells at $2.8\%$ and $0.4\%$) and rose modestly with the number of treatments (from $0.3\%$ at $m=2$ to roughly $0.9\%$ at $m=6$), while within every cell the two estimators differed by at most $0.8$ percentage points. Per-cell Wald coverage stayed near nominal across the full range of dimension studied (median $0.949$ at $m=2$ and $0.946$ at $m=6$, with full range $0.922$--$0.963$ and the lowest cell at $m=5$, small $n$, and strong imbalance). Analytic standard errors were well calibrated in every cell (per-cell ratio $0.95$--$1.10$, median $1.02$), confirming coverage tracks the small residual bias rather than any variance-estimation failure.

\section{Application results}\label{app:appresults}

This appendix reports the estimated counterfactual mean costs behind Figure~\ref{fig:application}, the contrast and standard-error checks cited in Section~\ref{sec:example}, and the eligibility patterns used for stratification.

\paragraph{Estimates.} Table~\ref{tab:application} gives the estimates for all nine values of $\delta$ and all four cost outcomes. For total cost, the contrast between the extreme scenarios, $\delta=2$ versus $\delta=0.5$, is $-\$1{,}693$ (SE \$279). This standard error cannot be recovered from Panels A and B, because it depends on the covariance between the estimates at $\delta=0.5$ and $\delta=2$. We computed it from the difference of the two estimated influence functions, which share one set of cross-fitted nuisance estimates.

\begin{table}[h]
\centering
\caption{Estimated counterfactual mean healthcare cost $\hat\psi(\bdelta) = \E[Y^{\Q_{\bdelta}}]$ (in dollars) under uniform odds-scaling interventions, for total, inpatient, and outpatient cost and, as a non-additive overlay, mental-health (MH) cost. Standard errors in parentheses: Panel~A gives the level at each $\delta$, and Panel~B the difference from $\delta=1$, whose standard error accounts for the strong positive correlation between estimates at different $\delta$ (all $\delta$ share the same cross-fitted nuisance estimates) and is therefore much smaller than the level standard errors. Total, inpatient, and outpatient cost are each estimated separately, so the components need not sum exactly to the total. MH cost is a non-additive overlay spanning both. $\delta = 1$ corresponds to no intervention. Estimates use the cross-fitted one-step estimator with Kullback--Leibler divergence and $n = 159{,}466$ beneficiaries.}
\label{tab:application}
\begin{tabular}{ccccc}
\hline
$\delta$ & Total cost & Inpatient cost & Outpatient cost & MH cost \\
\hline
\multicolumn{5}{l}{\emph{Panel A: counterfactual mean cost $\hat\psi(\bdelta)$ (SE)}} \\
0.500 & 34,080 (393) & 23,608 (384) & 10,480 (54) & 12,574 (177) \\
0.667 & 33,700 (364) & 23,172 (355) & 10,532 (54) & 12,450 (157) \\
0.833 & 33,407 (349) & 22,837 (339) & 10,572 (54) & 12,364 (149) \\
0.952 & 33,236 (343) & 22,641 (333) & 10,595 (54) & 12,317 (146) \\
1.000 & 33,174 (341) & 22,571 (331) & 10,603 (55) & 12,301 (146) \\
1.050 & 33,112 (340) & 22,501 (330) & 10,611 (55) & 12,285 (146) \\
1.200 & 32,948 (338) & 22,315 (328) & 10,632 (55) & 12,244 (147) \\
1.500 & 32,689 (340) & 22,022 (330) & 10,665 (55) & 12,182 (152) \\
2.000 & 32,387 (353) & 21,683 (344) & 10,703 (54) & 12,116 (166) \\
\hline
\multicolumn{5}{l}{\emph{Panel B: difference from $\delta = 1$, $\hat\psi(\bdelta) - \hat\psi(\mathbf 1)$ (SE)}} \\
0.500 & 906 (146) & 1,038 (144) & $-$124 (18) & 274 (82) \\
0.667 & 527 (84) & 601 (83) & $-$71 (10) & 149 (46) \\
0.833 & 234 (38) & 266 (37) & $-$31 (4) & 63 (20) \\
0.952 & 62 (10) & 70 (10) & $-$8 (1) & 16 (5) \\
1.000 & -- & -- & -- & -- \\
1.050 & $-$61 (10) & $-$70 (10) & 8 (1) & $-$16 (5) \\
1.200 & $-$225 (37) & $-$255 (37) & 29 (4) & $-$57 (20) \\
1.500 & $-$485 (83) & $-$548 (82) & 62 (9) & $-$118 (43) \\
2.000 & $-$787 (142) & $-$888 (140) & 100 (14) & $-$185 (73) \\
\hline
\end{tabular}

\end{table}

\paragraph{Standard-error check.} Table~\ref{tab:ifboot} compares the analytic standard errors for total cost with a bootstrap of the estimated influence function. We resampled beneficiaries with replacement $2{,}000$ times and recomputed the mean of the resampled influence-function values, holding the fitted nuisance estimates fixed. The bootstrap standard error is within $3\%$ of the analytic one for all seven quantities, and the endpoints of the percentile intervals are within \$35 of the Wald endpoints. Because the nuisance estimates are held fixed, this checks the variance of the influence function and not the variability due to nuisance estimation or cross-fitting. The estimated influence functions are also very heavy tailed (sample kurtosis between about $800$ and $4{,}600$), so the reported standard errors are asymptotic approximations.

\begin{table}[ht]
\centering
\caption{Total cost: analytic standard error versus influence-function bootstrap standard error ($2{,}000$ resamples, nuisance estimates held fixed). The last column is the bootstrap standard error divided by the analytic standard error.}
\label{tab:ifboot}
\begin{tabular}{lrrrr}
\hline
Quantity & Estimate & Analytic SE & Bootstrap SE & Ratio \\
\hline
Level, $\delta=0.5$ & 34{,}080 & 393 & 388 & 0.988 \\
Level, $\delta=1$ & 33{,}174 & 341 & 334 & 0.980 \\
Level, $\delta=1.5$ & 32{,}689 & 340 & 332 & 0.975 \\
Level, $\delta=2$ & 32{,}387 & 353 & 344 & 0.974 \\
Difference from $\delta=1$, $\delta=0.5$ & 906 & 146 & 144 & 0.983 \\
Difference from $\delta=1$, $\delta=1.5$ & $-$485 & 83 & 82 & 0.993 \\
Difference from $\delta=1$, $\delta=2$ & $-$787 & 142 & 141 & 0.997 \\
\hline
\end{tabular}
\end{table}

\paragraph{Eligibility patterns.} Table~\ref{tab:patterns} lists the $15$ eligibility patterns on which estimation is stratified. Within a pattern every beneficiary is eligible for the same subset of the six indicators, the intervention scales the odds of only those indicators, and the stratum estimates are pooled with weights proportional to the pattern sizes (Corollary~\ref{cor:onestep-fixed}). The two single-indicator patterns, cervical cancer screening only and antipsychotic adherence only, contain $55\%$ of the cohort ($88{,}144$ beneficiaries). For these patterns the intervention reduces to the scalar incremental intervention.

\begin{table}[ht]
\centering
\caption{The $15$ eligibility patterns among $n=159{,}466$ beneficiaries, ordered by size. A dot marks an indicator for which the pattern is eligible. CCS: cervical cancer screening; FUH30 and FUH7: follow-up after hospitalization within 30 and 7 days; AMM-c and AMM-a: antidepressant medication management, continuation and acute phases; SAA: adherence to antipsychotic medications. Percentages are of the cohort and may not sum to $100$ because of rounding.}
\label{tab:patterns}
\footnotesize
\begin{tabular}{rccccccrr}
\hline
Pattern & CCS & FUH30 & FUH7 & AMM-c & AMM-a & SAA & $N$ & \% \\
\hline
1 & $\bullet$ &  &  &  &  &  & 64{,}123 & 40.2 \\
2 &  &  &  &  &  & $\bullet$ & 24{,}021 & 15.1 \\
3 &  &  &  & $\bullet$ & $\bullet$ &  & 16{,}779 & 10.5 \\
4 & $\bullet$ &  &  & $\bullet$ & $\bullet$ &  & 16{,}437 & 10.3 \\
5 & $\bullet$ &  &  &  &  & $\bullet$ & 12{,}798 & 8.0 \\
6 &  & $\bullet$ & $\bullet$ &  &  & $\bullet$ & 6{,}570 & 4.1 \\
7 &  & $\bullet$ & $\bullet$ &  &  &  & 4{,}776 & 3.0 \\
8 & $\bullet$ & $\bullet$ & $\bullet$ &  &  &  & 3{,}097 & 1.9 \\
9 & $\bullet$ & $\bullet$ & $\bullet$ &  &  & $\bullet$ & 2{,}989 & 1.9 \\
10 &  &  &  & $\bullet$ & $\bullet$ & $\bullet$ & 1{,}985 & 1.2 \\
11 &  & $\bullet$ & $\bullet$ & $\bullet$ & $\bullet$ &  & 1{,}920 & 1.2 \\
12 &  & $\bullet$ & $\bullet$ & $\bullet$ & $\bullet$ & $\bullet$ & 1{,}141 & 0.7 \\
13 & $\bullet$ &  &  & $\bullet$ & $\bullet$ & $\bullet$ & 1{,}131 & 0.7 \\
14 & $\bullet$ & $\bullet$ & $\bullet$ & $\bullet$ & $\bullet$ &  & 1{,}091 & 0.7 \\
15 & $\bullet$ & $\bullet$ & $\bullet$ & $\bullet$ & $\bullet$ & $\bullet$ & 608 & 0.4 \\
\hline
\end{tabular}
\end{table}

\section{Baseline covariates}\label{app:covariates}

The analysis uses $66$ baseline covariates, listed by group in Table~\ref{tab:covariates}. All are constructed from Medicaid claims and enrollment data over the year before the index date, except the calendar date at the start of the baseline year. A diagnosis-based covariate requires a qualifying ICD-10 code on at least one primary acute inpatient discharge, or on at least two primary or secondary outpatient or emergency-department claims from different service dates within that year. Setting (inpatient versus outpatient or emergency department) is identified from New York State Office of Mental Health setting identifiers.
\begin{table}[h]
\centering
\caption{Baseline covariate groups, number of variables, and definitions.}
\label{tab:covariates}
\footnotesize
\begin{tabular}{p{0.23\textwidth}p{0.07\textwidth}p{0.60\textwidth}}
\hline
Group & Vars. & Definition \\
\hline
Age & 1 & Continuous, at the index date. \\
Sex & 1 & Female or male, at the index date. \\
Race/ethnicity & 1 & Black, Hispanic, White, other, unknown. \\
Aid category & 1 & Medicaid aid category, including SSI disability eligibility. \\
Acute psychiatric care & 4 & Any acute inpatient stay for a psychiatric diagnosis and total inpatient days. Any emergency-department visit for a psychiatric diagnosis and the number of visits. \\
Suicide attempt or self-harm & 1 & Any claim, in any setting, with a qualifying nonfatal suicide attempt or self-harm code. \\
Psychiatric comorbidity & 8 & Schizophrenia, bipolar disorder, severe major depressive disorder, other major depressive disorder, other psychotic conditions, other nonacute and nonorganic mental illness, substance use disorders, and a composite psychiatric comorbidity indicator. \\
Dementia & 1 & Dementias. \\
Organic disorders & 1 & Nonacute organic mental and cognitive disorders. \\
Chronic medical conditions & 21 & Dyslipidemia, hypertension, type 2 diabetes, coronary artery disease, ischemic heart failure, cerebrovascular disease, other vascular disease, remote cardiovascular disease, metabolic syndrome, obesity, prediabetes, gestational diabetes, tobacco use disorder, personal or family risk factors, a composite metabolic-risk indicator, type 1 diabetes, other diabetes (two indicators), conditions potentially caused by antipsychotic drugs, a composite chronic-comorbidity indicator, and non-diabetes conditions potentially treated with antidiabetic drugs. \\
Social risk & 4 & Homelessness or housing instability, lack of adequate food, extreme poverty, and all other ICD-10-CM Z55--Z65 codes. \\
Other service use & 6 & Number of psychiatric outpatient visits, number of non-psychiatric outpatient visits, non-psychiatric inpatient use (any use and days), and non-psychiatric emergency-department use (any use and number of visits), excluding visits that lead to a non-psychiatric inpatient stay. \\
Metabolic testing & 4 & At least one qualifying lipid test, at least one qualifying glucose test, both, and either. \\
Drugs with potential safety or effectiveness effects & 9 & At least one claim for a drug in each class: antidiabetic, antilipemic, cardiovascular or hypertension, weight or appetite, antipsychotic, mood stabilizer, antidepressant, other psychiatric, and other drugs. \\
Drugs with suicidality effects & 2 & At least one claim for a psychiatric drug with suicidality effects. At least one claim for a non-psychiatric drug with suicidality effects (the latter is zero for every beneficiary). \\
Calendar time & 1 & Calendar date at the start of the baseline year, to adjust for calendar time. \\
\hline
\end{tabular}
\end{table}

\end{document}